%% file: draft.tex
\documentclass[aps, 10pt, prl, twocolumn, superscriptaddress, nobibnotes, noshowpacs, preprintnumbers, floatfix]{revtex4-2}
\usepackage{style}
\usepackage{aas_macros}
\usepackage[normalem]{ulem}

\def\ttitle{Constraining Ultralight Scalars with Black Hole Binary Mergers in Galactic Nuclei}
\begin{document}
% \title{Probing Superradiance Clouds with Black Hole Binary Mergers in Galactic Nuclei}
\title{\ttitle}
\author{Majed Khalaf \orcidlink{0000-0001-5537-9992}}
\email{majed.khalaf@mail.huji.ac.il}
\affiliation{Racah Institute of Physics, Hebrew University of Jerusalem, 91904 Jerusalem, Israel}

\author{Eric Kuflik \orcidlink{0000-0003-0455-0467}}
\affiliation{Racah Institute of Physics, Hebrew University of Jerusalem, 91904 Jerusalem, Israel}

\author{Alessandro Lenoci \orcidlink{0000-0002-2209-9262}}
\email{alessandro.lenoci@mail.huji.ac.il}
\affiliation{Racah Institute of Physics, Hebrew University of Jerusalem, 91904 Jerusalem, Israel}

\author{Nicholas Chamberlain Stone \orcidlink{0000-0002-4337-9458}}
\affiliation{Racah Institute of Physics, Hebrew University of Jerusalem, 91904 Jerusalem, Israel}
\affiliation{Department of Astronomy, University of Wisconsin, Madison, WI 53706, USA}

\author{Ofri Telem \orcidlink{0000-0002-3120-0975}}
\affiliation{Racah Institute of Physics, Hebrew University of Jerusalem, 91904 Jerusalem, Israel}
% \date{\today}
\begin{abstract}
Ultralight scalars can form long-lived, macroscopic bound states around spinning black holes, known as superradiance clouds. These clouds provide an additional channel for energy dissipation during close encounters, enhancing the black hole binary formation and merger rates in dense environments such as galactic nuclei. 
We show that the rate of mergers with mass ratio below $3/10$ in the LIGO-Virgo-KAGRA GWTC-5 catalog can probe ultralight scalars in the mass range $[4\times 10^{-14},10^{-13}]$ eV, complementing existing strategies based on black hole spin measurements.
\end{abstract}
\maketitle

\textbf{{Introduction}}---Ultralight scalar particles with masses $\mu \ll {\rm eV}$ are well motivated in a broad class of extensions of the Standard Model of particle physics: dynamical solutions to the strong CP problem, such as the QCD axion~\cite{Peccei:1977hh,Kim:1979if,Weinberg:1977ma,Wilczek:1977pj,Shifman:1979if,Zhitnitsky:1980tq,Dine:1981rt}, as well as mechanisms addressing the electroweak hierarchy problem~\cite{Graham:2015cka,Arvanitaki:2016xds,Geller:2018xvz,Arkani-Hamed:2020yna,TitoDAgnolo:2021nhd,Chatrchyan:2022dpy,Csaki:2024ywk}. These particles also provide viable dark matter candidates~\cite{Preskill:1982cy,Abbott:1982af}. Their astrophysically large de Broglie wavelength can produce distinctive phenomenological signatures, as in wave or fuzzy dark matter~\cite{Hu:2000ke,Goodman:2000tg,Hui:2016ltb}, even if its interactions with ordinary matter are purely gravitational. 

Today ultralight particles are the focus of an extensive experimental search campaign, though their detection is notoriously challenging. These direct searches can be complemented by astrophysical probes relying on their intriguing interplay with spinning black holes (BHs). In particular, perturbations of massive bosonic fields are unstable in a background described by the Kerr metric. If the Compton wavelength $\mu^{-1}$ of an ultralight boson matches the size of the BH horizon, superradiant instability can generate a {superradiance cloud} (SC) around the BH \cite{Press:1972zz, Bardeen:1972fi}. This macroscopic state forms at the expense of the rest mass and angular momentum of the BH. SCs are unstable and are gradually depleted through their annihilation into {\it e.g.} gravitational waves (GWs)~\cite{Yoshino:2013ofa,Brito:2014wla}. 

Superradiance (SR) therefore predicts a depletion of  rapidly spinning BHs in specific regions of the BH mass--spin plane~\cite{Arvanitaki:2010sy}. 
Spin measurements can therefore be used to constrain boson masses: the observation of sufficiently old, rapidly spinning BHs in a region that would otherwise be depleted by SR rules out the corresponding boson mass range. Historically, these constraints relied on spin measurements of stellar-mass BHs in X-ray binaries \cite{Arvanitaki:2010sy,McClintock+14}, and supermassive BHs (SMBHs) in active galactic nuclei \cite{Reynolds13}; more recently, GW astronomy provides information on the spins of merging BHs~\cite{Caputo:2025oap,Aswathi:2025nxa,Roy:2025qaa,Ning:2026ebu,LIGOScientific:2026pwx,Kou:2026naz}.

\begin{figure}[t]
  \centering
\includegraphics[width=\linewidth]{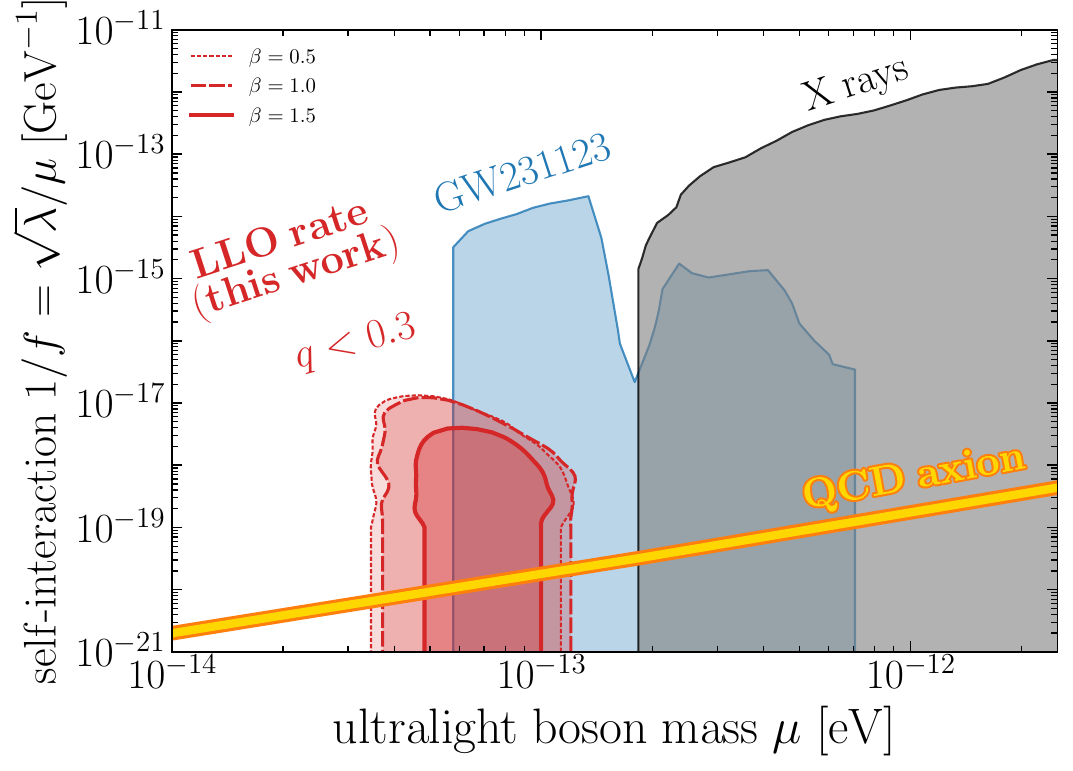}
  \caption{Exclusion plot in ultralight boson mass $\mu$ and self-interaction strength (expressed as $1/f = \sqrt{\lambda}/\mu$). The red contours represent the excluded region requiring the predicted rate of detections with mass ratio less than 3/10 to not exceed the counts from the GWTC-5 catalog~\cite{LIGOScientific:2026wfs} at 90\% confidence level. We assume a BH mass function in GNs $\propto M^{-\beta}$ with different values of $\beta=\{0.5,1,1.5\}$, increasing from the most aggressive to the most conservative, shown as different contours. We also present the constrains obtained from spin-down arguments, from a GW event~\cite{Caputo:2025oap,Aswathi:2025nxa} in blue and inferred from X-ray measurements~\cite{Witte:2024drg} in gray. }
\label{fig:summary_mu_inv_f}
\end{figure}

Beyond spin-down signatures, SCs may be probed through their direct or indirect effects on the BH environment. For example, Ref.~\cite{Khalaf:2024nwc} proposed using complementary spin measurements to search for evidence for the cloud's extended mass distribution. Likewise, the orbits of S-stars around Sgr A$^*$ are sensitive tracers of any extended potential \cite{AlushStone22}, providing a possible probe of an SC~\cite{GRAVITY:2023azi,GRAVITY:2023cjt,TomaselliCaputo26}. Continuous GW emission from boson annihilations offers a further avenue for detection~\cite{Arvanitaki:2016xds,Abbott22+,Ellis:2026gwt}.

Clouds can also affect the GW signal from binary BH inspirals. However, LIGO-Virgo-KAGRA (hereafter LVK) observes the final orbits of the inspiral, far inside the cloud's Bohr radius. Earlier stages may have already ionized most of the SC, leaving little impact on LVK-band GWs (however, see~\cite{Roy:2025qaa}). LISA~\cite{LISA:2017pwj,LISA:2022yao,LISA:2024hlh} could have the upper hand in direct SC detection \cite{Tomaselli:2024dbw,Tomaselli:2024bdd,Boskovic:2024fga,Boskovic:2025ixx,Kim:2025wwj,Boskovic:2026eth}, due to its remarkable phase accuracy and ability to probe larger radii, both in extreme mass ratio inspirals and in multi-band observations of stellar mass BH binaries \cite{Sesana:2016ljz}.

%az
%\al{This part can benefit from Nick's input: Describe the LVK catalog and the role of eccentric binaries and different binary formation channels}

The most recent LVK catalog, GWTC-5, covers observing runs through O4b and includes 390 candidate binary inspirals, the overwhelming majority of which are binary BH candidates \cite{LIGOScientific:2026wfs}. The binary BH population is multimodal in mass and has distinct subpopulations \cite{LIGOScientific:2026ctl}, the largest of which features low-mass BHs ($M \approx 10 M_\odot$) and comparable mass ratios ($q \equiv M_2/M_1\approx 1$) \cite{Guttman:2026cnv, Ray:2026uur}. A notable tail of more extreme events does exist, however, with the largest clearly measured masses exceeding $100 M_\odot$ \cite{LIGOScientific:2025rsn} and the most extreme mass ratios reaching as low as $q\approx 1/10$ \cite{LIGOScientific:2020zkf}. Galactic nuclei (GNs) could be ideal environments for the formation of these heavy BHs, {\it e.g.} in repeated (hierarchical) mergers \cite{Yang+19, Tagawa+21, Vaccaro:2026nkp}. Binaries can form via two-body captures mediated by GW emission (GW-capture), either in vacuum \cite{OLeary:2008myb, Gondan:2017wzd, Rasskazov:2019gjw} or in active \cite{Samsing+22} GN. The  majority of such binaries are characterized by large eccentricity at formation, and a smoking gun of such channel would be a BH inspiral with high eccentricity detectable in the LVK band $e(10\ {\rm Hz}) > 0.03$~\cite{OLeary:2008myb, Gondan:2017wzd}. Although several claims have been reported \cite{RomeroShaw+20, Romero-Shaw:2022xko}, there is as yet no conclusive evidence for individual events with measurable nonzero eccentricity in the LVK band, while recent work has suggested population-level evidence for an eccentric subpopulation \cite{Gupte+25}. 

In this work, we probe SCs through their effect on BH binary formation in GNs. Past work explored BH binary formation in GNs via GW-capture~\cite{OLeary:2008myb, Rasskazov:2019gjw}; the presence of SCs can enhance capture by providing an extra channel for energy loss~\cite{Zhang:2019eid,Tomaselli:2023ysb}. We quantify this enhancement, and, for the first time, establish clear predictions for LVK excluding some ultralight scalar parameter space. Compared to existing spin-based superradiance constraints, the scenario we develop here has a completely different set of strengths and weaknesses. For example, our arguments based on SC-mediated capture are unaffected by the BH Salpeter time and by the systematics that characterize X-ray or GW spin measurement techniques. However, our scenario carries its uncertainties, elaborated below and in the Supplemental Material (SupM).
 
 In Fig.~\ref{fig:summary_mu_inv_f}, we show the boson parameter space that is probed by comparing the SC-enhanced predictions of $q<0.3$ detection rates with LVK data. As usual with exclusions coming from the SR phenomenon, smaller values of $f$ (i.e. stronger self-interactions) are characterized by lighter SCs, therefore the constraints are stronger for weaker self-interactions. Our results are sensitive to the mass-function exponent $\beta$ of BHs residing in GN, defined as ${\rm d}N/{\rm d}M\propto M^{-\beta}$. The constraining power of our analysis comes mostly from heavy BHs, so smaller $\beta$ values enlarge the exclusion range. For fiducial $\beta$ choices, we exclude boson masses smaller than current spin constraints, either from GW~\cite{Caputo:2025oap,Aswathi:2025nxa,Ning:2026ebu,LIGOScientific:2026pwx} or X-ray astronomy \cite{Cardoso:2018tly}. Interestingly, our constraints based on detection rates are in tension with the values of $(\mu,f)$ typical for the QCD axion between $[4\times 10^{-14},10^{-13}]$ eV. 
% This is complementary to the work of Ref.~\cite{Ning:2026ebu}, where similar values of $\mu$ are constrained via a statistical analysis of spin measurements of inspiral components in the GWTC-5~\cite{LIGOScientific:2026wfs} catalog. 
Throughout this work, we use natural units $\hbar=c=1$.

\textbf{{Superradiance clouds}}---Consider a Kerr BH with mass $M_1$, gravitational radius $r_{\rm g}=GM_1$ and dimensionless spin $\chi=J/GM^2_1 < 1$. In this background, a scalar field of mass $\mu$ has a hydrogen-like spectrum of bound states, characterized by a fine-structure constant 
\begin{align}
  \alpha = GM_1 \mu = 0.075 \scalings{M_1}{100\ M_\odot}\scalings{\mu}{10^{-13}\ {\rm eV}}\ .
\end{align}
The states are labeled $|n\ell m\rangle$ with principal, angular and magnetic-like quantum numbers $n,\ell,m$, respectively. The in-going boundary condition at the event horizon implies a complex energy spectrum $\omega_{n\ell m}=E_{n\ell m}+i\Gamma_{n\ell m}$. In the non-relativistic approximation $\alpha\ll n/2$, the real part of the spectrum $E_{n\ell m}/\mu\simeq 1 - \alpha^2/2n^2 +{\cal O}(\alpha^4)$ closely resembles the spectrum of the hydrogen atom. The imaginary part is $\Gamma_{n\ell m}\propto (m\Omega_+-E_{n\ell m})$~\cite{Dolan:2007mj,Baumann:2019eav}, with $\Omega_+=\chi/[2r_{\rm g}(1+\sqrt{1-\chi^2})]$ the angular velocity of the outer horizon. If $\Gamma_{n\ell m}>0$ the state $\ket{n\ell m}$ is exponentially amplified by the SR instability over the timescale $\tau_{n\ell m}\sim \left(2\Gamma_{n\ell m}\right)^{-1}$. SR can occur only for states with $m>0$, i.e. the magnetic quantum number of the state is aligned with the direction of the BH spin, so that a transfer of angular momentum from the BH to the cloud is possible. As the BH spins down, $\Gamma_{n\ell m}$ decreases until the process saturates at $\Gamma_{n\ell m}\simeq 0$.
% this happens at a BH spin of $\chi_m = [\alpha_m/m]/ [\alpha^2_m/m^2+1/4]$. 
If instead $\Gamma_{n\ell m}<0$ to begin with, the population of the state $\ket{n\ell m}$ decays to the BH on a timescale $1/(2|\Gamma_{n\ell m}|)$.

Each bound state in an SC is described by a wave function $\psi_{n\ell m}({\bf r})$, that is hydrogen-like for $\alpha \ll n/2$. The probability density is peaked at radii $r_{\rm c} = n^2 r_{\rm g}/\alpha^2$ and the whole state has a mass $M_{n\ell m} = \mu \int d^3r|\psi_{n\ell m}|^2 $, that can reach up to 10\% of the BH mass depending on the initial BH mass and spin $M_{1,\rm i},\chi_{\rm i}$. Due to its hydrogen-like state, the SC enjoys a scale invariance property when radii are expressed as multiples of Bohr radii $r_{\rm B}\equiv(\mu \alpha)^{-1}=r_{\rm g}/\alpha^2$. For bound states with principal quantum number $n$, the SR production rate is the largest for the states with maximal magnetic quantum number, namely, $m=\ell=n-1$. 

In many models, ultralight scalars $\phi$ possess an attractive quartic self-interaction $(\lambda/4!)\,\phi^4$. For example, in axion-like models with a potential $\mu^2f^2 \cos (\phi/f)$, where $f$ is the scalar decay constant, expanding the potential around its minimum yields a quartic coupling $\lambda = \mu^2/f^2$. 
Self-interactions couple the system's eigenstates and transfer energy between modes. Owing to the macroscopically large occupation numbers $N_{n\ell m}$ in the SC, even extremely weak self-interactions can have significant effects. While the general dynamics are complex~\cite{Witte:2024drg}, they simplify considerably for $\alpha\lesssim0.2$~\cite{Baryakhtar:2020gao,Witte:2024drg}. We restrict our analysis to this regime, where the SC of self-interacting scalars can be effectively described as a lossy two-level system consisting of the two states $\ket{211}$ and $\ket{322}$.
% ---the only ones with sizable occupation numbers within a Hubble time.
These states can transition to unstable modes that decay into the BH, radiate energy to infinity, or annihilate into GWs. The system can be described by a set of ODEs~\cite{Baryakhtar:2020gao, Caputo:2025oap} for the occupation numbers $N_{211}, N_{322}$, the fine structure constant $\alpha$ and spin $\chi$. Given values of $\mu, f$ and initial conditions $\chi_{\rm i},\alpha_{\rm i}$, we solve these ODEs for the time-dependent occupation numbers and the BH parameters $M_1(t)=\alpha(t)/G\mu,\chi(t)$. We then define the SC mass ratio as $q_{\rm c}(t) = \max[M_{211}(t),M_{322}(t)]/M_1(t)$. 
% See the SupM for the SC ODEs and their numerical solution.

\textbf{{SC-enhanced capture}}---Consider a close (hyperbolic) encounter between two BHs of masses $M_2<M_1$, with impact parameter $b$ and relative velocity $w$. In GN environments, these encounters are almost always very close to parabolic~\cite{OLeary:2008myb}, and can form an (initially) highly eccentric binary ($e_0\simeq 1$), provided that they have efficient energy loss channels for capture.
% For this reason, we call these encounters ``capture candidates". \al{we dont}
Here we investigate how the existence of an SC around the heavier BH affects the capture rate in these events. 
The system is parameterized by its energy $E_0$ and angular momentum $L_0$ \textit{after} the first near-parabolic encounter. In particular, $E_0 = E_{\rm k} + \delta E$ and $L_0 = L_{\rm k} + \delta L$ where $E_{\rm k} = \mu_r w^2/2$ and $L_{\rm k} = \mu_r b w$ denote the (kinetic) energy and angular momentum, respectively, \textit{before} the encounter. The total and reduced masses are defined as $M_{\rm tot} = M_1 + M_2$ and $\mu_r = M_1 M_2/M_{\rm tot}$, respectively.
When $E_0<0$, the first encounter results in a capture. The angular-momentum loss $\delta L$ is negligible~\cite{OLeary:2008myb}, and we neglect it in our analysis.
% (see SupM for an estimate of $\delta L$ in the presence of an SC).
% To single out the effect of the SC on the capture, 
We split the energy transfer as $\delta E = \delta E_{\rm GW} + \delta E_{\rm c}$, with the first term denoting the losses to GWs and the the second indicating the energy transferred to the SC via tidal interactions. 
% \begin{figure}
%   \centering
% \includegraphics[width=\linewidth]{wmax_enhancement_combined.pdf}  \caption{\textbf{Left}: the maximum relative velocity that allows capture, $w_{\rm max} = \sqrt{2(1+q) |\delta E|/qM}$, as a function of normalized pericenter distance $r_{\rm p}/r_{\rm g}$ for an SC in $\ket{211}$ state (top left) or $\ket{322}$ state (bottom left). The gray shaded area shows the distances where the GW emission dominates the capture dynamics. \textbf{Right}: the $w_{\rm max}$ enhancement $\cal W$ over pure GW capture, for an SC in the $\ket{211}$ state (top right) or $\ket{322}$ state (bottom right) as a function of $r_{\rm p}/r_{\rm B}$. In all the plots we show the results varying the inclination of the orbit $\iota$ as visible in the color scale. Solid lines show where perturbativity $P_{\rm tot}<0.1$ is satisfied while dashed show when it is not.}
%   \label{fig:wmax}
% \end{figure}
The energy radiated via GWs is
% \begin{align}\label{eq:dE_GW}
${\delta E_{\rm GW}}/M_1 = -n_{\rm GW}q^2(1+q)^{1/2} \left( {r_{\rm g}}/{r_{\rm p}}\right)^{7/2}$~\cite{Peters:1963ux,Turner:1977ab},
% \end{align}
where $ n_{\rm GW}\equiv{85\pi}/{12\sqrt{2}}$, $q \equiv M_2/M_1$ is the mass ratio and $r_{\rm p}$ is the pericenter of the initial encounter
\begin{align}\label{eq:rp_b_w}
  r_{\rm p} = \bigg[\sqrt{\frac{1}{b^2}+ \frac{(GM_{\rm tot})^2}{b^4 w^4}} + \frac{GM_{\rm tot}}{b^2 w^2}\bigg]^{-1} \simeq \frac{b^2 w^2}{2GM_{\rm tot}}.
\end{align}
% The convention in Eq.~\eqref{eq:dE_GW} is such that energy lost from the binary has negative sign. 

The %second term is the 
total energy transferred to the cloud %, which 
 can be written in the simple form (see the SupM and \cite{Tomaselli:2023ysb})
\begin{align}
  \frac{\delta E_{\rm c}}{M_1} = -q_c \sum_{j}\frac{\Delta E_{js}}{\mu}P_{s j} \equiv -q_{\rm c}q^2 \alpha^2 {\cal E}_{s,q, \theta}\left(\frac{r_{\rm p}}{r_{\rm B}} \right),
 \end{align}
where the sum over $j$ denotes a sum over the bound states of the SC, and an integral over its unbound states. Here $q_c,M_1$ are implicit functions of time; we assume that the encounter interaction is effectively instantaneous over the timescales of SC evolution. $P_{sj}$ denotes the transition probability from $|s\rangle$ to a bound state $|j\rangle$, or the transition probability density to an unbound state $|j\rangle$.
% The formalism does not automatically make $P^{s}_{\rm tot}= \sum_j P_{sj}$ small, especially when $q\sim 1$. \al{Write about no-feedback assumption and no perturbation on the orbit}
The kernel ${\cal E}_{s,q, \theta}$ is an order-one function of the SR state $s=\{\ket{211},\ket{322}\}$, the mass ratio $q$ and the Euler angles of the orbital encounter $\theta=(\iota,\Omega, \omega)$: inclination $\iota$ (the angle between the spin direction of the primary BH and the orbit's angular momentum), the longitude of the ascending node $\Omega$ and argument of pericenter $\omega$. We computed ${\cal E}_{s,q,\theta}$ and tabulated it. Owing to the scaling properties of the cloud, the radial dependence enters through $r_{\rm p}/r_{\rm B}$.
Transitions to higher-energy states (with respect to $\ket{s}$) provide an additional energy sink, beyond GW, and therefore enhance capture. Conversely, transitions to lower-energy states transfer energy to the orbit and thus tend to suppress capture. The total net effect results in an enhanced capture.

In computing $\delta E_{\rm c}$, we employed first-order perturbation theory, neglecting both the feedback of other states on the cloud and perturbations to the secondary's orbit. The validity of this treatment requires the total transition probability, $P^{s}_{\rm tot}=\sum_j P_{sj}$, to remain small. As $r_{\rm p}$ approaches $r_{\rm c}$, this condition eventually breaks down, particularly for $q\sim1$. We (conservatively) restrict our analysis to the perturbative regime in which the self-consistency criterion $P^{s}_{\rm tot}\lesssim0.1$ is satisfied, while non-perturbative calculations (perhaps like in \cite{Guo:2025pea}) are left for future work.

We compute the differential capture cross section as follows. First, we parametrize our encounters using $(r_{\rm p}, w)$. For a given value of the pericenter $r_{\rm p}$, we determine the initial velocity $w_{\rm max}$ for which $0=E_0=\delta E (r_{\rm p}) + E_{\rm k}(w_{\rm max})$. Encounters with $w<w_{\rm max}$ lead to capture, while those with $w>w_{\rm max}$ do not. Expressing the result in terms of $M_1$ and $q$, we have $w_{\rm max}=\sqrt{2(1+q)|\delta E(r_{\rm p})|/qM_1}$. 
% This quantity is shown in the left panels of Fig.~\ref{fig:wmax} for some example configurations, where one can also see the impact of the SC in the binary formation relative to standard GW-capture.
Note that $\delta E(r_{\rm p})$, hence $w_{\rm max}$, depend on the Euler angles of the encounter, as well as on the BH age $t_{\rm BH}$ and initial BH spin $\chi_{\rm i}$ via the SC mass ratio $q_{\rm c}$. 
By inverting Eq.~\eqref{eq:rp_b_w} and differentiating, we can get the differential capture cross section at fixed $r_{\rm p}$,
\begin{align}\label{eq:diff_sigma}
  \frac{\d \sigma}{\d r_{\rm p}}=\frac{\d (\pi b^2)}{\d r_{\rm p}}\bigg|_{E_0<0}=\frac{2\pi GM_{\rm tot}}{w^2}\Theta[w_{\rm max}(r_{\rm p})-w]\,.
\end{align}
$w_{\rm max}$ is the sole quantity through which the SC effect enters. Over most of the $r_{\rm p}$ integration domain relevant to this work, capture is SC-dominated, such that $|\delta E_{\rm GW}| \ll |\delta E_{\rm c}|$. This can increase $w_{\rm max}$ by up to three orders of magnitude (see SupM). Consequently, at fixed $w$, the effective integration domain in $r_{\rm p}$, as selected by the Heaviside function in Eq.~\eqref{eq:diff_sigma}, is enlarged, leading to an enhanced capture cross section.

\textbf{{Detection rate of BH binary mergers}}---In order to set limits on SCs leading to enhanced binary capture rates, we now compute merger detection rates in GW interferometers such as LIGO, following a methodology similar to that used in~\cite{Gondan:2017wzd,Rasskazov:2019gjw}. More details on this calculation are shown in the SupM. To describe the binary formation, as a first step, we take an ensemble average $\left\langle \d \sigma/\d r_{\rm p}\right\rangle$ of the differential cross section over $t_{\rm BH}$, the initial spin $\chi_{\rm i}$ and the encounter Euler angles $(\iota,\Omega, \omega)$.
% These quantities determine the SC or the energy transfer between a perturber and the SC during binary formation. 
They are sampled as follows: the BH age $\log_{10}(t_{\rm BH}/{\rm Gyr})\in {\cal U}[-1,1]$ from a log-uniform distribution; the Euler angles isotropically over the sphere $(\cos\iota,\Omega, \omega) \in {\cal U}[-1,1]\times{\cal U}[0,2\pi)\times{\cal U}[0,2\pi) $, the initial spin (before SR) of the primary BH as $\chi_{\rm i}\in {\cal U}[0,0.998]$. 

% We compute the differential capture rate for a pair of BHs with masses $M_1$ and $M_2$ with relative velocity $w$, initial pericenter distance of $r_{\rm p}$ at distance $r$ from the SMBH. The stellar populations in the GN inhabit a quasi-spherical nuclear star cluster characterized by a number density profile $n_M(r) = r^{-a(M)}$, where $a(M)$ is a monotonically increasing representation of mass segregation \cite{Bahcall:1977ab,Freitag:2006qf}. Concretely, we use $a(M)=3/2+ p_0 M/M_{\rm max}$ for BHs, where $p_0\simeq{0.5}$~\cite{OLeary:2008myb} is the mass-segregation parameter and $M_{\rm max}$ is the highest possible BH mass in the GN. This description is assumed to be valid for $r\in [r_{\rm min}(M_1,M_2, M_{\rm SMBH}), r_{\rm max}(M_{\rm SMBH})]$ and $w<v_{\rm max}(r)$ where $v_{\rm max}$ is the escape velocity from the SMBH gravitational potential. $r_{\rm min}$ can be obtained from imposing the relaxation time to be of the order of the inspiral timescale~\cite{Gondan:2017wzd,Rasskazov:2019gjw} and $r_{\rm max}$ which is usually the radius of influence of the SMBH, i.e. the radius at which the mass of the SMBH equals twice the total included stellar mass. While commonly a relation between the SMBH mass and the velocity dispersion is used to infer the radius of influence~\cite{Rasskazov:2019gjw}, we exploit an empirical relation motivated by actual observations~\cite{Hannah+24}, and in good agreement with the literature~\cite{Thomas+16}, reading $r_{\rm max} = 2.25 ~{\rm pc} ~ \left( {M_{\rm SMBH}}/{4 \times 10^6 M_\odot} \right)^{0.75}$.
We compute the differential capture rate for a pair of BHs with masses $M_1$ and $M_2$, relative velocity $w$, and initial pericenter distance $r_{\rm p}$, at a distance $r$ from the SMBH. The stellar populations in the GN are assumed to reside in a quasi-spherical nuclear star cluster with a mass-dependent number-density profile $n_M(r)\propto r^{-a(M)}$, where the slope $a(M)$ increases monotonically with mass as a consequence of mass segregation~\cite{Bahcall:1977ab,Freitag:2006qf}. For the BH population, we adopt $a(M)=3/2+p_0 M/M_{\rm max}$, where $p_0\simeq 0.5$ is the mass-segregation parameter~\cite{OLeary:2008myb}, and $M_{\rm max}$ denotes the maximum BH mass present in the GN. We assume that this description is valid over the radial range $r\in[r_{\rm min}(M_1,M_2,M_{\rm SMBH}),r_{\rm max}(M_{\rm SMBH})]$ and for relative velocities $w<v_{\rm max}$, where $v_{\rm max}(r)$ is the local escape velocity from the SMBH potential. The inner cutoff $r_{\rm min}$ accounts for the depletion of compact objects sufficiently close to the SMBH, where GW-driven inspiral removes them faster than two-body relaxation can replenish the cusp; it is therefore determined by comparing the relaxation and inspiral timescales~\cite{Gondan:2017wzd,Rasskazov:2019gjw} (see SupM for the resulting expression). The outer cutoff $r_{\rm max}$ is identified with the SMBH radius of influence, conventionally defined as the radius within which the enclosed stellar mass is twice the SMBH mass. Rather than inferring this scale from an $M_{\rm SMBH}$--velocity-dispersion relation~\cite{Rasskazov:2019gjw}, we adopt the empirical relation motivated by observations~\cite{Hannah+24}, and consistent with the literature~\cite{Thomas+16}, $r_{\rm max}=3.57\,{\rm pc}\left[M_{\rm SMBH}/(4\times10^6\,M_\odot)\right]^{0.75}$.
Under the GN-environment assumptions, the differential rate can be written as
\begin{align}
  \frac{\d \Gamma_{M_1 M_2}}{\d r_{\rm p}} = 4\pi C_{M_1} C_{M_2}\int \d r\d w\,n_{M_1}n_{M_2} \left\langle \frac{\d \sigma}{\d r_{\rm p}}\right\rangle w.
\end{align}
The effect of the SC is fully contained in the Heaviside theta in the differential cross section term as in Eq.~\eqref{eq:diff_sigma}. Above, $C_{M} =C_{r}(M)C_{w}(M) N_{\rm BH}(M_{\rm SMBH}) F_{\beta}(M) $ contains normalization constants for a given $M$, the total number of BHs $N_{\rm BH}\simeq 2\times 10^{4}[M_{\rm SMBH}/(4\times 10^6 M_\odot)]$~\cite{Miralda-Escude:2000kqv}, and an empirical BH mass function $F_{\beta}\propto M^{-\beta}$~\cite{Alexander:2008tq} that gives unity when integrated in the range $M\in [M_{\rm min},M_{\rm max}]$. While $M_{\rm min}=5\ M_\odot$ is established~\cite{Ozel+10,LIGOScientific:2026wfs} both from X-ray and GW astronomy, $M_{\rm max}$ is uncertain. 
Given the dense environment of a GN favoring repeated mergers~\cite{Yang+19}, and the very composition of the LVK catalog~\cite{LIGOScientific:2025rsn,LIGOScientific:2026wfs}, including a substantial tail of components with large masses, we fiducially consider $M_{\rm max}=100\ M_\odot$. The exponent $\beta$ is also not well-established. It is reasonable to expect $\beta$ to be smaller than the exponent of the initial mass function (IMF) of massive stars~\footnote{The exponent may also be lower due to the top-heavy nature of \textit{in situ} star formation in the GN~\cite{Lu+13}.}, $\beta_{\rm IMF} = 2.35$, due to the repeated mergers and growth that naturally occur in active GN accretion disks, which would strongly flatten the distribution \cite{Gonglewski+26}, {\it e.g.} by $\Delta \beta \sim 1.3$~\cite{Yang+19}. We therefore explore values in $\beta\in [0.5, 1.5]$ with our fiducial value being $\beta = \beta_{\rm IMF}-\Delta\beta \simeq 1$. The astrophysical parameters $M_{\rm max}$, $\beta$, and the choice of $r_{\rm max}$ are our main sources of uncertainty, as they govern the density of heavy BH binaries in the GN, to which our capture rates are sensitive. This is intrinsic to the GN-environment and the GW-capture scenario suffers from the same uncertainties~\cite{Gondan:2017wzd,Rasskazov:2019gjw}.

Under the assumption that once a binary is formed, it is not disrupted by a third body encounter, which we find to be almost always the case if $r_{\rm p}\lesssim 10^4 GM_{\rm tot}$, the \textit{capture} rate is equivalent to a \textit{merger} rate. We ignore the impact of the SC on the binary evolution following the initial capture, which can change the inspiral timescale, eccentricity and GW waveforms (see \textit{e.g.}~\cite{Tomaselli:2024dbw,Tomaselli:2024bdd,Boskovic:2024fga,Tomaselli:2025jfo,Boskovic:2025ixx,Kim:2025wwj})---this is beyond the scope of the current analysis and we leave it for future work. Despite very eccentric orbits being far from understood, the recent literature hints to the SC-binary dynamics leading anyways to a merger~\cite{Guo:2025pea}. To find the total differential merger rate, we integrate the local rate over all the galaxies with GNs:
\begin{align}\label{eq:rate_all_GNs}
  \frac{\d {\cal R}_{M_1 M_2}}{\d r_{\rm p}} = \int_{10^{5}M_\odot}^{10^7M_\odot} \d M_{\rm SMBH} \frac{\d n_{\rm gal}}{\d M_{\rm SMBH}} \xi \frac{\d \Gamma_{M_1 M_2}}{\d r_{\rm p}}\ .
\end{align}
The integration range is chosen such that objects around such SMBHs have dynamically relaxed over less than a Hubble time~\cite{BarOr:2013ab}.
Here $\d n_{\rm gal}/\d M_{\rm SMBH}$ is the differential number density of galaxies per mass of their central SMBH, and we use the fit from Ref.~\cite{Aller:2002rp}. The local capture rate depends on several parameters, \textit{e.g.} $(\beta, M_{\rm max}, r_{\rm min}, r_{\rm max})$, that are GN-dependent and therefore scattered, as shown in observations~\cite{Hannah+24}. The formalism so far implicitly assumes that the used parameters are average values over all the GNs, thereby ignoring this variance. The parameter $\xi$ corrects for this intrinsic scatter. Considering the $M_{\rm SMBH}$--velocity-dispersion relation, \cite{Rasskazov:2019gjw} estimates $\xi =3$. We obtain $\xi\simeq 7.5$ by taking into account also the observed scatter in influence radii ~\cite{Hannah+24}, while conservatively neglecting denser outlier GN \cite{vanVelzen16} and other sources of intrinsic scatter (see SupM).

We can obtain the single binary \textit{detection} rate in a detector $\cal D$ by integrating Eq.~\eqref{eq:rate_all_GNs} over all the comoving volume over which such binary can be seen by the detector with signal-to-noise ratio (SNR) larger than some threshold. In particular, the maximum redshift for a binary to be detectable, dubbed $z_{\rm max}^{\cal D}(M_1,M_2, r_{\rm p})$, is determined by imposing the maximum luminosity distance of a given binary detected with $\rm SNR>8$ equal to the actual luminosity distance at a given redshift and (numerically) solving the equation (see \textit{e.g.}~\cite{OLeary:2008myb,Gondan:2017wzd,Rasskazov:2019gjw} and the SupM). 
Further integrating over all binary component masses and pericenter separations gives us the total \textit{detection} rate: 
\begin{align} \nonumber 
   {\cal R}_{\rm det}^{\cal D} = &
  \int_{5M_\odot}^{M_{\rm max}} \d M_1 \int _{5M_\odot}^{M_{1}}\d M_2 \int_{8 GM_{\rm tot}}^{10^4 GM_{\rm tot}} \d r_{\rm p} \\
  &\times \int_0^{z_{\rm max}^{\cal D}}\frac{\d z}{1+z}\frac{dV_{\rm C}}{\d z}
   \frac{\d {\cal R}_{M_1 M_2}}{\d r_{\rm p}}\,. 
\end{align}
The integration range in $r_{\rm p}$ discards the direct collisions and most binaries that can be disrupted by encounters with a third object.
$V_{\rm C}$ is the comoving volume in the $\rm\Lambda CDM$ cosmology~\cite{Planck:2018vyg}. The whole calculation of the detection rate is performed with a Monte Carlo technique using $4\times 10^6$ binary BH samples.
We also compute \textit{partial} detection rates for some sub-regions of the $(M_1,M_2,r_{\rm p})$ parameter space, such as the one for $M_1>50M_{\odot}$ and $q<0.3$: these outcomes are enhanced for SC-mediated capture but are observed rather infrequently in the existing GWTC-5 catalog \cite{LIGOScientific:2026wfs}.
 \begin{figure}
  \centering
\includegraphics[width=\linewidth]{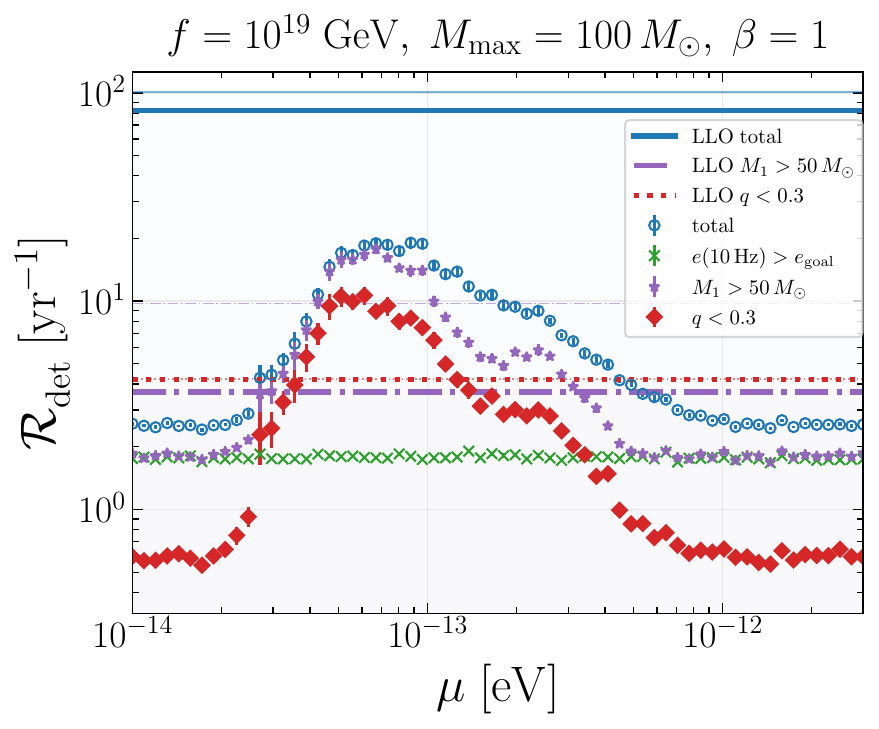}
  \caption{Detection rates obtained from the LLO detections in the LVK catalog (horizontal lines) compared to prediction of merger rates in LLO from SC-enhanced captures in GNs (scatter points) as a function of the ultralight scalar mass $\mu$. For each LLO rate the thin lines show 90\% upper bounds on this rates. For illustration, we fix $f=10^{19}$ GeV, $M_{\rm max}=100M_\odot$ and $\beta=1$. We show different partial rates with different colors. In all cases except in the $e(10\ {\rm Hz})>0.03$ channel for a given range of ultralight scalar masses $\mu$, the SC enhances the rates with respect to the standard GW-capture prediction. By comparing the points to the upper bounds, we set constraints on a certain $\mu$ interval. }
  \label{fig:det_rate}
\end{figure}

We compare the predicted rates to upper bounds from the GWTC-5 catalog. We extract exclusion regions on the ultralight scalar mass $\mu$ by requiring the predicted rate in a given detector, like the LIGO Livingston Observatory (LLO) to be larger than the upper bound on the observed one at 90\% confidence level. A full Bayesian analysis of the events is left for future work. We show the procedure in Fig.~\ref{fig:det_rate} for a specific benchmark and we defer to the SupM for details on the statistical analysis. The predicted rates at the edge of the shown $\mu$ range converge to the the GW-capture scenario making evident the role of the SC in enhancing the rates. 
% We note that heavy primary BHs ($M_1>50M_\odot$) dominate the rate in the SC-enhanced capture, and small mass ratios ($q<0.3$) contribute substantially, while in the LVK sample they represent roughly $2\mathrm{-}3\%$ \cite{LIGOScientific:2026wfs} of the events (up to $4\mathrm{-}8\%$ including the statistical uncertainties).
We repeat this analysis for other values of $f$: for smaller $f$, the cloud mass is usually lighter, weakening the constraints and shrinking the range of excluded values of $\mu$. We also vary the BH mass function exponent $\beta$. The excluded parameter space become larger for smaller $\beta$, consistent with the larger number of heavy BHs present in GN. On the other hand, larger $\beta$ imply a lighter population, hence worsening the constraints. The results are summarized in Fig.~\ref{fig:summary_mu_inv_f}, which illustrates how our fiducial models exclude significant and novel mass ranges for the QCD axion. 

\textbf{{Conclusion}}---In this work, we have shown that SCs can significantly enhance the formation of BH binaries through close encounters in GNs by providing an additional channel for energy dissipation. This enhancement translates into a potentially observable excess of mergers in terrestrial GW detectors, with the effect being particularly pronounced for binaries with heavy primaries and small mass ratios. Comparing our predictions with the GWTC-5 catalog, we find that the observed rate of mergers with $q<0.3$ can probe ultralight scalars in the mass range $\mu \in [4\times 10^{-14},10^{-13}]\,{\rm eV}$, reaching parameter space complementary to, and in some regions beyond, existing constraints based on BH spin measurements. While the precise reach depends on astrophysical uncertainties, most notably the high-mass end of the BH population in GN, the underlying mechanism provides a qualitatively new way of searching for ultralight fields: through their impact on binary formation rather than solely through BH spin down. Improved modeling of BH populations, post-capture binary evolution, and non-perturbative cloud--perturber interactions, together with the growing GW catalog, will further sharpen the accuracy and reliability of this probe.
 
\begin{acknowledgments}
\textbf{{Acknowledgments}}---We thank Silvia Gasparotto, Sivan Ginsburg, Yonit Hochberg, Dina Meylakh, Orion Ning, Rotem Ovadia, Nir Shaviv and Edoardo Vitagliano for insightful discussions. O.T. is supported in part by the ISF grant No. 3533/24 and the BSF grant No. 2024169. M.K., E.K., O.T. and A.L. are supported in part by grants No 2022713 and  2024091 from the US-Israel BSF/NSF and by grant No 2023711 from the US-Israel BSF. M.K. and A.L. are grateful to the Azrieli Foundation for the award of an Azrieli Fellowship. M.K. and A.L. are supported by an ERC STG grant (``Light-Dark,'' grant No. 101040019).
This project has received funding from the European Research Council (ERC) under the European Union’s Horizon Europe research and in- novation programme (grant agreement No. 101040019). Views and opinions expressed are however those of the author(s) only and do not necessarily reflect those of the European Union. The European Union cannot be held responsible for them.
\end{acknowledgments}

\bibliographystyle{apsrev4-2}
\bibliography{ref}
 
\input{SupM}

\end{document}

%% file: SupM.tex
\onecolumngrid
\newpage

\setcounter{section}{0}
\setcounter{equation}{0}
\setcounter{figure}{0}
\setcounter{table}{0}
\setcounter{page}{1}

\setcounter{secnumdepth}{2}
\setcounter{tocdepth}{2}

\renewcommand{\theequation}{S\arabic{equation}}
\renewcommand{\thefigure}{S\arabic{figure}}
\renewcommand{\thepage}{S\arabic{page}}

\begin{center}
	\textbf{\large Supplemental Material for the Letter:\\\vspace{0.1cm}
		{\em \ttitle}}
\end{center}
In this Supplemental Material, we provide additional explanations and complementary results that further support the content presented in the main text. In Section~\ref{app:evol_SC}, we discuss the physics of black hole (BH) superradiance (SR), and we explain how the superradiance cloud (SC) forms and evolves in time in the presence of self-interactions among ultralight scalars. Section~\ref{app:E_and_L} illustrates the formalism to compute energy and angular momentum exchanges between a BH dressed with the SC and a perturber (in this case the secondary BH), crucial to determine the conditions for binary formation. Section~\ref{app:rates} explains in detail how binary BH merger rates are computed, with a focus on the environment of galactic nuclei (GNs), and how these rates are translated into predictions for observable event rates in gravitational-wave (GW) detectors. Section~\ref{app:statistical} outlines the statistical analysis used to infer observed detection rates, derive 90\% confidence upper limits, and constrain the parameter space of ultralight scalars through comparison with the predicted rates. In Section~\ref{app:robustness}, we discuss in depth the various astrophysical uncertainties that characterize our environment of binary formation, galactic nuclei, and study the robustness of our results under variations of the fiducial model assumed in the main text. Finally, Section~\ref{app:perturbativity}
explores the parameter space of ultralight scalars that could potentially be probed once the interaction between a perturber and the SC is computed non-perturbatively.

% \tableofcontents
\section{Evolution of Superradiance clouds}\label{app:evol_SC}

In this Section, we briefly review the process of BH SR that results in the formation of the cloud. We survey the mechanisms that affect the time evolution of the cloud, namely, GW emission, decay into the BH, and possible axion-like quartic self-interactions. 

The underlying mechanism for SR is best illustrated with an analogy. Consider a dissipative cylinder rotating with frequency $\Omega_{\rm cyl}$, and an electromagnetic perturbation of the form $\Psi \sim e^{-i\omega t + i m \phi}$. In the rotating frame of the cylinder, the frequency of the mode registers as $\omega_{\rm rot} = \omega - m \Omega_{\rm cyl}$. If $\omega > m \Omega_{\rm cyl}$, corresponding to $\omega_{\rm rot} > 0$, then part of the wave is dissipated and the energy of the reflected wave is smaller than the impinging one. However, if the opposite relations holds so that $\omega_{\rm rot}<0$, then part of the rotational energy of the cylinder is converted to an energy of the electromagnetic mode, so that the cylinder slows down and the reflected mode is amplified~\cite{Zeld1971}. This is called rotational SR. 

Rotational SR is not unique to rotating cylinders or electromagnetic waves; it can also arise in purely gravitational settings. Indeed, it was shown in~\cite{Press:1972zz,Bardeen:1972fi} that energy and angular momentum can be extracted from a rotating (Kerr) BH through the amplification of reflected waves. This phenomenon is known as black hole SR, which we will refer to simply as SR below for brevity. For our purposes, we focus on scalar (spin-0) waves, although SR can occur for any bosonic field, whether massless or massive, including electromagnetic and gravitational waves. However, only massive fields lead to the formation of a cloud.

\subsection{Black hole superradiance}
Consider a Kerr BH of mass $M_1$ and dimensionless spin $\chi\equiv {J}/{G M_1^2}$, where $J$ is the BH's angular momentum. Denote the mass of the scalar field by $\mu$.
The rotational frequency of the outer horizon of the BH is given by $\Omega_+=\chi/[2r_{\rm g}(1+\sqrt{1-\chi^2})]$. Massive scalar fields have quasi-bound states around a BH, with complex eigenfrequencies (eigenvalues of the Hamiltonian). For a Schwarzschild BH, the imaginary part of the eigenfrequencies is always negative; this indicates that the corresponding quasi-bound state is unstable to decays into the BH. For a Kerr BH, however, there are eigenfrequencies with a positive imaginary part; this indicates an instability to growth rather than decay, known in the literature as a SR instability. In the non-relativistic small-coupling limit, the potential acting on the scalar field behaves as $1/r$, corresponding to a gravitational hydrogen-like atom. Specifically, the real part of the eigenfrequecies resembles the hydrogen-like atom, with the eigenstates labeled by hydrogenic quantum numbers $(n,l,m)$. Quantitatively, quasi-bound state eigenfrequencies are given by~\cite{Baumann:2019eav}
\begin{align}
	\omega_{n\ell m} &= E_{n\ell m} + i \Gamma_{n\ell m},\\
	E_{n\ell m} &= \mu \left(1-\frac{\alpha^2}{2 n^2}\right) + \mathcal{O} \left(\alpha^4\right),\\
	\Gamma_{n\ell m}\simeq2\varrho_+\,\alpha^{4\ell+5}\!\left(m\Omega_{+}-E_{n\ell m}\right)\frac{2^{4\ell+1}(n+\ell)!}{n^{2\ell+4}(n-\ell-1)!}&\left[\frac{\ell!}{(2\ell)!(2\ell+1)!}\right]^{2}\prod_{j=1}^{\ell}\left[j^{2}(1-\chi^{2})+\left(\chi m-2\varrho_+r_{\mathrm g}\,E_{n\ell m}\right)^{2}\right],
\end{align}
where $\alpha \equiv G M_1 \mu$ is the gravitational fine-structure constant, and $\varrho_+=1+\sqrt{1-\chi^{2}}$ is the dimensionless outer horizon. Clearly, states with $m<0$ have $\Gamma_{n\ell m}<0$ and are not superradiant. Importantly, when $E_{n\ell m}< m \Omega_+$, then $\Gamma_{n\ell m}>0$, and the mode $\ket{n\ell m}$ undergoes SR and exponentially grows while the BH spins-down.  
The exponential growth results in a macroscopic occupation number of the state, referred to as an SC. Note that the initial seed for the boson cloud need not be external; instead it can originate from quantum fluctuations in the Kerr background~\cite{Unruh:1974bw}.

In this work, we consider the regime $\alpha\ll 1$, where the SR growth rates exhibit a strong hierarchy among the different states. The $\ket{211}$ state has the highest rate, followed by $\ket{322}$, with higher $\ell$ states suppressed by even higher powers of $\alpha$. Consequently, a cloud seeded near the vacuum is initially dominated by $\ket{211}$ for a sufficiently low value of $\alpha$. In the absence of sufficiently strong self-interactions, the cloud grows, populating the $\ket{211}$ state while spinning-down the BH until the SR condition is saturated and SR shuts down. Meanwhile, self-annihilations of the $\ket{211}$ population into GWs occur through the process $\ket{211} + \ket{211} \rightarrow \rm{GWs}$. Eventually, after a timescale of $\left(2\Gamma_{322}\right)^{-1}$, SR kicks in for $\ket{322}$ and it spins-down the BH further until the SR condition is saturated, and so on. 

\subsection{Self-interactions}
We now discuss the case of non-negligible self-interactions between ultralight scalars. In particular, we focus on an attractive quartic self-interaction $\lambda \varphi^4/4!$, with $\lambda \equiv \mu^2/f^2$, arising in axion-like models. For the values of $\alpha$ considered here, $\left|\varphi\right|/f$ is always small, and so higher order interactions can be neglected~\cite{Baryakhtar:2020gao}. Furthermore, in the $\alpha\lesssim 0.2$ regime considered here, self-interactions can be treated perturbatively~\cite{Baryakhtar:2020gao}, avoiding non-perturbative effects like a bosenova~\cite{Arvanitaki:2010sy,Yoshino:2012kn}. In this case, cloud evolution can be effectively modeled as a lossy two-level system comprised of the two states $\left\{\ket{211},\ket{322}\right\}$. In addition to SR and GW annihilations, which are already present in the purely gravitational case, self-interactions enable processes that transfer population between the $\ket{211}$ and $\ket{322}$ levels. At leading order in $\alpha$, the dominant channels are $\ket{211}+\ket{211}\rightarrow\ket{322}+{\rm BH}$, in which two $\ket{211}$ quanta are converted into one $\ket{322}$ quantum while the remaining excitation is absorbed by the black hole, and $\ket{322}+\ket{322}\rightarrow\ket{211}+\infty$, in which two $\ket{322}$ quanta produce one $\ket{211}$ quantum and an unbound scalar wave that escapes to infinity. The first process can populate the $\ket{322}$ level rapidly despite its comparatively slow intrinsic superradiant growth, whereas the second transfers population back to $\ket{211}$ while depleting the cloud through scalar emission. 

Let $N_s$ denote the occupation number of the state $\ket{s}$, and define the ``normalized" occupation numbers $\varepsilon_s \equiv N_s/\left(G M_{\rm i}^2\right)$, where $M_{\rm i }$ is the initial mass of the BH. The evolution equations for $\varepsilon_s$, $\chi$ and $\alpha$ are given by~\cite{Baryakhtar:2020gao} 
\begin{subequations}\label{app:ceq}
	\begin{align}
		\frac{\dot{\varepsilon}_{211}}{ \mu}
		={}&
		\frac{2\Gamma_{211}}{\mu}
		\varepsilon_{211}
		-2\kappa_{211\times211}^{322\times{\rm BH}}
		\alpha^{11}
		\left(\frac{M_{{\rm pl}}}{f}\right)^{4}
		\varrho_+\,
		\varepsilon_{211}^{2}\varepsilon_{322}
		+\kappa_{322\times322}^{211\times\infty}
		\alpha^{8}
		\left(\frac{M_{{\rm pl}}}{f}\right)^{4}
		\varepsilon_{322}^{2}\varepsilon_{211}
		-2\kappa_{211\times211}^{{\rm GW}}
		\alpha^{14}\varepsilon_{211}^{2}
		+\cdots\\
		\frac{\dot{\varepsilon}_{322}}{\mu}{}={}&\frac{2\Gamma_{322}}{\mu}\varepsilon_{322}+\kappa_{211\times211}^{322\times{\rm BH}}\alpha^{11}\left(\frac{M_{{\rm pl}}}{f}\right)^{4}\varrho_+\varepsilon_{211}^{2}\varepsilon_{322}-2\kappa_{322\times322}^{211\times\infty}\alpha^{8}\left(\frac{M_{{\rm pl}}}{f}\right)^{4}\varepsilon_{322}^{2}\varepsilon_{211}+\cdots\\
		{\dot \chi} ={}& -2 \Gamma_{211} \varepsilon_{211} -4 \Gamma_{322} \varepsilon_{322}\\
		\frac{\dot \alpha}{ \alpha} ={}& -2 \Gamma_{211} \varepsilon_{211} - 2 \Gamma_{322} \varepsilon_{322} + \mu \kappa_{211\times211}^{322\times{\rm BH}}\alpha^{11}\left(\frac{M_{{\rm pl}}}{f}\right)^{4}\varrho_+\varepsilon_{211}^{2}\varepsilon_{322}
	\end{align}
\end{subequations}
where $M_{\rm pl}\equiv G^{-1/2}$ is the Planck mass, $\kappa_{211\times211}^{322\times{\rm BH}}\simeq 4 \times 10^{-7},\,\kappa_{322\times322}^{211\times\infty}\simeq 10^{-8}$, and $\kappa_{211\times211}^{{\rm GW}} \simeq 10^{-2}$. 
\begin{figure}
	\centering
	\includegraphics[width=0.70\linewidth]{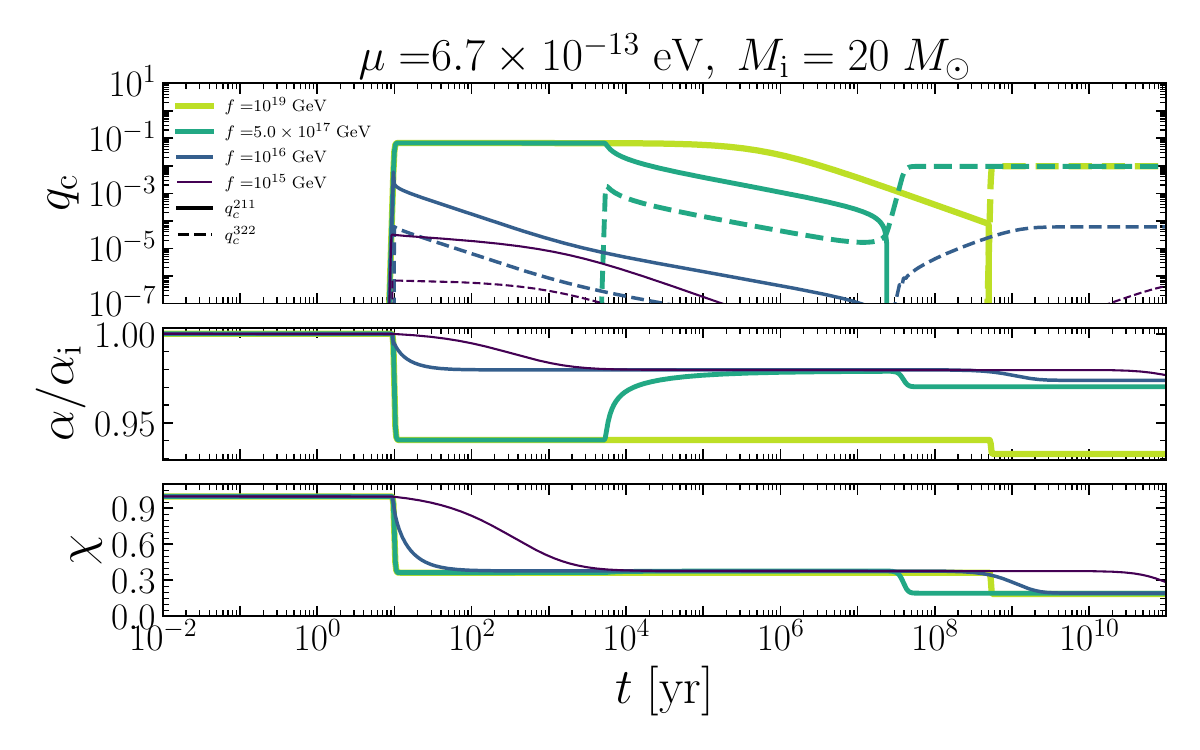}
	\caption{Solutions of the system of ODEs~\eqref{app:ceq} for a BH with initial mass $M_{\rm i}=20M_\odot$, initial spin $\chi_{\rm i}=0.998$ and ultralight boson mass $\mu=6.7\times 10^{-13}$ eV for different values of the self-interaction strength. The three panels show the cloud mass $q_{\rm c} = \alpha_{\rm i}^2\epsilon /\alpha$, normalized fine structure constant $\alpha/\alpha_{\rm i}$ and dimensionless spin $\chi$. We see that for weak self-interactions $f=10^{19}$ GeV, the 211 and 322 states arise sequentially while for stronger self-interactions there is a partial coexistence of the two states, with one dominating over the other. Note how the superradiant production of the $\ket{211}$ and $\ket{322}$ states is associated with substantial (and almost instantaneous) spin-down and mass loss of the BH, unless the self-interaction is strong. }
	\label{fig:cloudevolv}
\end{figure}
\begin{figure}
	\centering
	\includegraphics[width=0.9\linewidth]{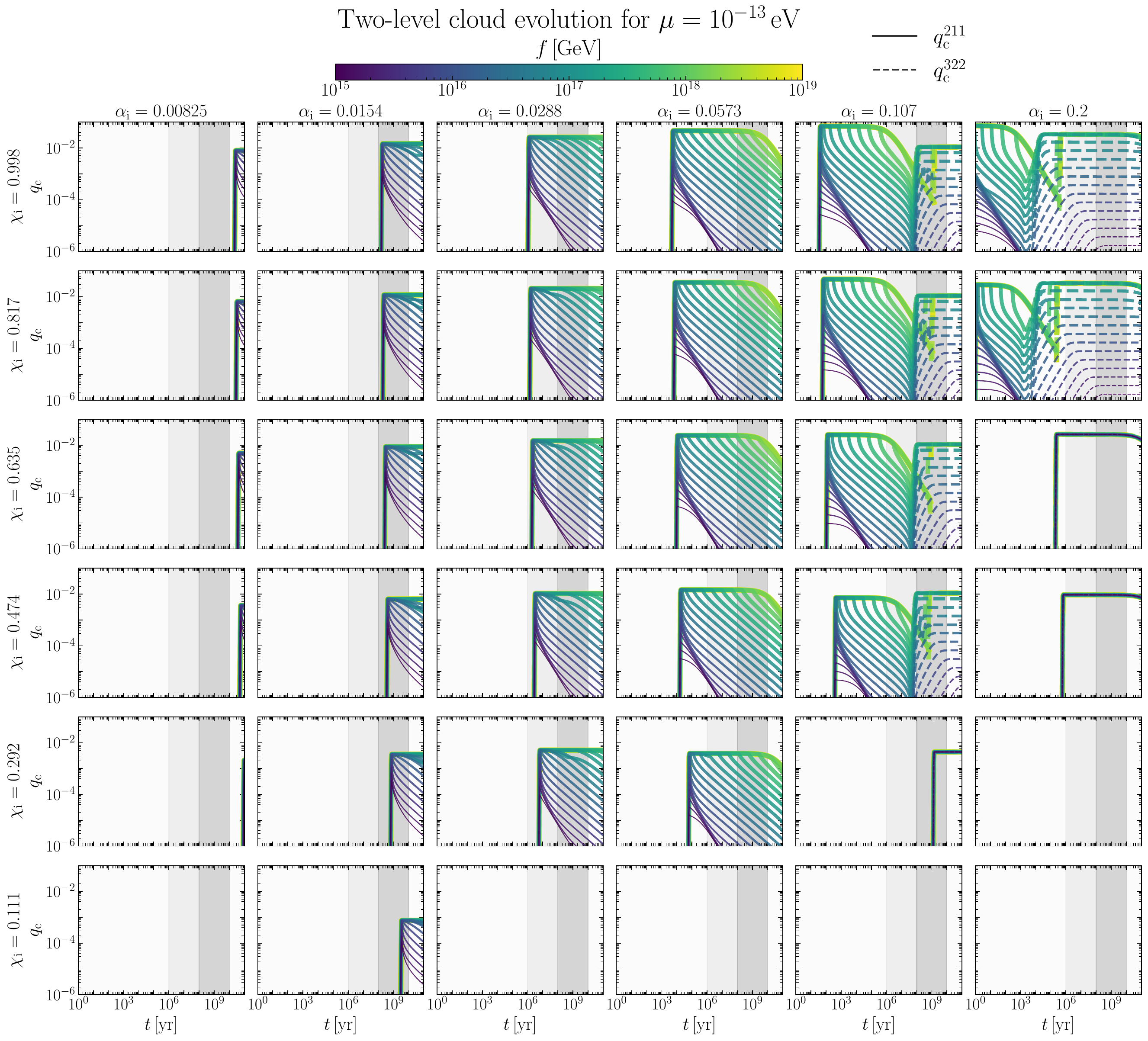}
	\caption{Representation of the cloud mass $q_{\rm c} \equiv \max[q_{\rm c}^{211},q_{\rm c}^{322}]$ obtained via solving the system of ODEs~\eqref{app:ceq} for different values of initial spin, fine structure constant and self-interaction strength for $\mu=10^{-13}$ eV. For illustration we randomly choose an array of $(\chi_{\rm i}, \alpha_{\rm i})$ randomly extracted between $[0,0.998]\times[0.005,0.2]$. For visibility the linewidth increases with $f$. Solid lines represent when the cloud is predominantly in the $\ket{211}$ state while dashed lines show when it is in the $\ket{322}$ state. The gray band shows the observationally relevant range of BH ages with the lighter shade showing a younger BH population. }
	\label{fig:cloudevol2}
\end{figure}

For each of the states $s\in \{211,322\}$, the corresponding cloud mass (normalized by the BH mass) is related to the normalized occupation numbers via
\begin{align}
	q^{s}_{\mathrm c}(t) = \frac{\alpha_{\mathrm i}^2} {\alpha(t)}\varepsilon_{s}(t)\,.
\end{align} 
As the cloud is usually dominated by a single most populated state, the total cloud mass is identified with the mass of that dominant state, namely $q_{\rm c} = \max\{q^{211}_{\rm c}, q^{322}_{\rm c}\}$. An example solution to Eqs.~\eqref{app:ceq}, obtained for fixed values of $\chi_{\mathrm i}$ and $\alpha_{\mathrm i}$, is shown in Fig.~\ref{fig:cloudevolv}. The figure explicitly displays the evolution of $q_{\mathrm c}^{211/322}$, $\alpha$, and $\chi$ for different values of $f$. In Fig.~\ref{fig:cloudevol2}, we additionally present $q_{\mathrm c}(t)$ for a range of values of $\chi_{\mathrm i}$ and $\alpha_{\mathrm i}$, illustrating the dependence of the cloud evolution on the initial parameters.

\section{Energy and angular momentum transfer between a cloud and a perturber}\label{app:E_and_L}
In this Section, we review the formalism used throughout this work to calculate the energy and angular momentum exchanged with the cloud by an encounter with an object on a parabolic orbit (see~\cite{Tomaselli:2023ysb} for a similar derivation). Specifically, we focus on a perturber causing transitions between cloud states.

\subsection{Perturbation of the superradiance cloud}
% Specifically, we focus on a perturber causing transitions between cloud states. 
% \al{The following sentence will probably already appear in the previous section, adapt it later}
% Consider a scalar field of mass $\mu$ forming a cloud around a BH of mass $M$. Denote the gravitational Bohr radius by $r_{\rm B}\equiv(\mu \alpha)^{-1}=r_{\rm g}/\alpha^2$, where $\alpha\equiv GM\mu$ denotes the gravitational fine-structure constant.

In the non-relativistic approximation, the states/wavefunctions of the cloud correspond to those of a hydrogen-like atom. Hence, we denote the states by $ \ket{n\ell m}$ and $\ket{k\ell m}$, corresponding to bound and unbound states, respectively. The states famously factor into radial and angular pieces,

\begin{align}
	\ket{n\ell m} &= \psi_{n\ell m} = R_{n\ell}(r) Y_{\ell m}(\theta,\phi), \\
	\ket{k\ell m}&= \psi_{k\ell m} = R_{k\ell}(r) Y_{\ell m}(\theta,\phi), 
\end{align}
where the radial wavefunctions are explicitly given by
\begin{align}
	R_{n\ell}(r) &= 
	\sqrt{\Big( \frac{2}{nr_{\rm B}} \Big)^3 \frac{(n-\ell-1)!}{2n(n+\ell)!}} 
	e^{- \frac{r}{n r_{\rm B}}}
	\Big(
	\frac{2r}{n r_{\rm B}}
	\Big)^{\ell}
	L^{2\ell+1}_{n-\ell-1}
	\Big( \frac{2r}{n r_{\rm B}} \Big), \\
	R_{k\ell}(r) &=\frac{1}{r_{\rm B}} \frac{2kr_{\rm B} \exp\left[\frac{\pi}{2kr_{\rm B}}\right] \left|\Gamma\left(1 + \ell - \frac{i}{kr_{\rm B}}\right)\right|}{(2\ell + 1)!} (2kr)^\ell e^{-ikr} {}_1F_1\left(\ell + 1 + \frac{i}{kr_{\rm B}}, 2\ell + 2, 2ikr\right),
\end{align}
where $L_n^k$ denotes the associated Laguerre polynomial and ${}_1F_1$ denotes the Kummer’s confluent hypergeometric function. Note that the states have different mass dimension, in light of the different normalizations of the states.

An external perturber interacting with the cloud can cause a transition between two bound states (bound-to-bound) and an ionization of the cloud (bound-to-unbound). If the initial state of the cloud is $\ket{s}$, then these cases correspond to $\ket{s} \to \ket{n\ell m}$ and $\ket{s}\to \ket{k\ell m}$, respectively. Denote the mass of the perturber by $M_*$ and its density by $M_* n_* ({\bf r}')$. The interaction potential with the cloud is given by (see \textit{e.g.}~\cite{Kim:2025wwj})
\begin{align}
	V_{\rm int} \left({\bf r}\right) =  - GM_*\mu \int \d^3 r'\,\, n_{\rm *}({\bf r}')\bigg[ \frac{1}{|{\bf r} - {\bf r}'|}-\frac{1}{r'} - \frac{{\bf r}\cdot {\bf r}'}{{{\bf r}'}^3}\bigg].
\end{align}
We can expand the potential in the spherical harmonics basis,
\begin{align}
	&\frac{1 }{|{\bf r} - {\bf r}'|}
	-\frac{1 }{ r'}
	- \frac{ {{\bf r}} \cdot {\bf r}' }{r'^3} = 
	\sum_{\ell m}
	\frac{4\pi}{2\ell+1}
	\bigg(
	\frac{r_<^{\ell}}{r_>^{\ell+1}} 
	- \frac{r^\ell}{r'^{\ell+1}}\delta_{\ell \leq 1}
	\bigg)
	Y_{\ell m}^*(\hat{{\bf r}}')
	Y_{\ell m}(\hat{{\bf r}}) \equiv 
	\sum_{\ell m}
	\frac{4\pi}{2\ell+1}
	F_{\ell}(r, r')
	Y_{\ell m}^*(\hat{{\bf r}}')
	Y_{\ell m}(\hat{{\bf r}}), 
\end{align}
where $r_< = \min(r,r')$ and $r_>\equiv \max (r,r')$. Starting from the state $\ket{s}$ at $t=-\infty$, the probability to transition to the state $\ket{j}$ by $t=\infty$ (probability density if $\ket{j}$ is unbound) is 
given, to first order in perturbation theory, by

\begin{align} \label{app:probden}
	P_{sj} = \left|\int_{-\infty}^{\infty}\gamma_{sj}(t) e^{i\Delta E_{js} t}\,\, dt\right|^2,
\end{align}
where $\Delta E_{js}\equiv E_j-E_s$ is the energy difference and $ \gamma_{sj} (t)$ is the transition amplitude, given by
\begin{align} \label{app:gammaamp}
	\gamma_{sj} (t) \equiv\bra{j}V_{\rm int} \ket{s} =  \int \d^3 r\ \psi_j^*({\bf r}) \psi_{s}({\bf r}) V_{\rm int} \left({\bf r}\right) = - \frac{ G M_{*} \mu }{ r_{\rm B}^{\kappa_j}}
	\sum_{\ell m}
	\frac{4\pi}{2\ell+1} I^r_{\ell m}(s,j) I^\Omega_{\ell m}(s,j). 
\end{align}
Here we have defined the integrals
\begin{align}
	I^r_{\ell m}(i,j)
	&= 
	r_{\rm B}^{\kappa_j} 
	\int_0^\infty \d r \, r^2 R_i R_j
	\int_0^\infty \d r' r'^2 F_\ell(r, r') \, n_{\ell m}(r') ,
	\label{Ir}
	\\
	I^\Omega_{\ell m} (i,j)
	&= \int \d\Omega \, 
	Y_{\ell_j m_j}^*(\hat{{\bf r}}) 
	Y_{\ell m} (\hat{{\bf r}}) 
	Y_{\ell_i m_i} (\hat{{\bf r}}),
\end{align}
where we used $n_{*}( {{\bf r}}) = \sum_{\ell m} n_{\ell m}(r) Y_{\ell m}(\theta,\phi)$, and $R_i = R_{n_i \ell_i}(r)$.
We introduced the symbol $\kappa_j$ so that $\kappa_j = 1$ if $\ket{j}$ is bound and $\kappa_j = 1/2$ if it is unbound. This difference between bound and unbound stems from the different normalizations of the states, and comes to ensure that $I^r_{\ell m}$ is dimensionless.
We note that $I^\Omega_{\ell m} (i,j)$ encodes a set of selection rules, e.g. $m= m_j - m_i$, $|\ell_i - \ell_j| \leq \ell \leq \ell_i + \ell_j$ and $\ell_i + \ell + \ell_j = 2p$ with $p\in {\mathbb Z}$~\cite{Baumann:2018vus}.

We now specialize to the case of a point-like perturber at position ${\bf r}_*$, so that $n_*\left({\bf r}'\right) = \delta^{3}({\bf r}'-{\bf r}_*)$. This corresponds to $n_{\ell m} \left(r'\right) = {r'}^{-2} \delta(r' - r_*)Y_{\ell m}^*(\hat{{\bf r}}_*)$, yielding
\begin{align}
	I^r_{\ell m}(i,j) = 
	Y_{\ell m}^*(\hat{{\bf r}}_*)
	r_{\rm B}^{\kappa_j} \int_0^\infty \d  r \, 
	r^2 R_i(r) R_j (r) F_\ell(r,r_*) \equiv Y_{\ell m}^*(\hat{\bf{r}}_*) J^r_{\ell}\left(\frac{r_*}{r_{\mathrm B}},i,j\right).
\end{align}
Note that $J_\ell^r$ depends on $r_*$ only through the ratio $r_*/r_{\mathrm B}$. If $\ket{j}$ is an unbound state with quantum number $k$, then the dependence on it is through the combination $k r_{\mathrm B}$.

\subsection{Encounter on a parabolic trajectory}

We now specify the parabolic trajectory. To do this, we fix the spin of the BH to lie on the $z$-axis. The parabolic trajectory is constrained to the plane orthogonal to the angular momentum $\bf{L}$ of the orbit. The orbit is characterized by its inclination $\iota$ (angle of $\bf{L}$ with respect to $z$-axis), the longitude of the pericenter $\omega$, and the longitude of the ascending node $\Omega$. In its plane of motion, the parabolic orbit is conveniently parametrized using $r_*(t)$ and $\Phi(t)$, defined as
\begin{subequations} \label{app:parorbit}
	\begin{align}
		r_* &= r_{\rm p}(1+x^2),\\
		t &= \tau\,(x+x^3/3),\\
		x &= \tan (\Phi/2),
	\end{align}
\end{subequations} 
where $\tau = \sqrt{\frac{2r_{\rm p}^3}{G(M+M_*)}}$, and $r_{\rm p}$ is the pericenter distance. In order to evaluate $P_{sj}$ using Eq.~\eqref{app:probden}, it is necessary to project the orbit back to the standard coordinate system (where the BH's spin is along the $z$-axis), as it appears in $Y_{\ell m}^*(\hat{\bf{r}}_*)$. This can be conveniently encapsulated using Wigner $D$-matrices, using which we obtain
\begin{align}
	Y_{\ell m}(\hat{\bf{r}}_*) =\sum_{m'=-\ell}^\ell D_{mm'}^{\ell}\left(\Omega-\frac{\pi}{2},\iota, \omega+\frac{\pi}{2}\right) Y_{\ell m'}\left(\frac{\pi}{2},0\right) e^{i m' \Phi (t)}.
\end{align}
With this, the transition amplitude reads
\begin{align}
	\gamma_{sj} = 
	- \frac{ q\alpha }{r_{\rm B}^{\kappa_j}}
	\sum_{\ell=0}^\infty\sum_{m=-\ell}^\ell \sum_{m'=-\ell}^\ell
	\frac{4\pi}{2\ell+1} J^r_{\ell}\left(\frac{r_*}{r_{\mathrm B}},s,j\right) I^\Omega_{\ell m}(s,j)  \left[D_{mm'}^{\ell}\left(\Omega-\frac{\pi}{2},\iota, \omega+\frac{\pi}{2}\right) \right]^* Y_{\ell m'}^*\left(\frac{\pi}{2},0\right) e^{-i m' \Phi(t)},
\end{align}
where we introduced the mass ratio $q\equiv M_*/M$. Plugging back in Eq.~\eqref{app:probden} and changing the integration variable to $x$ using Eqs.~\eqref{app:parorbit} results in 
\begin{align}
	P_{sj} =\frac{4 q^2 \alpha^2 \tau^2}{r_{\rm B}^{2\kappa_j}}\bigg|\sum_{\ell,m,m'} W_{\ell m m'}^{sj}(\iota, \Omega, \omega) {\cal F}_{\ell m m'}^{sj}\left(\frac{r_{\rm p}}{r_{\mathrm B}},\Delta E_{js} \tau\right) \bigg|^2,
\end{align}
where we defined
\begin{align}
	W_{\ell m m'}^{sj}(\iota, \Omega, \omega)\equiv \frac{4\pi}{2\ell+1} I_{\ell m}^\Omega \left[D_{mm'}^{\ell}\left(\Omega-\frac{\pi}{2},\iota, \omega+\frac{\pi}{2}\right) \right]^* Y_{\ell m'}^*\left(\frac{\pi}{2},0\right)
\end{align}
and the function
\begin{align}
	{\cal F}_{\ell m m'}^{sj}(a,z) \equiv \int_{0}^{\infty}\d x (1+x^2) J^r_{\ell}\left(a(1+x^2),s,j\right)\cos(z (x+x^3/3)-2m'\arctan x) . 
\end{align}
We note that, in practice, we take the state $s$ to be a pure hydrogen-like state rather than a superposition. In this case, the dependence of $\gamma_{sj}$ on the Euler angle $\Omega$ amounts only to a global phase, which drops out of $P_{sj}$. We therefore set $\Omega = 0$ throughout our calculations.

Finally, in our analysis, we use first-order perturbation theory, which requires the transition probabilities to remain small and implies no feedback on the orbits. It is therefore important to verify that the parameter space included in our analysis remains within the perturbative regime. To quantify this, we self-consistently require
\begin{align}
	P^{s}_{\rm tot} \lesssim 0.1,
\end{align}
where \(P^{s}_{\rm tot}\) is the total transition probability out of the state \(\ket{s}\), defined by
\begin{align}
	P^{s}_{\rm tot} \equiv \sum_j P_{sj} + \int \frac{d\rho}{2\pi}\,P_{s\rho}.
\end{align}
Here, the first term sums over transitions to bound states, while the second integrates over transitions to unbound states (ionization).

\subsection{Energy transfer}\label{app:SC_energy}
\begin{figure}
	\centering
	\includegraphics[width=0.75\linewidth]{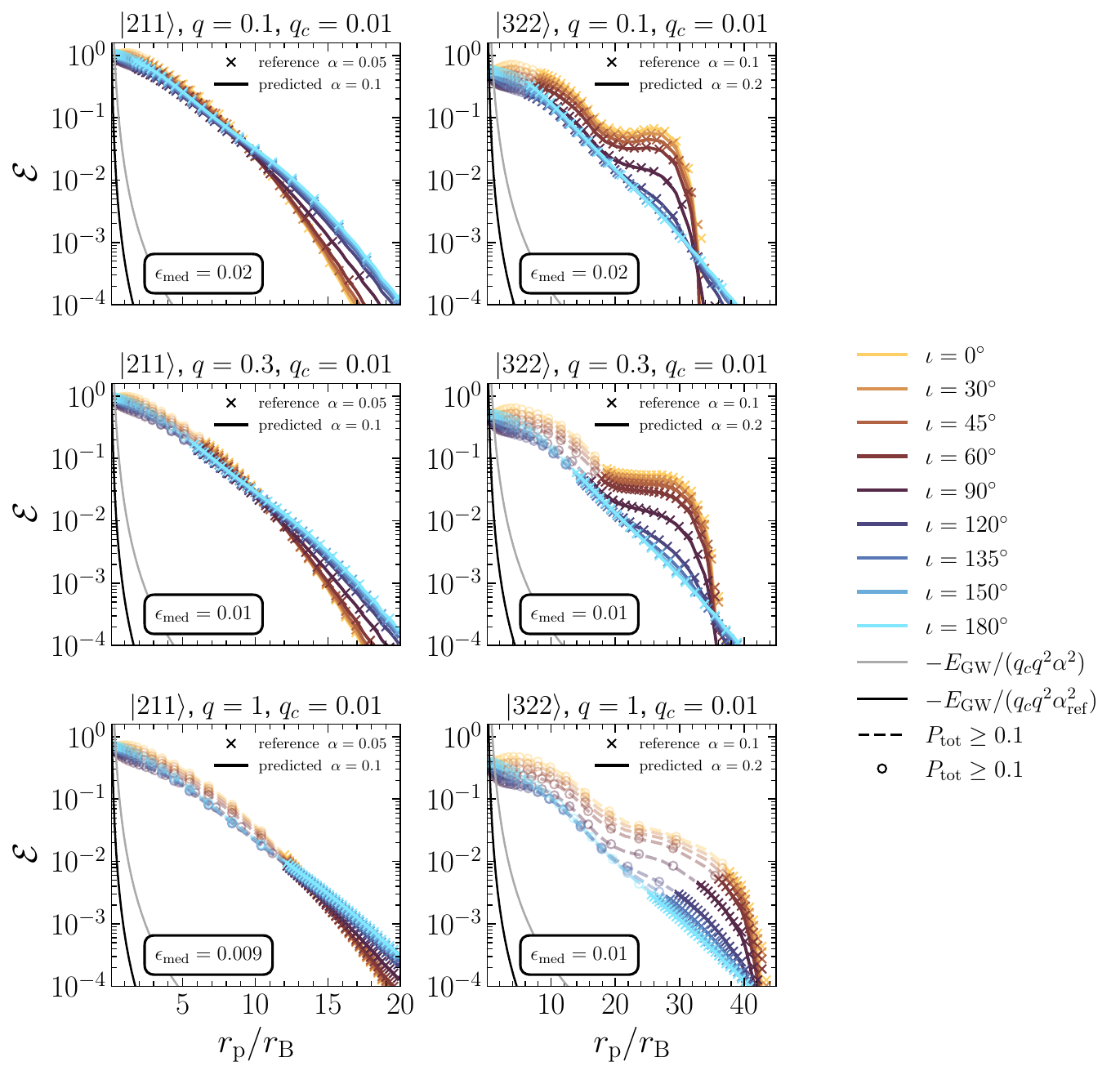}
	\caption{The energy transfer kernels $\cal E$ as function of pericenter distance in Bohr radii units $r_{\rm p}/r_{\rm B}$ computed for two different values of $\alpha$ and for the $\ket{211}$ (\textbf{left} panels) and $\ket{322}$ (\textbf{right} panels) SC states. We consider $\omega=0$, vary $\iota$, fix $q_{\rm c}=0.01$ and show $q=\{0.1,0.3,1\}$ from top to bottom. The colored lines show the kernel \textit{predicted} from the one computed at the reference value. The percent-level agreement is shown by overlapping the actual calculation with markers: $\epsilon$ is the median error. Transparent dashed lines and empty circles markers indicate the region where $P_{\rm tot}^s>0.1$ and the perturbative formalism is not valid anymore.  }
	\label{fig:dE_transfer}
\end{figure}
Finally, we can now express the energy and angular momentum transfer between the cloud and the perturbing object during their encounter. Starting with bound-to-bound transitions ($\kappa_j=1$), the total energy transfer from the perturber to other states of the cloud is then 
\begin{align} \label{app:ELbound}
	\frac{\delta E^{\rm b}_{\mathrm c}}{M} = - q_{\rm c}\sum_j \frac{\Delta E_{js}}{\mu} P_{sj} \approx-\frac{4q^2 \alpha^2 \tau^2}{r_{\rm B}^2} q_{\rm c}\sum_j\frac{\Delta E_{js}}{\mu} \bigg|\sum_{\ell,m,m'}W_{\ell m m'}^{sj} {\cal F}_{\ell m m'}^{sj}\left(\frac{r_{\rm p}}{r_{\mathrm B}},\Delta E_{js} \tau\right) \bigg|^2,
\end{align}
with $q_{\rm c}=M_{\rm c}/M$ the fractional cloud mass. The sign of the energy transfer $\delta E$ is chosen so that if $\delta E>0$, then the cloud transfers energy to the orbit, and vice versa. We can rewrite Eq.~\eqref{app:ELbound} in the following convenient form
\begin{align}
	\frac{\delta E^{\rm b}_{\mathrm c}}{M} &\approx - q_{\mathrm c} q^2 \alpha^2 \mathcal{E}^{\rm b}_{s,q,\theta} \left(\frac{r_{\mathrm p}}{r_{\mathrm B}}\right),\\
	\mathcal{E}^{\rm b}_{s,q,\theta} \left(\frac{r_{\mathrm p}}{r_{\mathrm B}}\right) &= \frac{8}{(1+q)} \frac{r_{\rm p}^3}{r_{\rm B}^3} \sum_j\frac{\Delta E_{js}}{\mu \alpha^2} \bigg|\sum_{\ell,m,m'}W_{\ell m m'}^{sj} {\cal F}_{\ell m m'}^{sj}\left(\frac{r_{\rm p}}{r_{\mathrm B}},\Delta E_{js} \tau\right) \bigg|^2.
\end{align}
The symbol $\theta$ denotes the dependence on the three angles $\theta=(\iota,\Omega,\omega)$ together. Note that to a good approximation, $\mathcal{E}^{\rm b}_{s,q,\theta}$ is independent of $\alpha$, since $\frac{\Delta E_{js}}{\mu \alpha^2} = \frac{1}{2n_s^2} - \frac{1}{2n_j^2} + \mathcal{O}(\alpha^2)$.

For bound-to-unbound transitions ($\kappa_j=1/2$), the analogous expression is
\begin{align} \label{app:ELunbound}
	\frac{\delta E^{\rm u}_{\mathrm c}}{M} \approx -\frac{4q^2 \alpha^2 \tau^2}{r_{\rm B}^2} q_{\rm c} \int \frac{ \d \rho}{2\pi}\sum_{\ell_k, m_k}\frac{\Delta E_{\rho s}}{\mu} \bigg|\sum_{\ell,m,m'}W_{\ell m m'}^{s,\rho, \ell_k, m_k} {\cal F}_{\ell m m'}^{s,\rho,\ell_k, m_k}\left(\frac{r_{\rm p}}{r_{\mathrm B}},\Delta E_{\rho s} \tau\right) \bigg|^2
\end{align}
where $\rho\equiv k r_{\rm B}$. Here, we have expanded the notation so that $j \to (\rho,\ell_k,m_k)$ explicitly denotes the complete set of quantum numbers characterizing the final states. Similarly to the bound case, we rewrite Eq.~\eqref{app:ELunbound} in a more convenient form
\begin{align}
	\frac{\delta E^{\rm u}_{\mathrm c}}{M} &\approx - q_{\mathrm c} q^2 \alpha^2 \mathcal{E}^{\rm u}_{s,q,\theta} \left(\frac{r_{\mathrm p}}{r_{\mathrm B}}\right),\\
	\mathcal{E}^{\rm u}_{s,q,\theta} \left(\frac{r_{\mathrm p}}{r_{\mathrm B}}\right) &= \frac{8}{(1+q)} \frac{r_{\rm p}^3}{r_{\rm B}^3} \int \frac{ \d \rho}{2\pi}\sum_{\ell_k, m_k}\frac{\Delta E_{\rho s}}{\mu \alpha^2} \bigg|\sum_{\ell,m,m'}W_{\ell m m'}^{s,\rho, \ell_k, m_k} {\cal F}_{\ell m m'}^{s,\rho,\ell_k, m_k}\left(\frac{r_{\rm p}}{r_{\mathrm B}},\Delta E_{\rho s} \tau\right) \bigg|^2.
\end{align}

We can now add both contributions and make contact with the main text
\begin{align}
	\frac{\delta E_{\mathrm c}}{M} = \frac{\delta E^{\rm b}_{\mathrm c}}{M} + \frac{\delta E^{\rm u}_{\mathrm c}}{M} \approx - q_{\mathrm c} q^2 \alpha^2 \mathcal{E}_{s,q,\theta} \left(\frac{r_{\mathrm p}}{r_{\mathrm B}}\right),
\end{align}
where we defined
\begin{align}
	\mathcal{E}_{s,q,\theta} \left(\frac{r_{\mathrm p}}{r_{\mathrm B}}\right) \equiv \mathcal{E}^{\rm b}_{s,q,\theta} \left(\frac{r_{\mathrm p}}{r_{\mathrm B}}\right) + \mathcal{E}^{\rm u}_{s,q,\theta} \left(\frac{r_{\mathrm p}}{r_{\mathrm B}}\right). 
\end{align}

The validity of the approximate $\alpha$-independence of $\mathcal{E}_{s,q,\theta}$ for $s\in\{211,322\}$ is demonstrated in Fig.~\ref{fig:dE_transfer}. Specifically, we calibrate $\mathcal{E}_{211,q,\theta}$ at $\alpha=0.05$ and $\mathcal{E}_{322,q,\theta}$ at $\alpha=0.1$, and use the resulting calibrations to predict their values at $\alpha=0.1$ and $\alpha=0.2$, respectively. The prediction is then compared with the value obtained from a direct calculation at $\alpha=0.1$ and $\alpha=0.2$, yielding a median error of order $\sim 10^{-2}$. Importantly, we also distinguish between regions in which the perturbative formalism adopted here is expected to remain valid, corresponding to a total transition probability in the cloud of $P^s_{\rm tot}<0.1$, and those in which it is expected to break down, with $P^s_{\rm tot}>0.1$. As one might expect, as $q\to1$, this perturbative region becomes smaller. We only include the perturbative regions in our analysis. 

All the energy-transfer kernels ${\cal E}_{s,q,\theta}$ depend on the angles $\theta=(\iota, \omega)$, that can produce order 50\% variations. To take this dependence into account we tabulate the kernels for 9 angles in $\iota = \{0^\circ,30^\circ, 45^\circ, 60^\circ, 90^\circ, 120^\circ, 135^\circ, 150^\circ, 180^\circ \}$ and 8 angles in $\omega\in [0,2\pi)$. In all the ensemble averages, we assume a uniform distribution of angles in $\cos\iota$ and in $\omega$, under the reasonable assumption of isotropy: effectively we average only over the computed angles. The dependence on $q$ is dealt with by tabulating the kernels for $q=\{0.001,0.01,0.1,0.2,0.3,0.5,0.7,1\}$. Since rate calculations will deal with an integration of $q$ in the range $(0,1]$, the kernels for other values of $q$ are obtained via linear interpolation.

\subsection{Angular momentum transfer}\label{app:SC_angular_momentum}
The $z$-axis angular momentum transfer due to the SC, labeled $\delta L^z_{\rm c}$, can be computed in a similar way: for bound-to-bound
\begin{align}
	\frac{\delta L_{\mathrm c}^{z;\rm b}}{GM^2} =-\frac{M_{\rm c}}{GM^2\mu}\sum_j (m_j-m_s) P_{sj} \approx -\frac{4q^2 \alpha \tau^2}{r_{\rm B}^2}q_{\rm c}\sum_j{(m_j-m_s)}\bigg|\sum_{\ell,m,m'}W_{\ell m m'}^{sj} {\cal F}_{\ell m m'}^{sj}\left(\frac{r_{\rm p}}{r_{\mathrm B}},\Delta E_{js} \tau\right) \bigg|^2,
\end{align}
and for bound-to-unbound
\begin{align}
	\frac{\delta L_{\mathrm c}^{z;\rm u}}{GM^2} = -\frac{4q^2 \alpha^2 \tau^2}{r_{\rm B}^2} q_{\rm c} \int \frac{ \d \rho}{2\pi}\sum_{\ell_k, m_k}{(m_k-m_s)} \bigg|\sum_{\ell,m,m'}W_{\ell m m'}^{s,\rho, \ell_k, m_k} {\cal F}_{\ell m m'}^{s,\rho,\ell_k, m_k}\left(\frac{r_{\rm p}}{r_{\mathrm B}},\Delta E_{\rho s} \tau\right) \bigg|^2\ .
\end{align} 

\begin{figure}
	\centering
	\includegraphics[width=0.75\linewidth]{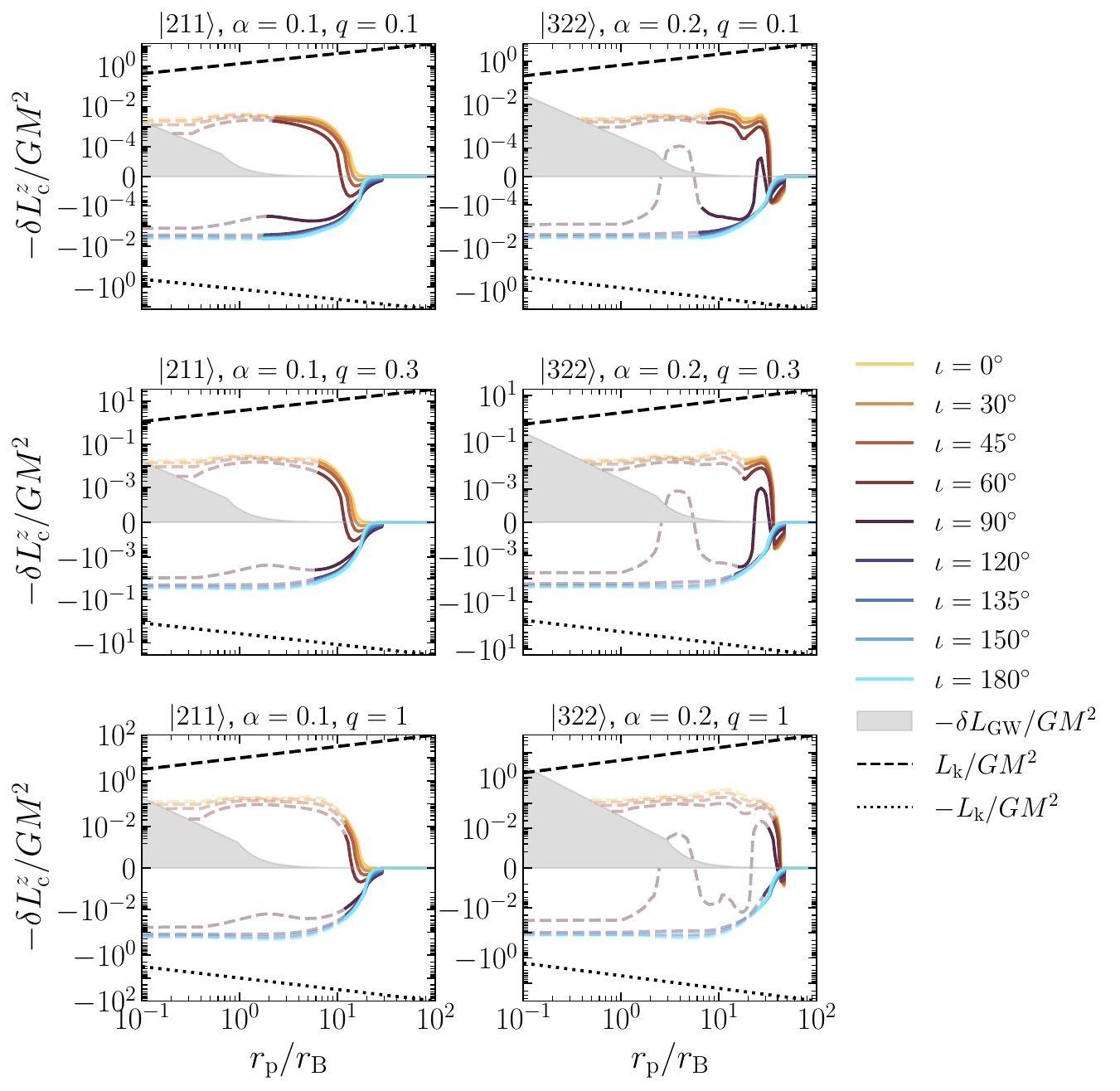}
	\caption{The angular momentum transfer in the interaction between the SC and the perturber in units of $GM^2$ as function of pericenter distance in Bohr radii units $r_{\rm p}/r_{\rm B}$ computed for two different values of $\alpha$ and for the $\ket{211}$ (\textbf{left} panels) and $\ket{322}$ (\textbf{right} panels) SC states. We consider $\omega=0$, vary $\iota$, fix $q_{\rm c}=0.01$ and show $q=\{0.1,0.3,1\}$ from top to bottom. As in Fig.~\ref{fig:dE_transfer}, transparent dashed lines indicate the region where $P_{\rm tot}^s>0.1$ and the perturbative formalism is not valid anymore. The gray-filled area shows where the angular momentum loss by GW emission is larger than the SC-mediated one while the black lines (dashed and dotted) show $|L_{\rm k}|/GM^2$. The SC angular momentum exchange is always within those two lines.}
	\label{fig:dL_summary}
\end{figure}

Similarly to the GW-capture case, the SC-exchanged angular momentum has little to no impact on the binary formation process, and for this reason the angular momentum transfer is neglected altogether~\cite{OLeary:2008myb}. This is because the SC-transfered angular momentum is negligible with respect to the kinetic one during the encounter between objects of masses $M=M_1$ and $M_2$: $\delta L_{\rm c}^z \ll L_{\rm k} = (M_1M_2/M_{\rm tot}) bw $ with $b\approx \sqrt{2 GM_{\rm tot} r_{\rm p}}/w$. We prove this by direct comparison in Fig.~\ref{fig:dL_summary}.

\subsection{Capture condition and binary formation}
After the parabolic encounter between the two objects $M_1$ (with SC) and $M_2$, a binary is formed if the total final energy $E_0 = E_{\rm k}+\delta E_{\rm GW} +\delta E_{\rm c}$ including the GW and SC contribution on top of the kinetic energy $E_{\rm k} = M_1 M_2 w^2 /(2M_{\rm tot})$ is negative. The energy and angular momentum $E_0, L_0\simeq L_{\rm k}$ determine the initial eccentricity of the bound system as 
\begin{align}
	e_0 = \sqrt{1 + \left(\frac{1+q}{q^3}\right)\left(\frac{2E_0}{M_1}\right)\left(\frac{L_0}{GM_1^2}\right)^2}\,
\end{align}
Clearly $e_0<1$ if $E_0<0$. Since for an SC-enhanced capture $E_0$ is \textit{more negative} than in the pure GW-capture case, the resulting initial eccentricities are smaller and further away from unity. 
\begin{figure}
	\centering
	\includegraphics[width=0.7\linewidth]{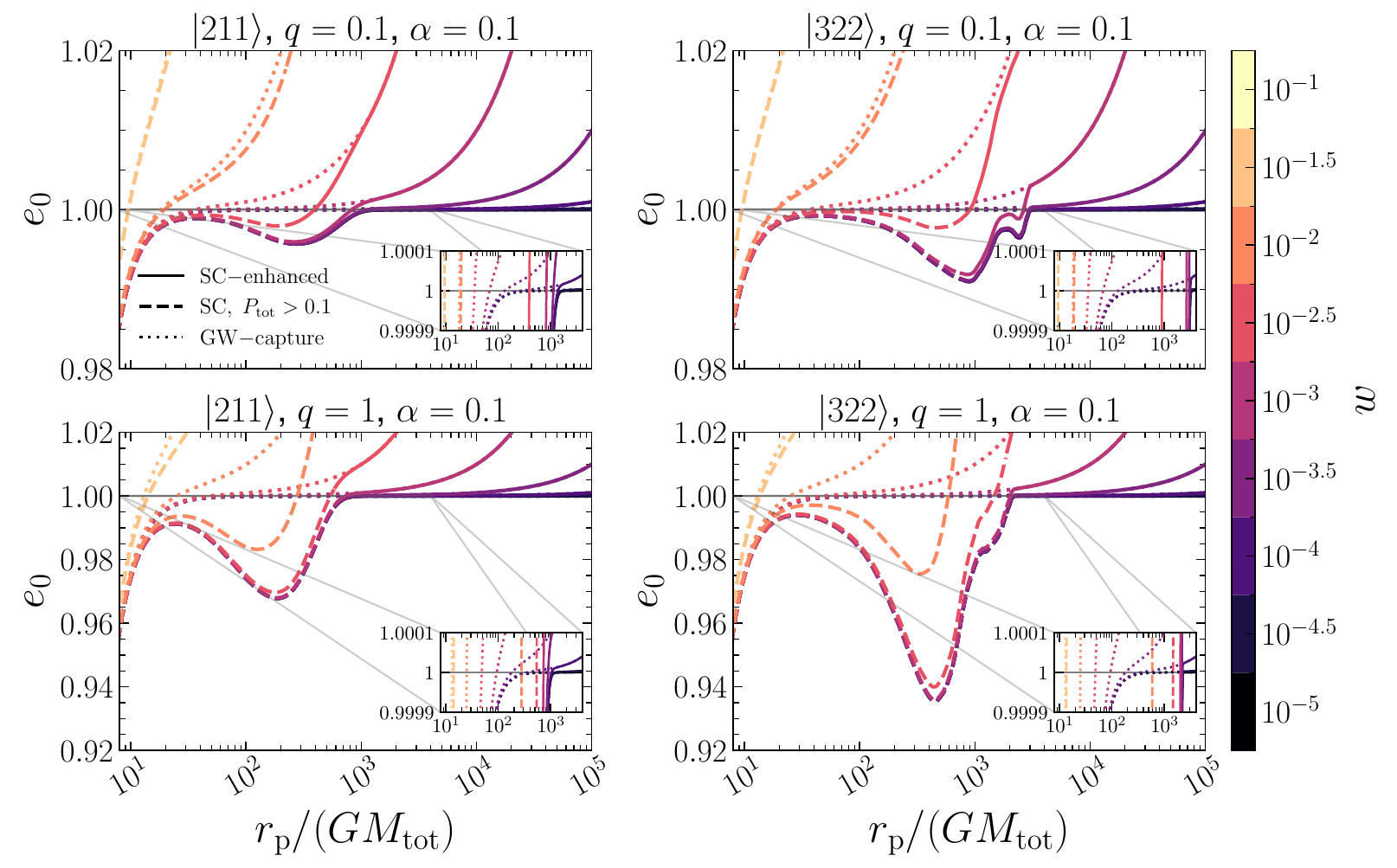}
	\caption{The eccentricity of the BH-perturber system after the parabolic encounter as a function of the normalized pericenter distance. \textbf{Left} panels show the encounter when the SC of the primary BH is in the $\ket{211}$ state, \textbf{right} show the case of $\ket{322}$ state. We show examples for $q=0.1$ (\textbf{top} panels) and $q=1$ (\textbf{bottom} panels). The angles of the encounter are fixed to $(\omega,\iota,\Omega)=(0,0,0)$. The relative velocity is varied according to the color map. Solid lines show the eccentricity when the SC-enhanced case dominates the energy transfer with dashed lines showing when this violates the perturbativity requirements. Dotted lines show the classic GW-capture case. Inset panels show the behavior of the curves very close to $e_0=1$.  }
	\label{fig:e_summary}
\end{figure}
We illustrate this in Fig.~\ref{fig:e_summary}: we argue that the presence of the SC supports the common choice of setting the initial eccentricity of newly formed binaries to $e_0=0.99$. We note that this has been purely a choice of computational convenience to calculate the signal-to-noise ratio in detection forecasts in the case of pure GW-capture~\cite{OLeary:2008myb,Gondan:2017wzd}, given that $e_0\simeq1$ even up to a part per million. 

\begin{figure}
	\centering
	\includegraphics[width=0.6\linewidth]{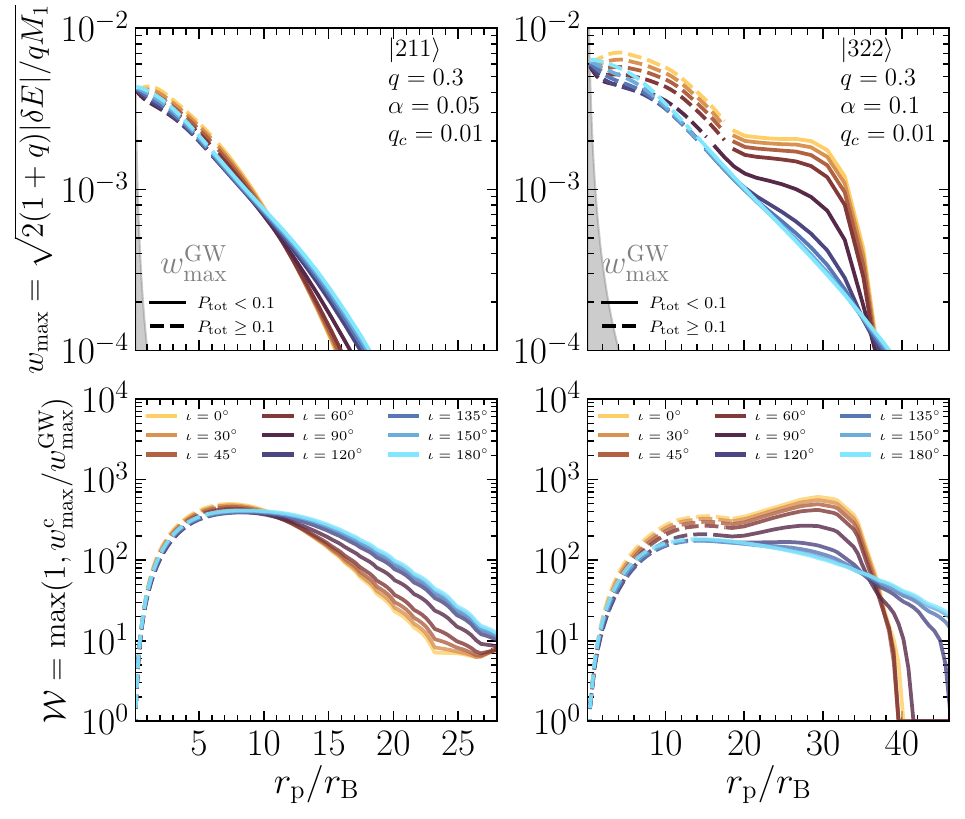}  \caption{\textbf{Left}: the maximum relative velocity that allows capture, $w_{\rm max}$, as a function of normalized pericenter distance $r_{\rm p}/r_{\rm B}$ for a SC in $\ket{211}$ state (top left) or $\ket{322}$ state (bottom left). The gray shaded area shows the distances where the GW emission dominates the capture dynamics. \textbf{Right}: the $w_{\rm max}$ enhancement $\cal W$ over pure GW capture, for a SC in the $\ket{211}$ state (top right) or $\ket{322}$ state (bottom right) as a function of $r_{\rm p}/r_{\rm B}$. In all the plots we show the results varying the inclination of the orbit $\iota$ as visible in the color scale. Solid lines show where $P_{\rm tot}<0.1$,while dashed show $P_{\rm tot}>0.1$.}
	\label{fig:wmax}
\end{figure}

For both GW-capture and SC-mediated capture, the capture condition can be expressed in a general way by computing the maximum relative velocity between the BH and its perturber that allows for $E_0<0$. This value of the relative velocity is given by, for a given pericenter distance $r_{\rm p0}$,
\begin{align}
	w_{\rm max}(r_{\rm p0},\dots) = \sqrt{\frac{2(1+q) |\delta E_{\rm GW}(r_{\rm p0})+\delta E_{\rm c}(r_{\rm p0},\dots)| }{qM_1}}\ .
\end{align}
We show in Fig.~\ref{fig:wmax} this maximum relative velocity and the enhancement in the SC-enhanced case with respect to the GW-capture value ${\cal W}=\max(1, {\cal W}_{\rm c}) $, explicitly
\begin{align}\label{eq:W}
	{\cal W}_{\rm c}\equiv \frac{{w_{\rm max}^{\rm c}}}{{w_{\rm max}^{\rm GW}}} = \left|\frac{
		12\sqrt{2}q_{\rm c}(r_{\rm p}/r_{\rm B})^{7/2}{\cal E}_{s,q, \theta}\left(r_{\rm p}/r_{\rm B}\right)
	}{85\pi(1+q)^{1/2}\alpha^5}\right|^{1/2}.
\end{align}
This quantity will be of primary importance for the computation of the rates in Sec.~\ref{app:rates}, as $ w_{\rm max} $ contains all the effects of the SC on the binary formation process. 

\section{Calculation of merger and detection rates}\label{app:rates}
In this Section, we show in details how the different rates of binary BH mergers are calculated, from the local system of a single galactic nucleus to the integration over all galaxies and the projection onto the events that can be seen in a gravitational wave interferometer.

\subsection{Local binary merger rate in a GN}
Nuclear star clusters surround most supermassive black holes in GNs~\cite{Neumayer+20}. Empirically, these systems are quasi-spherical, and can be characterized most simply by their {\it influence radius} $r_0$, the radius interior to which the enclosed stellar mass is double that of the SMBH: $M_{\rm enc}(r<r_0) = 2 M_{\rm SMBH}$. For $M_{\rm SMBH} \lesssim 10^7 M_\odot$, stellar distributions will lose memory of their initial conditions and relax~\cite{BarOr:2013ab} (via weak, diffusive two-body scatterings) into a stationary state. Theoretical calculations find that such stationary states feature power-law radial density profiles~\cite{Bahcall:1976aa} and quasi-isotropic velocity fields~\cite{CohnKulsrud78, LezhninVasiliev15}.

We start from encounters in a generic GN environment dominated by the SMBH. We roughly follow the procedure in~\cite{Gondan:2017wzd} and extend it to the case where an SC is present around black holes. The differential rate of the merger between two objects $1,2$ with masses $M_1,M_2$ with the same phase space distribution $f_1=f_2=f$ is given by
\begin{align}
	\d^{12}\Gamma_{M_1,M_2} = \sigma w f({\bf r}_1, {\bf v}_1,M_1)f({\bf r}_2, {\bf v}_2,M_2) \d^3r_1\d^3r_2\d^3v_1\d^3v_2\ .
\end{align}
with $w=|{\bf v}_1 -{\bf v}_2|$ their relative velocity. One can express the interaction cross section $\sigma$ using the impact parameter $\sigma = \pi b^2$. We are interested in the case of a relatively short-range encounter so we multiply the above expression by $\d^3r \delta^{(3)} ({\bf r} -{\bf r}_1)\delta^{(3)} ({\bf r} -{\bf r}_2)$ and integrate over $\d^3r_1\d^3r_2$. Further assuming spherical symmetry in $r$, we obtain
\begin{align}
	\d^{7}\Gamma_{M_1,M_2} = 4\pi^2 w b^2 f({\bf r}, {\bf v}_1,M_1)f({\bf r}, {\bf v}_2,M_2)  r^2 \d r \d^3v_1\d^3v_2\ .
\end{align}
Let us characterize the phase space distribution $f$. If we consider a spherical isotropic distribution of BHs around the SMBH, under the assumption of relaxation, the phase distribution is given by the Bahcall-Wolf~\cite{Bahcall:1976aa,Bahcall:1977ab,Keshet:2009vu,OLeary:2008myb} model 
\begin{align}
	f({\bf r},{\bf v},M) = F(M) \left[\frac{GM_{\rm SMBH}}{r} - \frac{v^2}{2}\right]^{p_0 M/M_{\rm max}} =F(M) \left[\frac{v_{\rm max}^2}{2}-\frac{v^2}{2}\right]^{p_0 M/M_{\rm max}}\ .
\end{align}
$F(M)$ is a normalization constant that is set enforcing the normalization of the distribution $N_{\rm BH} = \int \d M \d^3r \d^3 v f$. We assume this distribution is valid for $r\in [r_{\rm min}, r_{\rm max}]$ and $v\leq v_{\rm max}(r)$ with $v_{\rm max}=\sqrt{2GM_{\rm SMBH}/r}$ the local escape velocity. We take $r_{\rm max}=r_0$, i.e. the radius of influence of the SMBH and $r_{\rm min}\propto M^{0.28} M_{\rm SMBH}^{0.69} r_0^{0.31}$ from Eq. (29b) in Ref.~\cite{Rasskazov:2019gjw}, defined as the radius where the GW inspiral time falls below the relaxation time:
\begin{align}
	r_{\rm min}(M,M_{\rm SMBH}) &=6.9\times 10^{-5} {\rm pc} \left( \frac{M_{\rm SMBH}}{4\times 10^6\ M_\odot}\right)^{0.69} \left(\frac{r_{\rm max}(M_{\rm SMBH})}{3\ {\rm pc}} \right)^{0.31} \left(\frac{M}{20\ M_\odot} \right)^{0.28} . 
\end{align}
Since $r_{\min}$ is BH-mass dependent, we take $r_{\rm min}\equiv\max[r_{\min}(M_1),r_{\rm min}(M_2)]$. 
% The prescription for the influence radius can be the one in~\cite{Rasskazov:2019gjw} Eq.~(33) based on the $M_{\rm SMBH}-\sigma$ relation
% \begin{align}
	%   r_0=r_{\rm max}(M_{\rm SMBH}) &= 3.14\ {\rm pc}\left( \frac{M_{\rm SMBH}}{4\times 10^6\ M_\odot}\right)^{0.543}
	% \end{align}
% or a more empirical one, e.g. the one from~\cite{Hannah+24}~\al{Elaborate or reference previous elaboration}
% \begin{equation}
	%   r_0 =r_{\rm max}(M_{\rm SMBH}) = 2.25 ~{\rm pc} ~ \left( \frac{M_{\rm SMBH}}{4 \times 10^6 M_\odot} \right)^{0.75}
	% \end{equation}
% resulting in a steeper dependence on $M_{\rm SMBH}$.

Once the range of validity of our approximations is established, the spherically-symmetric phase space distribution of BHs in the SMBH-dominated environment of a GN can be written as 
\begin{align}
	f(r, \overline v, M) = C(M) n(r, M) [1-\overline v^2]^{p_0M/M_{\rm  max}},
\end{align}
with $\overline v=v/v_{\rm max}$, $n(r,M)\equiv r^{-a(M)}$, $a(M)= 3/2 + p_0 M/M_{\rm max}$ and $C(M)$ a normalization function:
\begin{align}
	C(M) = N_{\rm BH} {F_\beta(M)} C_r(M) C_{\overline v}(M) \ . 
\end{align}
Each of the factors reads as follows: $N_{\rm BH} = 2\times 10^4(M_{\rm SMBH}/[4.3\times 10^6 M_\odot])$ is an estimate of the number of BHs segregated into the GN~\cite{Miralda-Escude:2000kqv}; $F_\beta(M)$ is the BH present-day mass function with power law $\beta$ for a 10 Gyr of continuous star formation~\cite{Alexander:2008tq}. Limiting the distribution to BHs in the interval $M\in [M_{\rm min},M_{\rm max}]$,
\begin{equation}
	F_\beta(M) = \begin{dcases} \frac{(1-\beta)M^{-\beta}}{{M_{\rm max}}^{1-\beta}-{M_{\rm min}}^{1-\beta}} & \beta\neq 1\\
		\frac{1}{M(\log M_{\rm max}-M_{\rm min})}& \beta = 1
	\end{dcases}\ .
\end{equation}
Finally $C_r,C_{\overline v}$ are normalization constants defined such that $C_r n$ and $C_{\overline v} (1-\overline v^2)^{p_0M/M_{\rm max}}$ have unit integrals over $4\pi r^2\d r$ and $4\pi \overline v^2 \d \overline v$, respectively, for a fixed BH mass $M$. They read
\begin{align}
	C_r(M) = &\, \frac{3-a(M)}{4\pi}\frac{1}{r_{\rm max}^{3-a(M)}-r_{\rm min}^{3-a(M)}}, \\
	C_{\overline v}(M) = &\, \frac{4^{a(M)-1}a(M)[2a(M)-1]\Gamma^2[a(M)]}{\pi^2\Gamma[2a(M)]} = \frac{\Gamma[a(M)+1]}{\pi^{3/2}\Gamma[a(M)-1/2]}.
\end{align}
With this notation, the differential rate reads:
\begin{align}
	\d^{7}\Gamma_{M_1,M_2} = 4\pi^2 w b^2 C(M_1) C(M_2) n(r, M_1)n(r,M_2) [1-\overline v_1^2]^{p_0M_1/M_{\rm max}}[1-\overline v_2^2]^{p_0M_2/M_{\rm max}} r^2 \d r \d^3\overline v_1\d^3\overline v_2\ .
\end{align}
Continuing our calculation of the merger rate, we enforce that the capture process, both in the GW-only and SC-mediated cases, depends on ${\bf v}_1$ and ${\bf v}_2$ only via the combination $w=|{\bf v}_1 -{\bf v}_2|$. Therefore we multiply by $\d w \delta(w-|{\bf v}_1 -{\bf v}_2|)$ and integrate over $\d^3v_1$ and $\d^3 v_2$. We define 
\begin{align}
	U_{M_1,M_2}(w) \equiv \int \d^3v_1 [1-\overline v_1^2]^{p_0M_1/M_{\rm max}} \int \d^3 v_2 [1-\overline v_2^2]^{p_0M_2/M_{\rm max}} \delta(w-|{\bf v}_1 -{\bf v}_2|)\ .
\end{align}
As shown in~\cite{Gondan:2017wzd}, this relative velocity distribution can be expressed as 
\begin{align}
	U_{M_1,M_2}(\overline w) = \frac{4\pi^2 \overline w}{p_0 M_2/M_{\rm max}+1}\int_{\overline w-1}^{1}d\overline v_1\overline v_1(1-\overline v_1^2)^{p_0 M_1/M_{\rm max}}[1-(\overline v_1 -\overline w)^2]^{p_0 M_2/M_{\rm max}+1},
\end{align}
where $\overline w = w/v_{\rm max}$. Using this distribution, our differential rate reads
\begin{align}
	\d^2 \Gamma_{M_1,M_2} = 4\pi^2 w b^2 r^2 n(r,M_1) n(r,M_2) C(M_1)C(M_2) U_{M_1,M_2}(\overline w)  \d r \d \overline w . 
\end{align}

We can further differentiate this rate in the normalized pericenter distance $\rho_{\rm p0}=r_{\rm p0}/GM_{\rm tot}$. This is useful, since the loudness of the binary in a GW detector will depend explicitly on $\rho_{\rm p0}$ via the signal-to-noise ratio. The dependence on the pericenter distance arises once we express the impact parameter $b$ as a function of the pericenter distance at fixed relative velocity. 
Differentiating $b^2$ over $\rho_{\rm p0}$ by using
\begin{align}
	\frac{\d\sigma}{\d\rho_{\rm p0}}=\frac{\d(\pi b^2)}{\d\rho_{\rm p0}}=\frac{2\pi (GM_{\rm tot})^2}{w^2}\Theta[w_{\rm max}(\rho_{\rm p0})-w]\,,
\end{align}
we obtain 
\begin{align}
	\d^3 \Gamma_{M_1,M_2} = 8\pi^2\frac{(GM_{\rm tot})^2r^2}{v_{\rm max}(r)}  n(r,M_1) n(r,M_2) C(M_1)C(M_2) \frac{U_{M_1,M_2}(\overline w)}{\overline w}\Theta[\overline w_{\rm max}(\rho_{\rm p0}, r)-\overline w] \d \rho_{\rm p0} \d r \d \overline w . 
\end{align}
Therefore the binary merger rate in the GN given the masses and pericenter distance is given by
\begin{align}
	\frac{\d \Gamma_{M_1,M_2}}{\d \rho_{\rm p0}}= 8\pi^2{(GM_{\rm tot})^2}C(M_1)C(M_2)\int_{r_{\rm min}}^{r_{\rm max}} \d r \ r^2 \frac{n(r,M_1)n(r,M_2)}{v_{\rm max}(r)}\int_0^2 \frac{\d \overline w}{\overline w} 
	U_{M_1,M_2}(\overline w)\Theta[\overline w_{\rm max}(\rho_{\rm p0},r)-\overline w]\ .
\end{align}

In our treatment of the problem, the presence of an SC around a BH alters the capture condition, and in particular the maximum relative velocity that allows capture: in general an SC will increase $w_{\rm max}$, thereby increasing the local merger rate. The properties of the SC depend on (i) external parameters which should be considered fixed in nature, but unknown to us (that is the case of the ultralight boson mass $\mu$ and the self-interaction coupling $\lambda$) and (ii) on statistical parameters $\Xi=\{t_{\rm BH}, \chi_{\rm i}, \iota, \omega\}$, such as the age of the primary BH, its initial spin---that determine the SC mass $M_{\rm c}$---and the Euler angles of the specific encounter $\iota, \omega$---that set specifically the energy transfer and hence $w_{\rm max}$. Given this interpretation, we can write the local rate as function of external parameters $(\mu,f=\mu/\sqrt{\lambda})$ and average over all the internal or statistical parameters that contribute to changing $\overline w_{\rm max}$. Explicitly,
\begin{align}
	& \frac{\d \Gamma_{M_1,M_2}^{\rm (SC)}}{\d \rho_{\rm p0}}(\mu, f) = 8\pi^2{(GM_{\rm tot})^2}C(M_1)C(M_2)\int_{r_{\rm min}}^{r_{\rm max}} \d r \ r^2 \frac{n(r,M_1)n(r,M_2)}{v_{\rm max}(r)}\\\nonumber
	&\times \int \d t_{\rm BH}\, p(t_{\rm BH}) \int \d\chi_{\rm i}\, p(\chi_{\rm i})\int_{-1}^{1} \d \cos\iota\, {\cal U}(\cos\iota) \int_{0}^{2\pi} \d \omega \ {\cal U}(\omega) \int_0^2 \frac{\d \overline w}{\overline w} 
	U_{M_1,M_2}(\overline w)\Theta(\overline w_{\rm max}(\rho_{\rm p0},r ; \mu, f, \Xi)-\overline w) \ .
\end{align}
It is evident here that the role of the SC is to change $\overline w_{\rm max}$. 
While the Euler angles are distributed isotropically, the BH age distribution $p(t_{\rm BH})$ and the initial spin distribution $p(\chi_{\rm i})$ are less obviously determined. Our fiducial model assumes a log-uniform distribution $p(t_{\rm BH})={\cal U}(\log_{10} t_{\rm BH})$ between $10^8$ and $10^{10}$ yr for the BH age at the time of the encounter, and a uniform distribution of the initial spin $p(\chi_{\rm i})={\cal U}(\chi_{\rm i})$ in $[0 ,0.998]$. Section~\ref{app:robustness} discusses different choices of this distributions.

\subsection{Total binary merger rate}
To obtain the total binary merger rate we need to integrate it over all the galactic nuclei. This is done by integrating the \textit{average} local merger rate with the distribution of the GNs: 
\begin{align}
	{\cal R}_{\rm merg}^{(\rm SC)}(M_1,M_2, \rho_{\rm p0}) = \int^{M_{\rm SMBH}^{\rm max}}_{M_{\rm SMBH}^{\rm min}} \d M_{\rm SMBH} \frac{\d n_{\rm gal}}{\d M_{\rm SMBH}} \left\langle \frac{\d \Gamma_{M_1,M_2}^{(\rm SC)}}{\d \rho_{\rm p0}}\right\rangle\ . 
\end{align}
Here the mass distribution of SMBHs is given by a fitting function~\cite{Gondan:2017wzd}
\begin{align}
	\frac{\d n_{\rm gal}}{\d M_{\rm SMBH}} = \Phi_* \left( \frac{M_{\rm SMBH}}{M_*} \right)^{-1.25}\exp(-M_{\rm SMBH}/M_*) \ ,
\end{align}
with $M_*=1.3\times 10^8\ M_{\odot}$, $\Phi_*=3.2\times 10^{-11}\ M_{\odot}^{-1}{\rm Mpc}^{-3}$. The upper bound ${M_{\rm SMBH}^{\rm max}}=10^{7}M_\odot$ is obtained by requiring that the relaxation time inside the influence radius is shorter than the Hubble time~\cite{Gondan:2017wzd}. The lower bound is taken ${M_{\rm SMBH}^{\rm min}}=10^{5}M_\odot$~\cite{Gondan:2017wzd,Rasskazov:2019gjw}. The average local merger rate is a function of several properties of the GN, which are assumed to depend \textit{mainly} on the SMBH mass, but in reality vary from galaxy to galaxy: among this we include the exponent of the BH mass function in the GN $\beta$, the minimum and maximum BH masses in the GN $(M_{\rm min},M_{\rm max})$, the radius of influence $r_0$ determining $r_{\rm max}$ and $r_{\rm min}$ in the BH distribution, the segregation parameter $p_0$ and so on. As explained in~\cite{Rasskazov:2019gjw}, the variance can change, usually increasing, the average merger rate compared to its value when instead the average of the parameters is used. This can be taken into account by a quantity $\xi = \prod_i \xi_i$ such that if the dependence in the merger rate on these parameters $x_1,x_2,\dots$ is separable $\Gamma = \prod_i \gamma_i(x_i)$ and with no internal correlations, we can write the average merger rate
\begin{align}
	\langle \Gamma_{M_1,M_2}(x_1,x_2,\dots) \rangle =\left\langle \prod_i \xi_i \gamma_i(x_i)\right\rangle = \xi \Gamma_{M_1,M_2}( \langle{x_1}\rangle, \langle{x_2}\rangle, \dots)\ .
\end{align}
We compute the rate only over averaged parameters, as our understanding requires. The same reasoning applies to the differential rate in the pericenter distance.
% This factor is fixed to $\xi=3$ as estimated in Ref.~\cite{Rasskazov:2019gjw}\al{improve the discussion here}. 
Following~\cite{Rasskazov:2019gjw}, we consider $\xi = \xi_{\sigma} \xi_{\rm inf} \xi_{\rm other}$, where the three terms reflect intrinsic GN variance in the $M_{\rm SMBH}-\sigma$ relationship, the $r_0 - M_{\rm SMBH}$ relationship, and other model parameters.  We follow~\cite{Rasskazov:2019gjw} in setting $\xi_\sigma=3$ and (conservatively) $\xi_{\rm other}=1$, but we account for variance in the distribution of influence radii using the measured scatter~\cite{Hannah+24} and the formalism in Appendix D of Ref.~\cite{Rasskazov:2019gjw} to estimate $\xi_{\rm infl}=2.5$.  Combining the above, we take our fiducial as $\xi = 7.5$.  Although this total $\xi$ is larger than the value used in Ref.~\cite{Rasskazov:2019gjw}, it is much smaller than other values previously considered~\cite{OLeary:2008myb}. The sample of GN that we use to estimate scatter in $r_0$ does not include any post-starburst galaxies~\cite{Hannah+24}, which are empirically seen to (i) have substantially denser GN than regular galaxies~\cite{vanVelzen16}, and (ii) dominate the rate of other density-dependent transients, such as tidal disruption events~\cite{French16} and quasi-periodic eruptions~\cite{Wevers24}. Thus we may be {\it underestimating} the true scatter (and thus the true value of $\xi_{\rm infl}$) and {\it overestimating} the mean $r_0$; both of these biases may have reduced the volume-integrated rate of 2-body captures from GW emission and SC interaction.

With this in mind, we obtain
\begin{align}
	{\cal R}_{\rm merg}^{(\rm SC)}(M_1,M_2, \rho_{\rm p0}) = \int^{M_{\rm SMBH}^{\rm max}}_{M_{\rm SMBH}^{\rm min}} \d M_{\rm SMBH} \frac{\d n_{\rm gal}}{\d M_{\rm SMBH}}  \xi \frac{\d \Gamma_{M_1,M_2}^{(\rm SC)}}{\d \rho_{\rm p0}} \ . 
\end{align}
\subsection{Detection rate}
We write the detection rate per binary by integrating the total merger rate per binary over the comoving volume
\begin{align} 
	{\cal R}_{\rm det}^{{(\rm SC)},\cal D}(M_1,M_2, \rho_{\rm p0}) = \int_0^{z_{\rm max}^{\cal D}(M_1,M_2, \rho_{\rm p0})}\frac{\d z}{1+z}\frac{\d V_{\rm C}}{\d z} {\cal R}_{\rm merg}^{(\rm SC)}(M_1,M_2,\rho_{\rm p0})\ .
\end{align}
The explicit $1/(1+z)$ factor is the source-time to observer-time dilation. The Jacobian transforming the integration from the differential comoving volume to redshift is given by 
\begin{align}
	\frac{\d V_{\rm C}}{\d z} = \frac{4\pi}{(1+z)^2 H_0} \frac{d_{\rm L}^2(z)}{\sqrt{\Omega_{m}(1+z)^3 + \Omega_{\Lambda}}}\ ,
\end{align}
with $H_0=67.5$ km/s/Mpc, $\Omega_m = 0.315$, $\Omega_\Lambda=1-\Omega_m$ cosmological parameters in the $\Lambda$CDM model~\cite{Planck:2018vyg} and the luminosity distance given by
\begin{align}
	d_{\rm L}(z) = \frac{1+z}{H_0} \int_{0}^z \frac{\d z'}{\sqrt{\Omega_{m}(1+z')^3 + \Omega_{\Lambda}}}\ .
\end{align}
The value of $z_{\rm max}^{\cal D}$ corresponds to the maximum redshift our detector $\cal D$ is sensitive to. To find this value we use the formula for the maximum luminosity distance of a binary that can be detected by a detector $\cal D$ with a given SNR~\cite{OLeary:2008myb}:
\begin{align}
	d_{\rm L,\rm max}^{\cal D}(z) = \sqrt{\frac{48}{95} \frac{\eta M_{\rm tot}^3 (1+z)^3 \rho_{\rm p0}^2}{\rm SNR^2}\int_{e_{\rm LSO}}^{e_0} \sum_{n=1}^{n_{\rm max}(e_0)} \frac{\d e}{e} \frac{g(n,e) s(e,e_0)}{S_h^{\cal D}(f_n)}}\ .
\end{align}
Here $\eta = M_1M_2/M_{\rm tot}^2$. The value of $z_{\rm max}^{\cal D}$ is found by solving the equation $d_{\rm L,max}^{\cal D}(z_{\rm max}^{\cal D}) = d_{\rm L}(z_{\rm max}^{\cal D})$. Here $\rho_{\rm p0}$ and $e_0$ are the dimensionless pericenter distance and eccentricity at binary formation, respectively. As in~\cite{OLeary:2008myb,Gondan:2017wzd} we fix $e_0=0.99$, but note that this choice is more motivated if the SC is present as seen in Section~\ref{app:SC_angular_momentum}: the SC lowers the initial eccentricity from values much closer to unity to about a few percent lower. Once the binary is formed, we assume pericenter distance and eccentricity $(\rho_{\rm p},e)$ evolve \textit{only} via GW emission as in~\cite{Peters:1963ux}. Concretely,

\begin{align}
	\rho_{\rm p}(e) = C_0(e_0,\rho_{\rm p0}) \frac{e^{12/19}}{1+e} \left( 1+\frac{121}{304}e^2\right)^{870/2299}\ ,
\end{align}
with $C_0$ a constant set from initial conditions. The evolution is meaningful until the last stable orbit (LSO). The eccentricity at LSO can be approximated by numerically solving $\rho_{\rm p}(e_{\rm LSO}) = (6+2e_{\rm LSO})/(1+e_{\rm LSO})$.
The function $f_n$ represents the frequency of the $n$-th harmonic and it is given by 
\begin{align}
	f_n(e,z) = \frac{n(1-e)^{3/2}}{2\pi \rho_{\rm p}^{3/2} M_{\rm tot} (1+z)}\ .
\end{align}
Harmonics are summed until $n_{\rm max}(e_0) = {\rm floor}[5(1+e_0)^{1/2}/(1-e_0)^{3/2}]$ so that 99\% of the signal power is accounted for~\cite{OLeary:2008myb}. We fix the $\rm SNR=8$ and use the known expressions for $g(n,e), s(e,e_0)$ (Eq. 52 and 56 in~\cite{OLeary:2008myb}). Finally $S_h^{\cal D}$ is the one-sided spectral density of the specific detector under consideration. 

\begin{figure}
	\centering
	\includegraphics[width=0.5\linewidth]{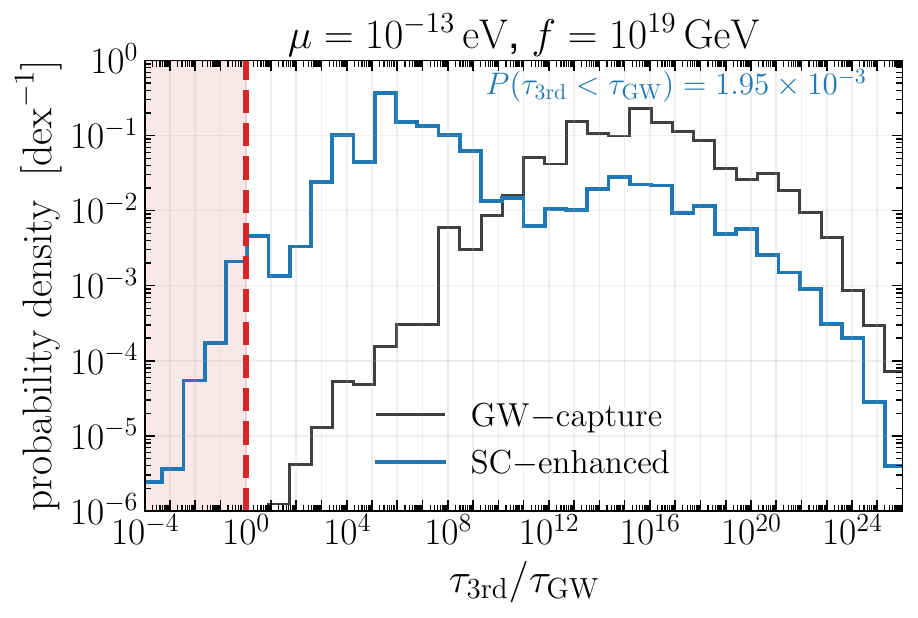}
	\caption{Histogram of $\tau_{\rm 3rd}/\tau_{\rm GW}$ of two simulations with 200000 binaries showing that about than 1 binary in a thousand can get ionized by an encounter with a third body in the case of SC-enhanced capture, and less than one per million in the GW-capture case. Even at the high densities of a GN, the timescale for binary ionization is several orders of magnitude larger than the time to merger.}
	\label{fig:third_body}
\end{figure}

In practice $z_{\rm max}^{\cal D}$ is precomputed for a grid of parameters $(\rho_{\rm p0},M_{\rm tot}, \eta)$, so that the integration over redshift is trivialized. Trivially, one swaps $M_{\rm tot},\eta$ for $M_1,M_2$. The total detection rate is then obtained by
\begin{align} 
	{\cal R}_{\rm det}^{{(\rm SC)},\cal D}=\int_{M_{\rm min}}^{M_{\rm max}} \d M_1 \int_{M_{\rm min}}^{M_1} \d M_2 \int_8^{\rho_{\rm p0}^{\rm max}} \d \rho_{\rm p0}\int_0^{z_{\rm max }^{\cal D}(M_1,M_2, \rho_{\rm p0})}\frac{\d z}{1+z}\frac{\d V_{\rm C}}{\d z} {\cal R}_{\rm merg}^{(\rm SC)}(M_1,M_2,\rho_{\rm p0})\ .
\end{align}
The integration range over $\rho_{\rm p0}$ is not arbitrary. Neglecting direct collisions, i.e. with impact parameter $b<b_{\rm DI} =4 GM_{\rm tot}/w $ imposes $\rho_{\rm p0}>8$. At the same time, the GW inspiral timescale grows with $\rho_{\rm p0}$, and it can be larger than the timescale over which the binary is perturbed by a third body, depending on the density of the environment we are considering. We verified that with $\rho_{\rm p0}^{\rm max}=10^4$ only a small fraction of binaries can get perturbed by a third body. This holds even in the presence of an SC, despite mergers with large $\rho_{\rm p0}$, i.e. with a longer merger time, being possible. This is illustrated in Fig.~\ref{fig:third_body} for an example benchmark, where we compare the typical timescale of an encounter between the binary and a third body 
\begin{align}
	\tau_{3\rm rd}\simeq \frac{1}{n \pi a^2 v(r)}
\end{align}
(with $a=r_{\rm p}/(1-e)$ the binary semi-major axis and $v(r)$ the Keplerian velocity in the GN) with the time to merger $\tau_{\rm GW}$~\cite{Peters:1963ux}.
While in the GW-capture case less than 1 binary per million can be ionized by a third-body encounter, with the SC the perturbed fraction can reach $0.1$\%, due to many more binaries forming with larger $\rho_{\rm p0}$.

The integration ranges for $M_1,M_2$ are dictated by the BH mass function and avoiding double counting. Further conditions or \textit{channels} $\cal C$ can be imposed via a simple Heaviside theta. For instance, if we want to count only the events with $q=M_1/M_2<0.3$, we can multiply the merger rate with $\Theta(0.3-q)$. In a general notation,
\begin{align} 
	{\cal R}_{\rm det}^{{(\rm SC)},\cal D}({\cal C}) =\int_{M_{\rm min}}^{M_{\rm max}} \d M_1 \int_{M_{\rm min}}^{M_1} \d M_2 \int_8^{\rho_{\rm p0}^{\rm max}} \d \rho_{\rm p0}\int_0^{z_{\rm max }^{\cal D}(M_1,M_2, \rho_{\rm p0})}\frac{\d z}{1+z}\frac{\d V_{\rm C}}{\d z} {\cal R}_{\rm merg}^{{(\rm SC)}}(M_1,M_2,\rho_{\rm p0})\Theta({\cal C})\ .
\end{align}

\section{Statistical analysis}\label{app:statistical}
\begin{figure}
	\centering
	\includegraphics[width=\linewidth]{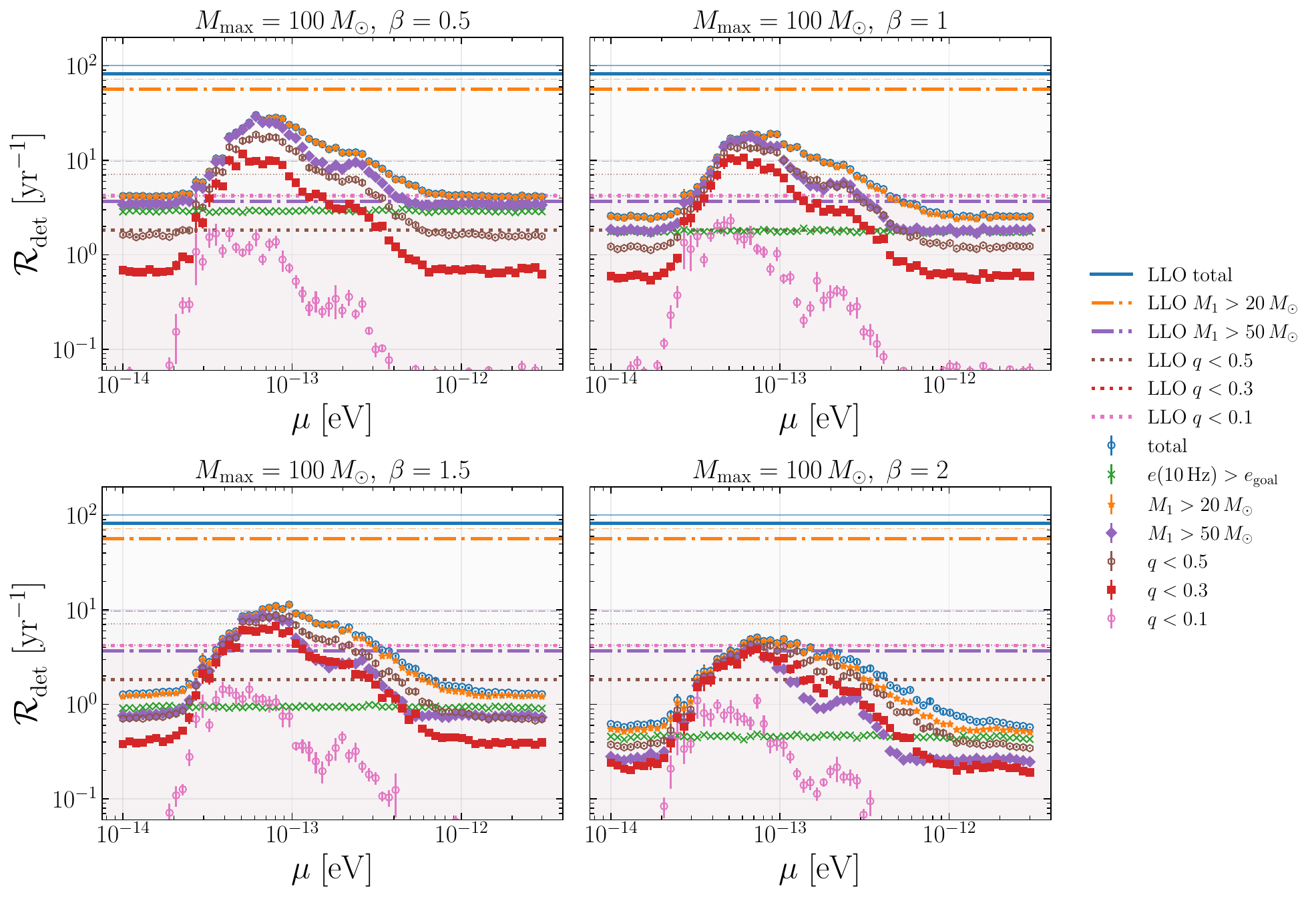}
	\caption{Detection rates obtained from the LLO detections in the GWTC-5 catalog (horizontal lines) compared to prediction of merger rates in LLO from SC-enhanced captures in GNs (scatter points) as a function of the ultralight scalar mass $\mu$. For each LLO rate the thin lines show 90\% upper bounds on this rates. We fix $f=10^{19}$ GeV, $M_{\rm max}=100M_\odot$ and vary $\beta=\{0.5,1,1.5,2\}$. We show different partial rates with different colors. We can exclude intervals in the ultralight scalar mass $\mu$ by comparing the points to the horizontal lines that represent upper bounds on the observed detection rates. }
	\label{fig:det_rate_summary}
\end{figure}

We compare the predicted detection rates computed in Sec.~\ref{app:rates} with the ones from the GWTC-5 catalog~\cite{LIGOScientific:2026ctl,LIGOScientific:2026wfs, LIGOScientific:2026pwx}. 
The \textit{observed} detection rates by a detector ${\cal D}=\{\rm LLO, LHO\}$, i.e. the LIGO Livingston Observatory (LLO) or LIGO Hanford Observatory (LHO), are computed as counts $N_{\cal C}^{\cal D}$ of events in the GWTC-5 (O4b) catalog satisfying the partial rate condition $\cal C$ (\textit{e.g.} ${\cal C}$ can be $e(10\,{\rm Hz})>0.03$, $M_1>50M_{\odot}$ or $q<0.3$ and so on), divided by the observation time of $T_{\rm obs}^{\cal D}=293.1\, {\cal A}_{\cal D}$ days with ${\cal A}_{\cal D}=\{0.68, 0.49\}$ the active fraction of the detector~\cite{LIGOScientific:2026sit} for LLO and LHO respectively. An event in the catalog is counted only if $\rm SNR>8$ in all the analysis pipelines and satisfies the condition $\cal C$ at 90\% confidence level, meaning that the whole reported confidence interval satisfies the condition $\cal C$. The rate is then found as ${\cal R}_{\rm det}^{\cal D}({\cal C})=N_{\cal C}^{\cal D}/T_{\rm obs}^{\cal D}$. The 90\% upper limit ${\cal R}_{\rm det}^{\cal D}<{\cal R}_{\rm 1-\delta}^{\cal D}({\cal C})$ is built using $\chi^2$ quantiles with Poisson statistics. With $\delta=0.1$ for a 90\% confidence interval, 
\begin{align}
	{\cal R}_{\rm 90\%}^{\cal D}({\cal C})&=\frac{1}{2}\frac{\chi^2_{1-\delta,2(N_{\cal C}^{\cal D}+1)}}{T_{\rm obs}^{\cal D}}
\end{align} 
Here $\chi^2_{1-\delta,N}$ is the $1-\delta$ percentile of the $\chi^2$ distribution with $N$ degrees of freedom. We show the results and the comparison between observed and predicted rates for LLO in Fig.~\ref{fig:det_rate_summary}. 

\begin{figure}
	\centering
	\includegraphics[width=0.9\linewidth]{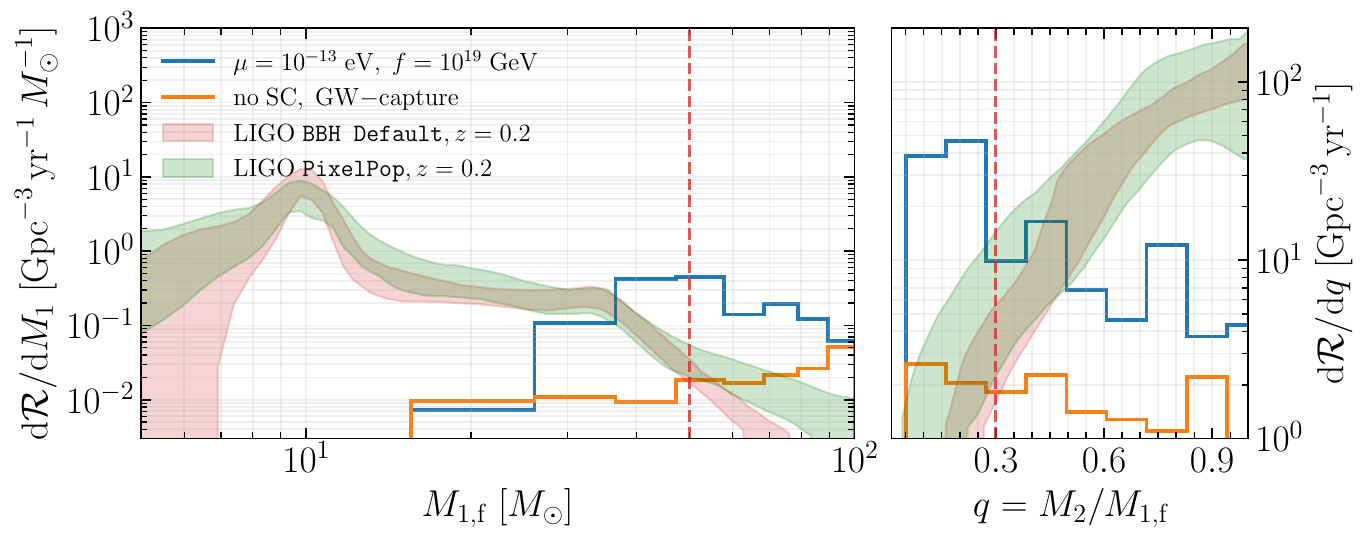}
	\begin{tabular}{c||c|c|c}
		& ${\cal R}_{\rm total}$ [Gpc$^{-3}$yr$^{-1}$] & ${\cal R}(M_1>50 M_\odot)$ [Gpc$^{-3}$yr$^{-1}$] & ${\cal R}(q<0.3)$ [Gpc$^{-3}$yr$^{-1}$]\\ \hline\hline
		LIGO \texttt{BBH Default} GWTC-5  $z=0.2$    &  $16-48$ & $0.11-0.30$& $0.04-0.45$\\
		LIGO \texttt{PixelPop} GWTC-5  $z=0.2$ & $16-65$  & $0.31-1.16$& $0.12-1.70$\\
		\hline
		SC with $\mu = 10^{-13}$ eV, $f=10^{19}$ GeV & {\color{Green}$12\pm 1$} & {\color{Red}$8.462 \pm 0.001$} & {\color{Red}$5.65 \pm 0.002$} \\
		no SC, pure GW-capture &{\color{Green}$1.376 \pm 0.001$}&{\color{Orange}$1.4070 \pm 0.0004$}&{\color{Green}$0.5835 \pm 0.0001$}
	\end{tabular}
	\caption{\textbf{Top} panels: the  differential merger rate (in $M_1$, \textbf{left} and $q$, \textbf{right}) histograms from a binary BH population of $10^{6}$ samples in the SC-enhanced capture (blue) and GW-capture scenarios (orange) compared to the populations shown in the GWTC-5 paper~\cite{LIGOScientific:2026ctl}. The table at the \textbf{bottom} shows integrated merger rates in the different cases. The data extracted from the plot in~\cite{LIGOScientific:2026ctl} is represented as a 90\% confidence interval while the results from our Monte Carlo simulation are shown with the associated Monte Carlo error over 10 batches.  The fields compatible with the LIGO rates are shown in {\color{Green}green}, tensions in {\color{Orange}orange}, while the ones overshooting the LIGO merger rates are shown in {\color{Red}red} and allow to constrain the ultralight scalar. Note that the shown histogram and results correspond to our fiducial choice of astrophysical parameters ($\beta=1$, $M_{\rm max}=100M_\odot$, $p_0=0.5$, radius of influence from~\cite{Hannah+24}, $\chi\in {\cal U}[0,0.998]$, $\log_{10}(t_{\rm BH}/{\rm yr})\in {\cal U}[8,10]$).   }
	\label{fig:LIGO_merger}
\end{figure}

Another possible way to find constraints on the ultralight scalar parameter space is to compare differential merger rates with the ones shown in Ref.~\cite{LIGOScientific:2026ctl}. Integrating these differential merger rates over the appropriate intervals (i.e. $q>0.3$ or $M_1>50M_\odot$), we can exclude cases where the predicted partial merger rate is larger than the one shown in Ref.~\cite{LIGOScientific:2026ctl}. We present this procedure in Fig.~\ref{fig:LIGO_merger} and its attached table. It is clear that the population predicted from SC-enhanced capture and GW-capture are different from the ones inferred by the LVK collaboration: the distributions favor equal-mass inspirals $q=1$ and lighter $\sim 10M_\odot$ BHs with a small feature around $40M_\odot$. Instead, distribution of inspirals originating in GNs are top-heavier and flatter in the mass ratio; compared to GW-capture, SC-enhanced distributions show a bias towards smaller values of $q$ (mostly because of our conservative choice of avoiding non-perturbative interactions at large $q$) and even larger $M_1$. Overall, the case of pure GW-capture looks compatible with the LVK data as a sub-relevant population. The table below the plot shows the total and partial rates in the considered cases: the total SC-enhanced merger rate approaches from below the 90\% confidence interval of the GWTC-5 merger rate with the shown pipelines \texttt{BBH Default} and $\texttt{PixelPop}$, while it completely overshoots the partial rates ${\cal R}(M_1>50 M_\odot)$
and ${\cal R}(q<0.3)$. Note how instead the standard GW-capture lies below or within the GWTC-5 catalog's partial rates; the slight tension for the $M_1>50M_\odot$ rate in the GW-capture case is not representative of the full astrophysical uncertainties, which can affect this rate by one order of magnitude, as we show in other sections of this Supplemental Material. 

% Although SC-mediated two-body capture does not produce observably eccentric inspirals within the LIGO band, this may be a promising avenue for constraining ultralight bosons with future detectors such as Einstein Telescope or LISA~\cite{Sesana:2016ljz} that will probe lower frequencies in the stellar mass BH inspiral. \ns{Done?}

\section{Astrophysical uncertainties}\label{app:robustness}
As in earlier investigations of GW-mediated two-body capture~\cite{OLeary:2008myb, Gondan:2017wzd, Rasskazov:2019gjw}, our results depend on the astrophysical properties and populations of the nuclear star clusters that surround SMBHs. While these inputs are known to some extent, there are both inherent uncertainties as well as substantial variance between GN that must be accounted for. In this section, we discuss two key uncertainties: the structural properties of GN star clusters, and the populations of stellar mass BHs contained therein. In this Section we explore several of these uncertainties and we justify the choice of our fiducial benchmark. 
\subsection{Mass segregation}
In multi-mass systems, energy exchange between heavy and light particles (e.g. stars and stellar mass BHs) leads to mass segregation, with the heavy species attaining a steeper radial profile. When the heavy species is numerous enough to dominate overall rates of energy diffusion (said rates for species $i$ scale as $\propto N_i M_i^2$), a simple form of weak segregation results, with the number density of the heaviest species $n \propto r^{-7/4}$ while lighter species scale as a weaker power law, with asymptotic behavior to $n \propto r^{-3/2}$ for the lightest species~\cite{Bahcall:1977ab}. If the heavy species is too rare to dominate energy diffusion rates, a stronger mass segregation regime will result, although the nature of this regime remains an active topic of theoretical study. The original works on strong segregation paired analytic arguments with both Fokker-Planck~\cite{Alexander:2008tq} and N-body~\cite{Preto+10} simulations to argue that heavy species could follow much steeper profiles, up to $n \propto r^{-11/4}$. More recent analytic work has challenged these results, arguing for a more complex picture in which individual species may follow even steeper power laws at large radii before flattening to an $n \propto r^{-7/4}$ profile at small radii~\cite{LinialSari22}. In light of these complexities, we continue to follow the approximate mass segregation parametrization of Refs.~\cite{OLeary:2008myb, Rasskazov:2019gjw} as a fiducial, somewhat conservative choice: $p_0=0.5$ implies that the heaviest species will have a profile $\propto r^{-2}$, in rough agreement with the milder end of the strong segregation regime, while the lightest species will scale $\propto r^{-3/2}$, in line with weak segregation. 

\begin{figure}
	\centering
	\includegraphics[width=0.48\linewidth]{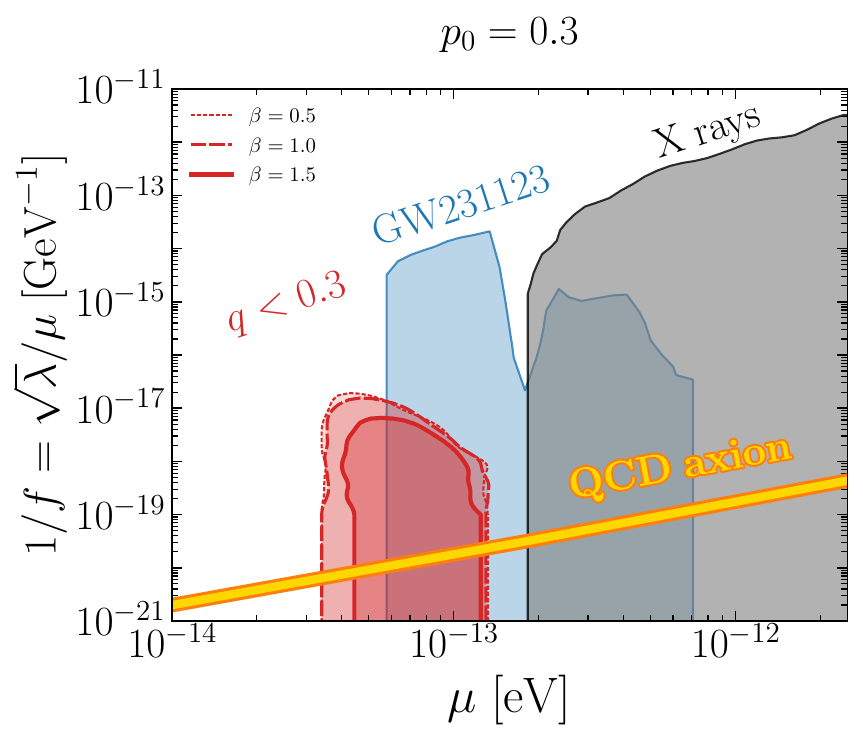}
	\includegraphics[width=0.48\linewidth]{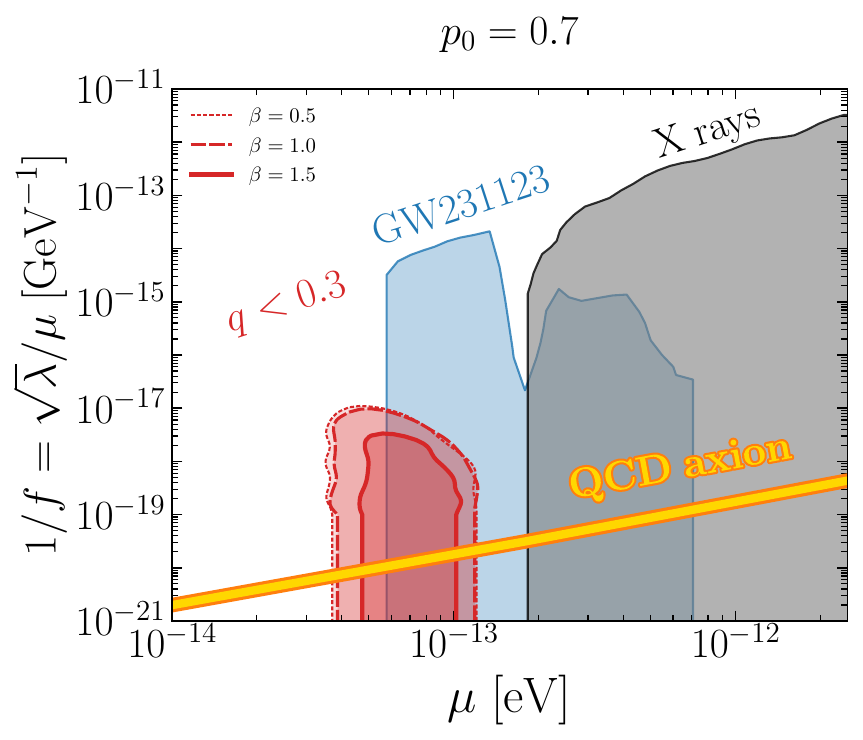}
	\caption{Exclusion plot in ultralight boson mass $\mu$ and self-interaction strength (expressed as $1/f = \sqrt{\lambda}/\mu$) for different values of the segregation parameter $p_0=0.3$ (\textbf{left}) and $p_0 = 0.7$ (\textbf{right}). The red contours represent the excluded region requiring the predicted rate of detections with mass ratio less than 3/10 to not exceed the LLO counts from the GWTC-5 catalog~\cite{LIGOScientific:2026wfs} at 95\% confidence level. We assume a BH mass function in GNs $\propto M^{-\beta}$ with different values of $\beta=\{0.5,1,1.5\}$, shown as different contours.  }
	\label{fig:p0_robustness_exclusion}
\end{figure}

\begin{figure}
	\centering
	\includegraphics[width=0.9\linewidth]{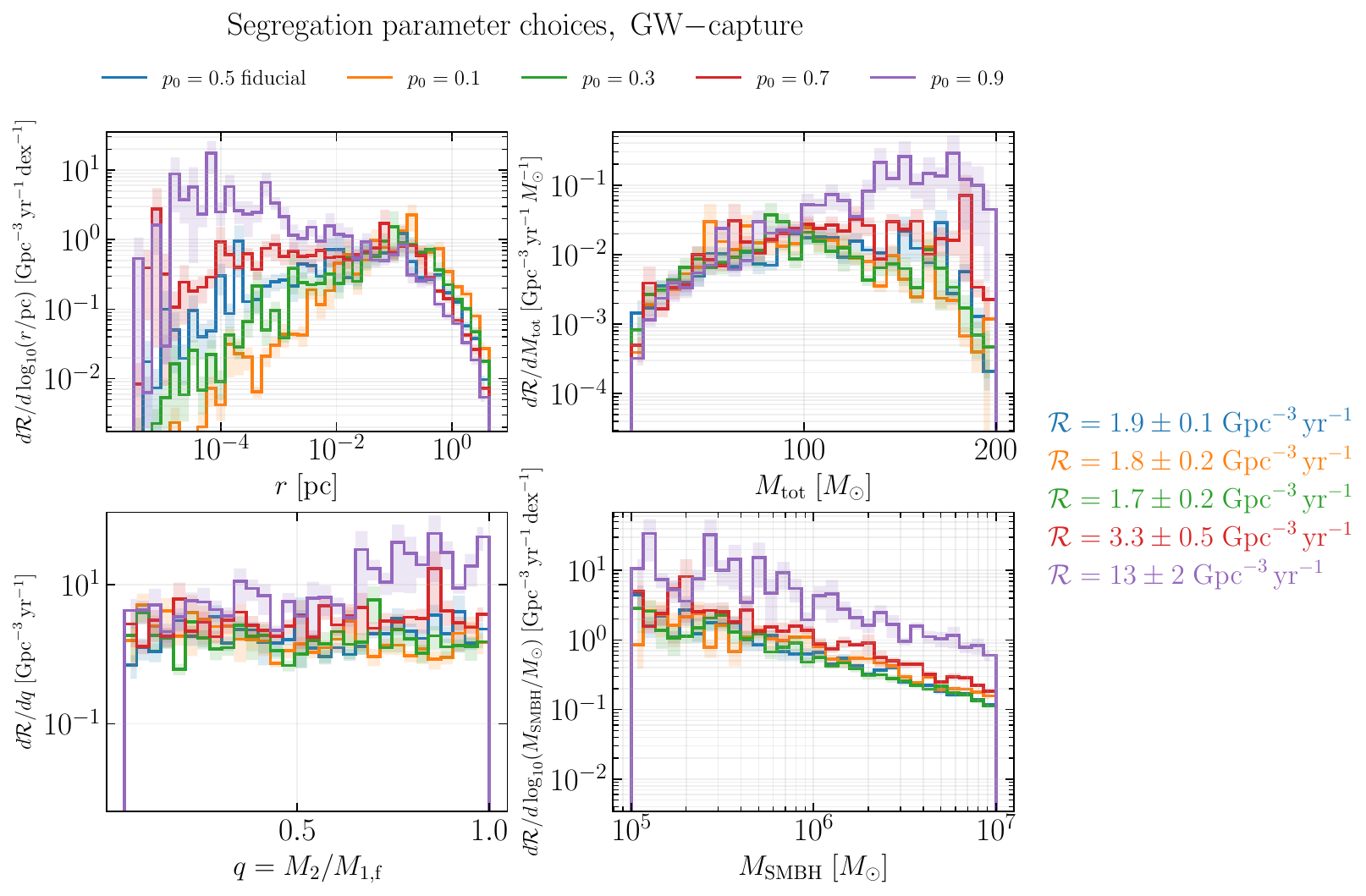}
	\includegraphics[width=0.9\linewidth]{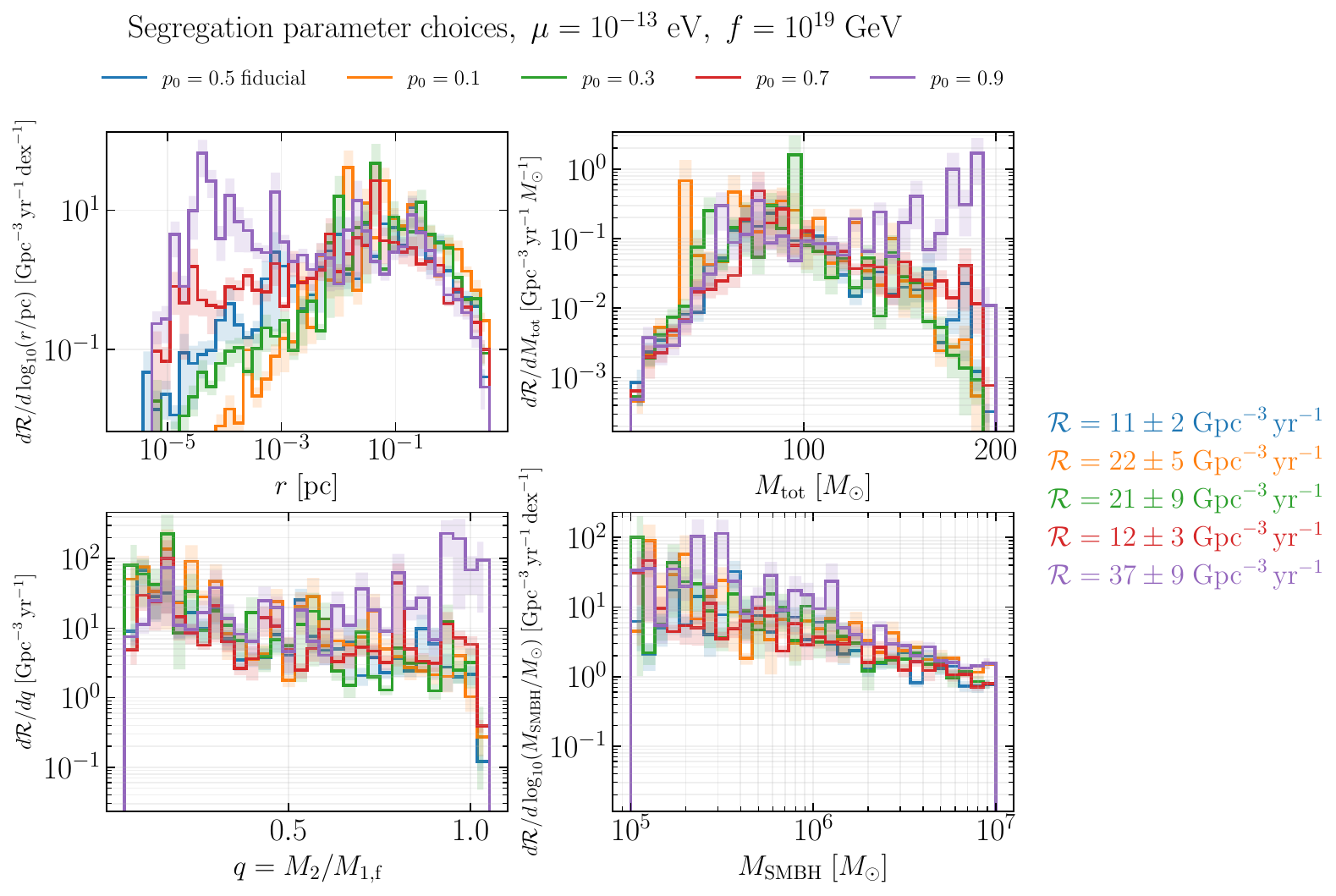}
	\caption{Differential merger rates (in distance from the center of the GN, total mass, mass ratio and SMBH mass) for different choices of the segregation parameter $p_0=\{0.5,0.1,0.3,0.7,0.9\}$ obtained in a MonteCarlo fashion with 200000 binary samples. We consider the vanilla GW-capture case (\textbf{top panels}) and the case of SC-enhanced capture with $\mu=10^{-13}$ eV and $f=10^{19}$ GeV (\textbf{bottom panels}). On the side of the plots we show the total merger rates (integrated over all the parameters).}
	\label{fig:p0_distributions}
\end{figure}

%We choose $p_0=0.5$ as our fiducial value throughout this work. 
Fig.~\ref{fig:p0_robustness_exclusion} shows how the exclusion contours in the $(\mu, 1/f)$ parameter space for ultralight scalars changes with a lower and a larger value of $p_0$. 
We explore the effect of the segregation parameter on various differential merger rates in detail in Fig.~\ref{fig:p0_distributions}. While for the pure GW-capture case a larger segregation parameter simply implies a larger merger rate, the SC-mediated calculations shows that also low values of $p_0<0.5$ can lead to a merger rate larger than in the $p_0=0.5$ by roughly a factor of two, as already hinted by Fig.~\ref{fig:p0_robustness_exclusion}.

\subsection{Influence radius}
Aside from the density profile shape, a key component of our GN modeling is density profile normalization, which depends on the actual spatial extent of the GN. This is usually described in terms of the radius of influence, i.e. the radius $r_0$ where the enclosed total stellar mass equals twice the SMBH mass $2M_{\rm SMBH} =  M_{\rm enc}(r<r_0)$. We then identify $r_{\rm max}\equiv r_0$. 
In the literature it is common 
to use an empirical calibration of the $M_{\rm SMBH}-\sigma$ relationship between SMBH mass and velocity dispersion, as in~\cite{Rasskazov:2019gjw}, where the authors use
\begin{equation}
	M_{\rm SMBH} = 3.097 \times 10^8 M_\odot \left( \frac{\sigma}{200\ {\rm km/s}} \right)^{4.384},
\end{equation}
giving (under the assumption of a flat galactic velocity dispersion profile $\sigma$) the radius of influence
\begin{align}
	r_0 = \frac{GM_{\rm SMBH}}{\sigma^2} = 3.14\ {\rm pc} \left( \frac{M_{\rm SMBH}}{4 \times 10^6 M_\odot} \right)^{0.543}\ .
\end{align}
However, rather than assuming that $r_0 = G M_{\rm SMBH} / \sigma^2$, we take a data-driven approach and empirically calibrate the influence radius from a large sample of GNs observed with the Hubble Space Telescope. 
We specifically employ Ref.~\cite{Hannah+24}, and in particular their analysis of galaxies with robust dynamical mass estimates for SMBHs.  Note that in Ref.~\cite{Hannah+24}, the definition of the influence radius was slightly different than the one used in this paper (i.e. Ref.~\cite{Hannah+24} defined $r_0$ as the radius enclosing $1M_{\rm SMBH}$ rather than $2M_{\rm SMBH}$).  We have increased the prefactor on $r_0$ to match our definition under the assumption of a typical $\gamma=7/4$ stellar cusp, so we use
\begin{equation}
	r_0 = 3.57 ~{\rm pc} ~ \left( \frac{M_{\rm SMBH}}{4 \times 10^6 M_\odot} \right)^{0.75}.
\end{equation}
This relationship is ultimately somewhat steeper than that in Ref.~\cite{Rasskazov:2019gjw}, which will decrease two-body capture rates for large $M_{\rm SMBH}$ but increase rates for small $M_{\rm SMBH}$. At the same time this relation is less steep than the results of~\cite{Thomas+16}, where $r_0 = 1.63{\ \rm pc}\ ( {M_{\rm SMBH}}/{[4 \times 10^6 M_\odot]} )^{0.812}$. 

\begin{figure}
	\centering
	\includegraphics[width=0.9\linewidth]{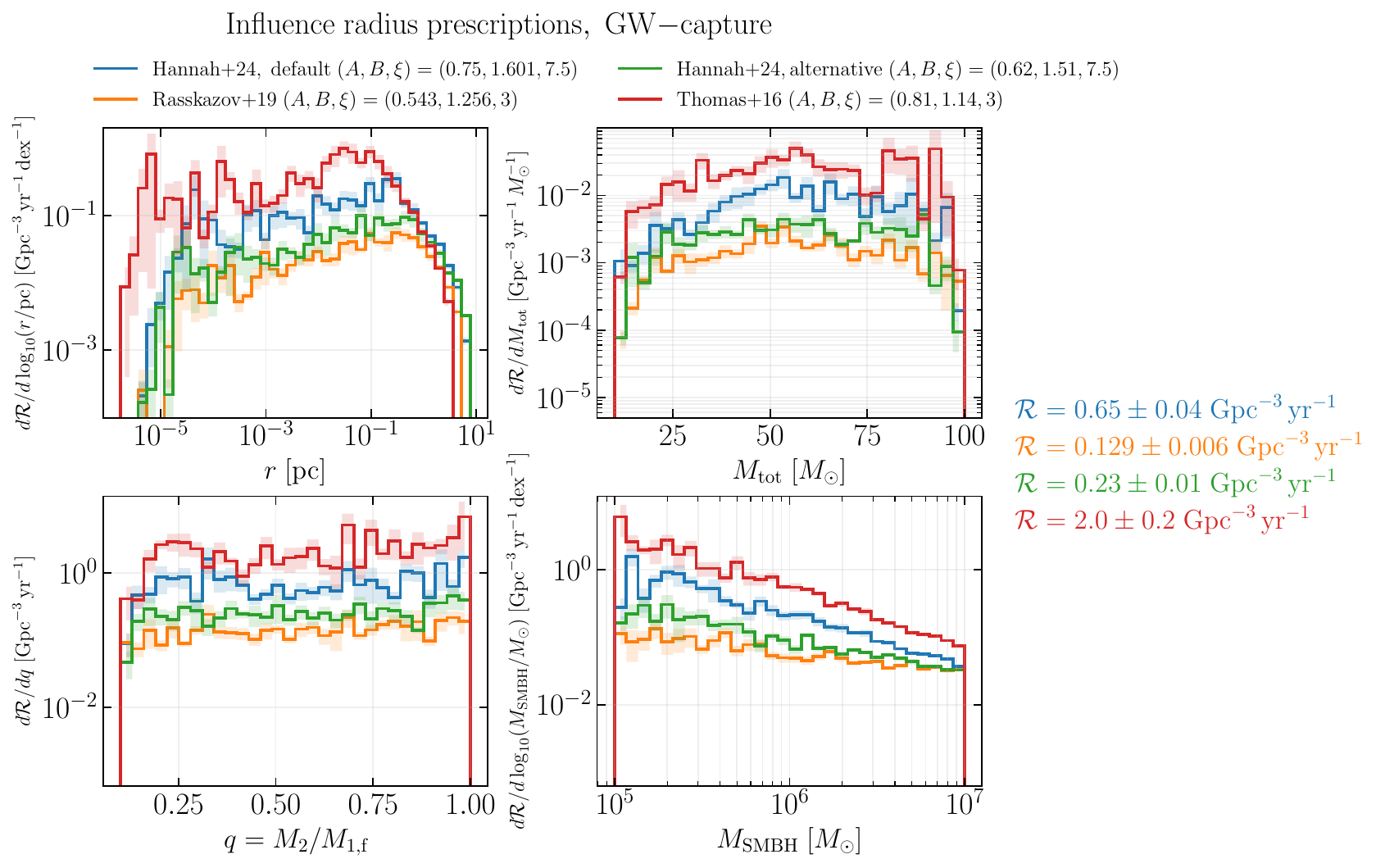}
	\includegraphics[width=0.9\linewidth]{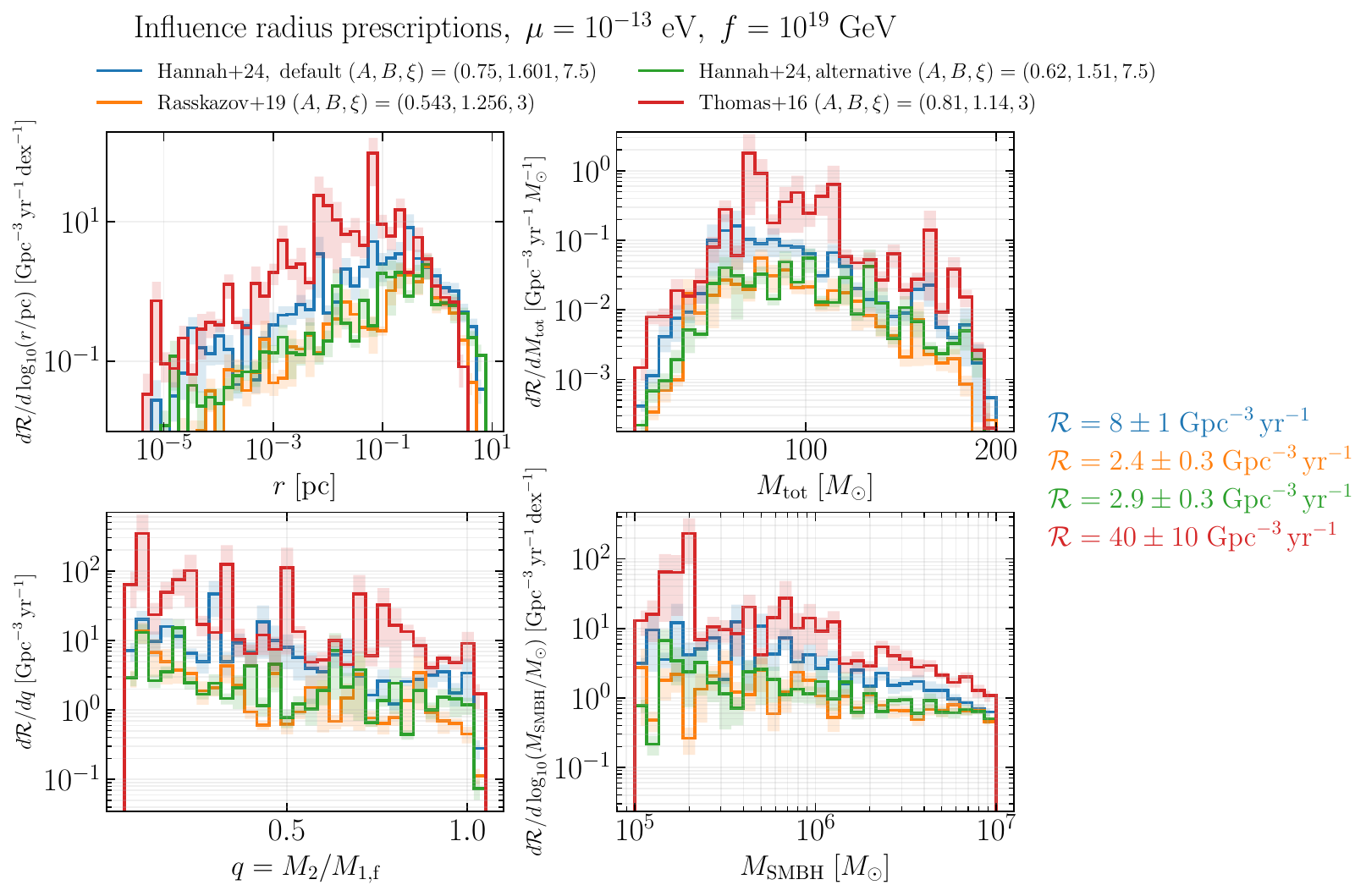}
	\caption{Same as Fig.~\ref{fig:summary_mu_inv_f}, but we vary the prescriptions for the influence radius $r_0$ which determines the $r_{\rm max}$ in the GN BH number density. We indicate in the legend the values of $(A,B)$ to substitute in Eq.~\eqref{eq:rinfl_prescriptions} to obtain the different prescriptions. Hannah+24~\cite{Hannah+24} (in blue) is our fiducial prescription while the one in green refers to a prescription obtained from a larger dataset that includes estimates from the $M_{\rm SMBH}-\sigma$ relation rather than direct measurements of $M_{\rm SMBH}$. Rasskazov+19 (orange) is the prescription adopted in Ref.~\cite{Rasskazov:2019gjw} and Thomas+16 is the prescription obtained from the results presented in Ref.~\cite{Thomas+16}. The rates depend sensibly on the $r_0$ prescription due to the different number density. }
	\label{fig:rinfl_distributions}
\end{figure}

To compare results with different $r_0$ prescription, we represent them in the form
\begin{align}\label{eq:rinfl_prescriptions}
	\log_{10}\frac{r_0}{\rm pc} = A \log_{10}\left( \frac{M_{\rm SMBH}}{10^{8}M_\odot}\right) +B
\end{align}
for some values $A,B$. In Fig.~\ref{fig:rinfl_distributions} we compare (differential) merger rates under different influence radius prescriptions both in the GW-capture case and in one SC-enhanced capture scenario. We see that the choice of the prescription has a relevant impact on the merger rates.

\subsection{Black hole mass function}

Another important astrophysical uncertainty comes from the choice of stellar mass BH mass function. Throughout this work we assume that ${\rm d}N/{\rm d}M \propto M^{-\beta}$ between a minimum mass $M_{\rm min}$ and a maximum mass $M_{\rm max}$.

\begin{figure}
	\centering
	\includegraphics[width=0.48\linewidth]{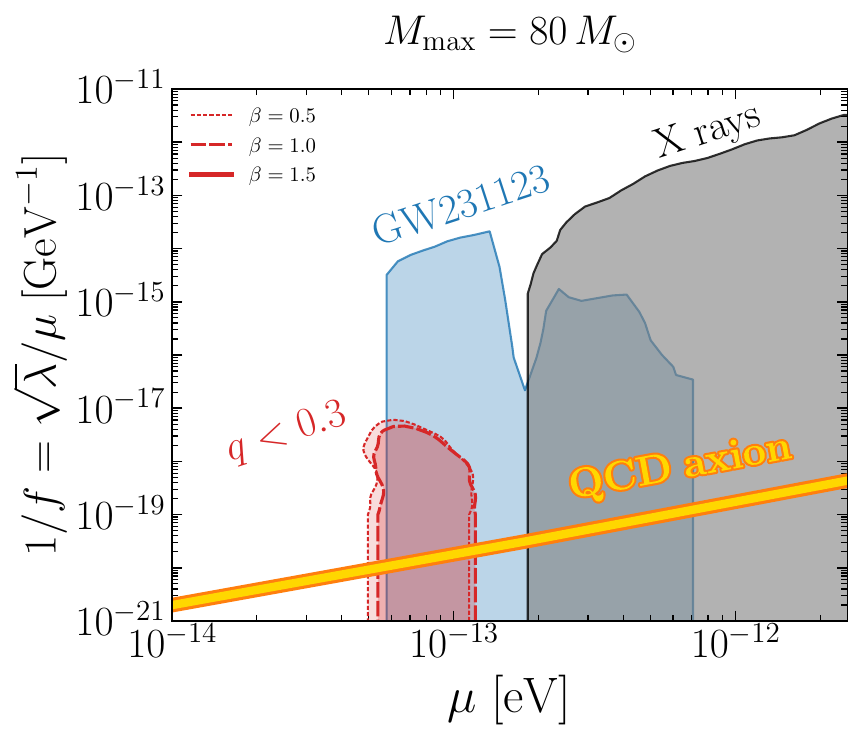}
	\caption{Same as Fig.~\ref{fig:summary_mu_inv_f}, but we now vary $M_{\rm max} = 80 M_\odot$ instead of the fiducial value of $M_{\rm max} = 100 M_\odot$. Comparing this to the summary plot in Fig.~\ref{fig:summary_mu_inv_f} in the main text, we see that the excluded parameter space is reduced.}
	\label{fig:Mmax_80_exclusion}
\end{figure}
We take $M_{\rm min} = 5 M_\odot$, in approximate agreement with both electromagnetic observations of BH X-ray binaries~\cite{Ozel+10} and the more recent LVK sample~\cite{LIGOScientific:2026wfs}. The choice of $M_{\rm max}$ is much less clear. A low value of $M_{\rm max} \approx (20-50) M_\odot$ could be motivated by theoretical models for isolated massive star evolution, which find that for the high-metallicity populations prevalent in GN~\cite{Neumayer+20}, powerful line-driven winds cap BH masses in this approximate range~\cite{SperaMapelli17, Ugolini+25}. Even in low-metallicity stellar populations, larger masses are expected to be suppressed by the onset of destructive pair instability episodes during core carbon burning phases, creating a pair instability mass gap~\cite{RakavyShaviv67, Barkat+67}. Although the exact width of the pair instability mass gap remains debated, it is expected to begin at a BH mass of roughly $45-60 M_\odot$~\cite{Farmer+19}. Nevertheless, we consider a substantially larger value of $M_{\rm max}=100 M_\odot$ as our fiducial case. Our reasoning here is two-fold. Empirically, the LVK sample includes a substantial tail of mergers with progenitor component masses out beyond $100 M_\odot$~\cite{LIGOScientific:2026wfs}; the largest such event, GW231123, has a primary with mass $M_1 = 137_{-18}^{+23} M_\odot$~\cite{LIGOScientific:2025rsn}. The existence of such high-mass BHs, in (and possibly beyond) the pair instability mass gap, is most commonly attributed to hierarchical mergers and growth, which can only occur in dense environments where binaries are dynamically assembled and merger products can be retained despite large ($\sim 10^{2-3}~{\rm km~s}^{-1}$) GW recoil kicks. Although hierarchical growth can occur in globular clusters, the large escape velocities of galactic nuclei may favor them as sites for multiple generations of mergers. Theoretical calculations of repeated mergers in active galactic nuclei (the phase of GN evolution when {\it in situ} star formation occurs) indicate that masses far beyond $100 M_\odot$ can be achieved~\cite{Yang+19, Gonglewski+26}.

We show the impact of a lower $M_{\rm max}$ on our results in Fig.~\ref{fig:Mmax_80_exclusion}. Since the merger rate for SC-enhanced capture is dominated by the heaviest binaries, the excluded area in the $(\mu,1/f)$ plane shrinks considerably. 

Finally, there is the question of $\beta$. Isolated stellar evolution-through supernova calculations find~\cite{Ugolini+25} $\beta \approx \beta_{\rm IMF}$, where $\beta_{\rm IMF} $ is the high mass slope of the stellar initial mass function. For field populations, $\beta_{\rm IMF} \approx 2.3$ is a near-universal value, however, {\it in situ} star formation in GN likely produces a flatter slope, with $\beta_{\rm IMF} \approx 1.7$ measured in our own Galactic Center~\cite{Lu+13}. However, the same GN environments that create nuclear populations of stellar mass BHs also catalyze repeated mergers and growth in the dissipative environments of active galactic nucleus accretion disks, likely flattening~\cite{Gonglewski+26} the BH mass function by an amount $\Delta \beta \approx 1.3$~\cite{Yang+19}. We therefore consider a range of $\beta \approx 0.4 - 1$ as our fiducial case, although we stress that this is uncertain and larger $\beta$ values may also be achieved in nature.
Given the crucial role of the parameter $\beta$ we always vary it between 0.5 and 1.5 when showing our results, namely the exclusion regions in the $(\mu, 1/f)$ parameter space of Fig.~\ref{fig:summary_mu_inv_f} in the main text and in Figs.~\ref{fig:p0_robustness_exclusion},~\ref{fig:chi_i_robustness_exclusion},~\ref{fig:Mmax_80_exclusion} and~\ref{fig:perturbativity_exclusion} in this Supplemental Material.

\subsection{Initial spin distribution}
\begin{figure}
	\centering
	\includegraphics[width=0.6\linewidth]{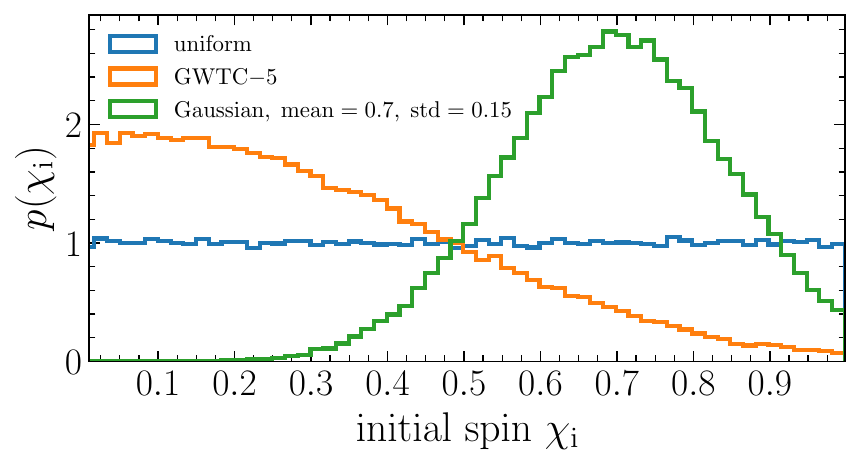}
	\caption{Histograms of the three initial spin distributions considered in our work.}
	\label{fig:spin_distributions}
\end{figure}

\begin{figure}
	\centering
	\includegraphics[width=0.48\linewidth]{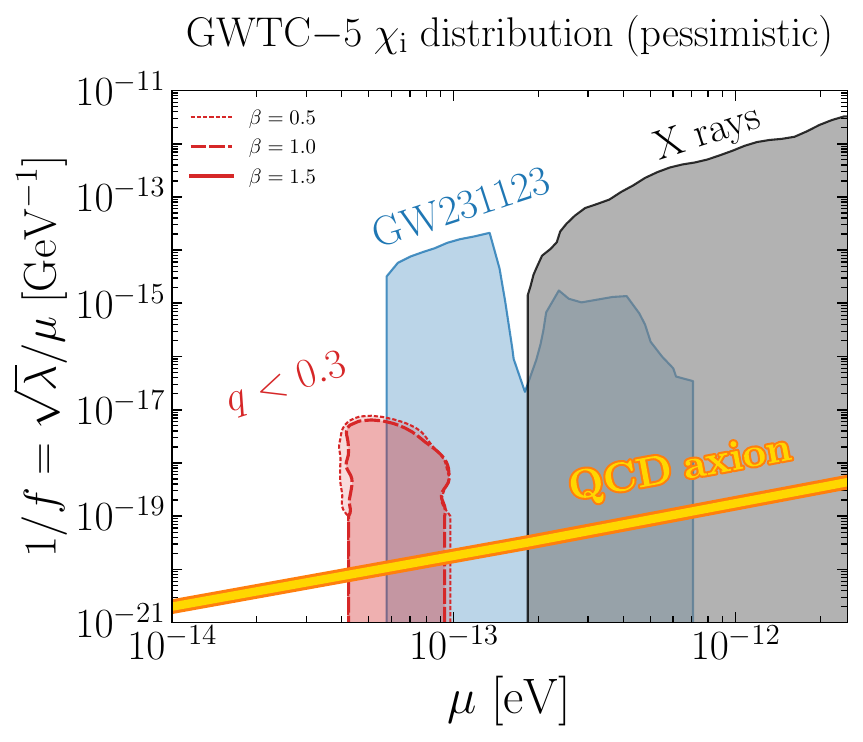}
	\includegraphics[width=0.48\linewidth]{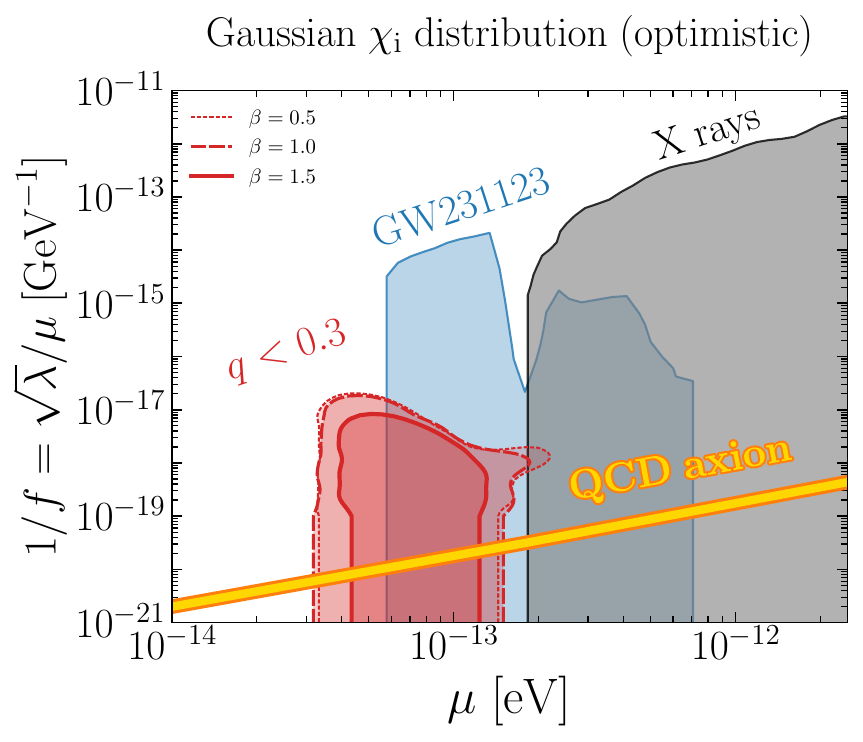}
	\caption{Same as Fig.~\ref{fig:summary_mu_inv_f}, but we compare the exclusion plots obtained under different assumptions for the initial spin distribution. We consider the GWTC-5 empirical distribution from the LVK binaries and a simple Gaussian with mean 0.7 and standard deviation of 0.15 as the pessimistic (\textbf{left}) and optimistic case (\textbf{right}). As evident from Fig.~\ref{fig:summary_mu_inv_f} in the main text, the agnostic uniform distribution represents a conservative choice in the middle of the ones presented here. The distributions are shown in Fig.~\ref{fig:spin_distributions}.}
	\label{fig:chi_i_robustness_exclusion}
\end{figure}

Unlike the previous uncertainties, the BH spin distribution characterizes the SC-enhanced capture process, since in this case the energy transfer is dependent on the presence and mass of the SC, which is sensitive to the BH spin. 
In practice, as we have seen in Sec.~\ref{app:evol_SC}, the value of the initial spin $\chi_{\rm i}$ sets what is the maximum value of the SC mass, for a given initial BH mass and ultralight scalar mass. Therefore, changing the distribution of $\chi_{\rm i}$ in the GN will change the impact of the SC-enhanced capture. To explore the effect, we consider (i) an agnostic and rather conservative uniform spin distribution ${\cal U}[0,0.998]$, (ii) the spin distribution extracted from the binaries in the GWTC-5 catalog~\cite{LIGOScientific:2026ctl} as an example of rather pessimistic and skewed towards lower spins, (iii) a Gaussian distribution peaked about $\chi_{\rm i}=0.7$ and with standard deviation of 0.15, as an example of a more realistic, but still optimistic spin distribution of heavy BHs in a GN. These three distributions are shown in Fig.~\ref{fig:spin_distributions}. Our logic is the following: the larger BH number densities in GNs are driven by mass segregation, and this environment favors repeated and hierarchical mergers. %These have been shown to favor larger spin remnants\al{Nick help with this part and references}. 
The spin angular momentum of BH merger products is generally dominated by the orbital angular momentum of the progenitor binary at the point of merger, and as a result, the merger product $\chi$ distribution is tightly clustered around $\chi \approx 0.7$~\cite{Gerosa21}.  While this type of distribution would be more suited with the population of BHs that end up dominating the SC-enhanced capture channel, we decide to be conservative and to choose the agnostic uniform distribution as our reference initial spin distribution. 
In Fig.~\ref{fig:chi_i_robustness_exclusion} we show the impact of the different initial spin distributions on the constraints in the ($\mu,1/f)$ plane. Intuitively, a distribution skewed towards lower spins does not allow for a large number of BHs to be dressed with an SC, thereby limiting the constraining power of merger rates. On the contrary a distribution skewed towards faster spinning-BHs, which is more likely to be the true case in GNs, considerably increases the constrained area of parameter space. Nevertheless, to be conservative, the results in the main text refer to the more agnostic uniform distribution.
\subsection{Black hole age distribution}
\begin{figure}
	\centering
	\includegraphics[width=0.48\linewidth]{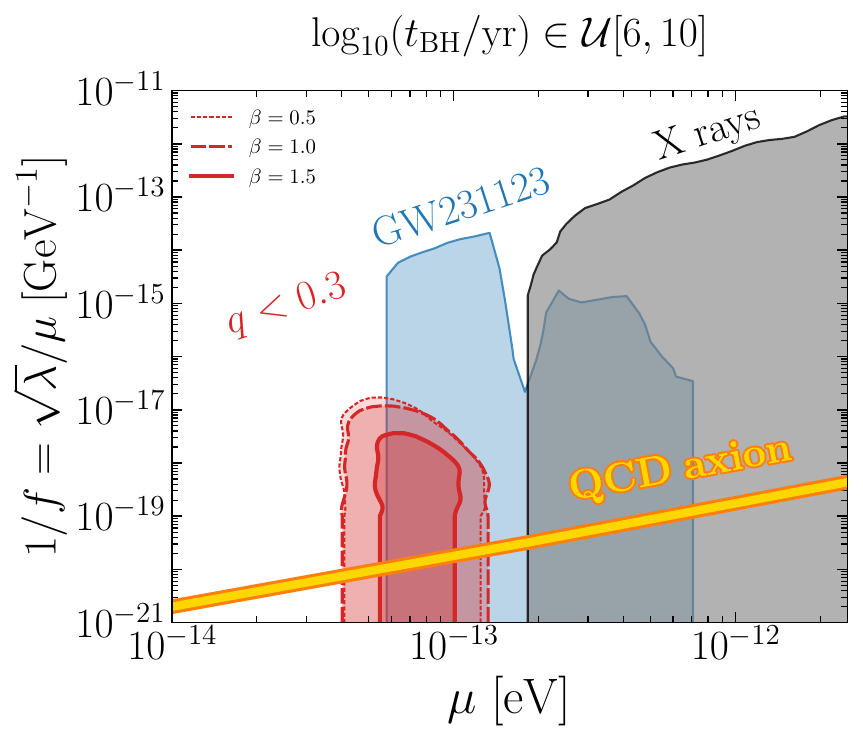}
	\caption{Same as Fig.~\ref{fig:summary_mu_inv_f}, but the distribution of black hole ages is extended to a log-uniform distribution between $10^6$ yr and $10^{10}$ yr. The impact of this modification on the exclusion contours is minor.}
	\label{fig:tBH_6_10_robustness}
\end{figure}

Another uncertainty that characterizes the SC-enhanced capture process only is the BH age distribution, since the cloud state and mass depends on the age of the BH. This is well depicted in Figs.~\ref{fig:cloudevolv} and~\ref{fig:cloudevol2} that show $q_{\rm c}(t)$. Our fiducial choice is to take a log-uniform distribution of black hole ages between $10^8$ yr and $10^{10}$ yr, as most GNs formed the large majority of their stellar population over 1 Gyr ago~\cite{Neumayer+20}, with a tail of high-mass star formation extending to the present day~\cite{Levin03}. In Fig.~\ref{fig:tBH_6_10_robustness} we show the exclusion contours in the ultralight scalar parameter space if we allow for younger BHs with a log-uniform  distribution between $10^6$ and $10^{10}$ yr, $\log_{10}(t_{\rm BH}/{\rm yr}) \in {\cal U}[6,10]$. By comparing Fig.~\ref{fig:tBH_6_10_robustness} with Fig.~\ref{fig:summary_mu_inv_f}, we see that the BH age distribution does not have a large impact on the results.
\section{Perturbativity}\label{app:perturbativity}
In this Section, we show what our results would look like if we were not restricting the energy transfer calculation presented in Sec.~\ref{app:SC_energy} to the perturbative result enforced by the condition on the total transition probability from the SR state $s$ $P_{\rm tot}^s<0.1$. This condition affects mainly cases in which the perturber and the primary BH with the SC have similar mass, $q>0.3$. We highlight that the $P_{\rm tot}^s<0.1$ requirement is \textit{necessary} in our formalism since we do not consider the time evolution of the orbit as a response to the interaction between the perturber and the cloud. However, we want to illustrate, and perhaps motivate a more in-depth, non-perturbative (and inevitably complicated) treatment of the system by showing a raw extrapolation of our calculation, basically ignoring the perturbativity constraint altogether.

\begin{figure}
	\centering
	\includegraphics[width=0.32\linewidth]{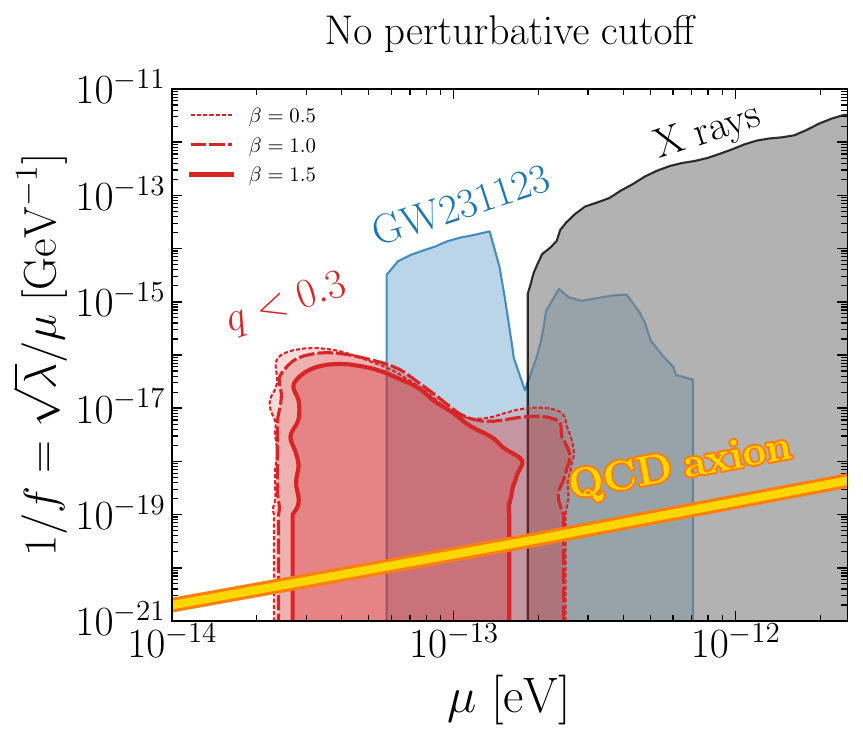}
	\includegraphics[width=0.32\linewidth]{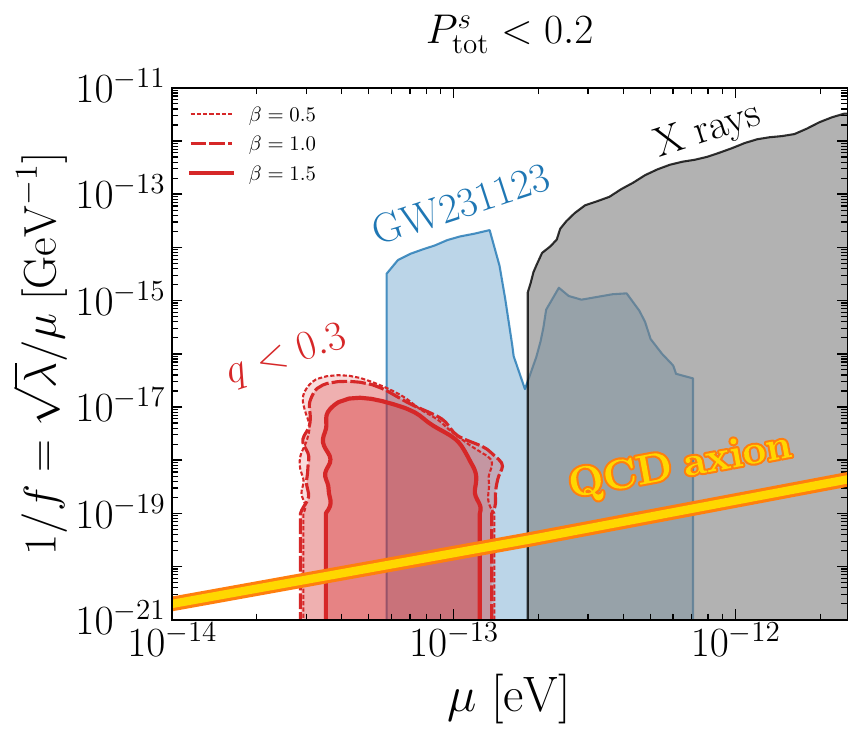}
	\includegraphics[width=0.32\linewidth]{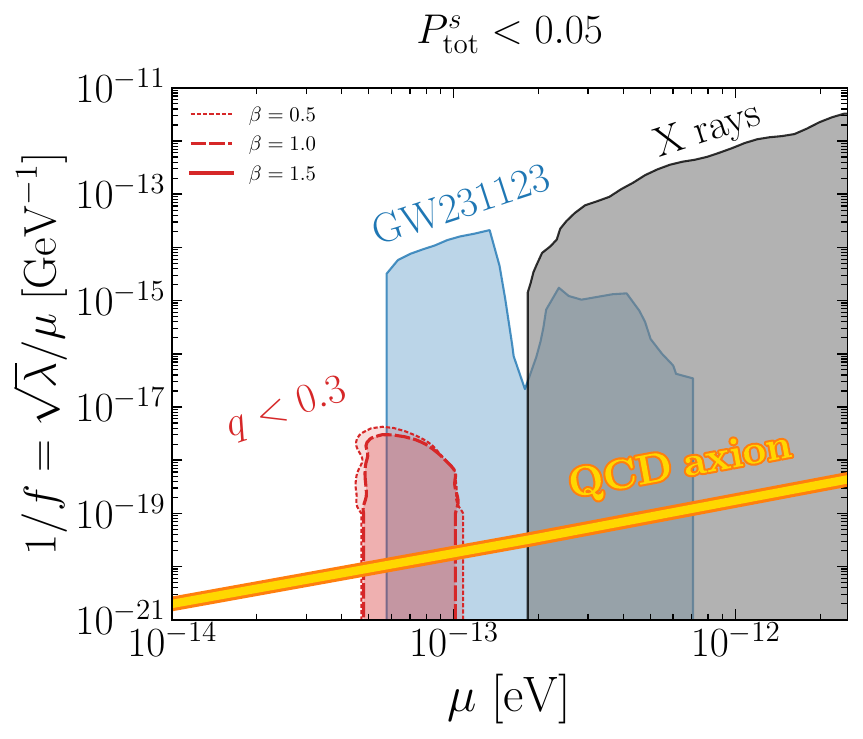}
	\caption{Same as Fig.~\ref{fig:summary_mu_inv_f},, but we adopt the fiducial model without enforcing the perturbativity condition on the total transition probability $P_{\rm tot}^s<0.1$ (\textbf{left} panel). As explained in the main text, this is \textit{not} a correct assumption and we show this results just for illustration of the capability of the method. In the other panels we vary the cutoff by a factor of 2 $P_{\rm tot}^s<0.2$ (\textbf{center}, relaxed cutoff ) and $P_{\rm tot}^s<0.05$ (\textbf{right}, aggressive cutoff) to show how the exclusion contours are affected by the cutoff choice. }
	\label{fig:perturbativity_exclusion}
\end{figure}

The results are shown in Fig.~\ref{fig:perturbativity_exclusion}. For our fiducial case, but with the kernel functions not restricted by the $P_{\rm tot}^s<0.1$ condition, the excluded parameter space is substantially larger, as it is evident if we compare Fig.~\ref{fig:perturbativity_exclusion} to Fig.~\ref{fig:summary_mu_inv_f} in the main text. We also show in Fig.~\ref{fig:perturbativity_exclusion} how the exclusion contours vary if we change the cutoff value to $P_{\rm tot}^s<0.2$ (relaxed cutoff) or to $P_{\rm tot}^s<0.05$ (aggressive cutoff). 

Our results motivate the need of a non-perturbative calculation of the binary formation event in case of equal (or close to) mass BHs, perhaps via numerical simulations, similarly to the case of study in Ref.~\cite{Guo:2025pea}, but to much higher eccentricities, closer to the open orbits.

%% file: draft.bbl
\begin{thebibliography}{111}%
\makeatletter
\providecommand \@ifxundefined [1]{%
 \@ifx{#1\undefined}
}%
\providecommand \@ifnum [1]{%
 \ifnum #1\expandafter \@firstoftwo
 \else \expandafter \@secondoftwo
 \fi
}%
\providecommand \@ifx [1]{%
 \ifx #1\expandafter \@firstoftwo
 \else \expandafter \@secondoftwo
 \fi
}%
\providecommand \natexlab [1]{#1}%
\providecommand \enquote  [1]{``#1''}%
\providecommand \bibnamefont  [1]{#1}%
\providecommand \bibfnamefont [1]{#1}%
\providecommand \citenamefont [1]{#1}%
\providecommand \href@noop [0]{\@secondoftwo}%
\providecommand \href [0]{\begingroup \@sanitize@url \@href}%
\providecommand \@href[1]{\@@startlink{#1}\@@href}%
\providecommand \@@href[1]{\endgroup#1\@@endlink}%
\providecommand \@sanitize@url [0]{\catcode `\\12\catcode `\$12\catcode
  `\&12\catcode `\#12\catcode `\^12\catcode `\_12\catcode `\%12\relax}%
\providecommand \@@startlink[1]{}%
\providecommand \@@endlink[0]{}%
\providecommand \url  [0]{\begingroup\@sanitize@url \@url }%
\providecommand \@url [1]{\endgroup\@href {#1}{\urlprefix }}%
\providecommand \urlprefix  [0]{URL }%
\providecommand \Eprint [0]{\href }%
\providecommand \doibase [0]{https://doi.org/}%
\providecommand \selectlanguage [0]{\@gobble}%
\providecommand \bibinfo  [0]{\@secondoftwo}%
\providecommand \bibfield  [0]{\@secondoftwo}%
\providecommand \translation [1]{[#1]}%
\providecommand \BibitemOpen [0]{}%
\providecommand \bibitemStop [0]{}%
\providecommand \bibitemNoStop [0]{.\EOS\space}%
\providecommand \EOS [0]{\spacefactor3000\relax}%
\providecommand \BibitemShut  [1]{\csname bibitem#1\endcsname}%
\let\auto@bib@innerbib\@empty
%</preamble>
\bibitem [{\citenamefont {Peccei}\ and\ \citenamefont
  {Quinn}(1977)}]{Peccei:1977hh}%
  \BibitemOpen
  \bibfield  {author} {\bibinfo {author} {\bibfnamefont {R.~D.}\ \bibnamefont
  {Peccei}}\ and\ \bibinfo {author} {\bibfnamefont {H.~R.}\ \bibnamefont
  {Quinn}},\ }\href {https://doi.org/10.1103/PhysRevLett.38.1440} {\bibfield
  {journal} {\bibinfo  {journal} {Phys. Rev. Lett.}\ }\textbf {\bibinfo
  {volume} {38}},\ \bibinfo {pages} {1440} (\bibinfo {year}
  {1977})}\BibitemShut {NoStop}%
\bibitem [{\citenamefont {Kim}(1979)}]{Kim:1979if}%
  \BibitemOpen
  \bibfield  {author} {\bibinfo {author} {\bibfnamefont {J.~E.}\ \bibnamefont
  {Kim}},\ }\href {https://doi.org/10.1103/PhysRevLett.43.103} {\bibfield
  {journal} {\bibinfo  {journal} {Phys. Rev. Lett.}\ }\textbf {\bibinfo
  {volume} {43}},\ \bibinfo {pages} {103} (\bibinfo {year} {1979})}\BibitemShut
  {NoStop}%
\bibitem [{\citenamefont {Weinberg}(1978)}]{Weinberg:1977ma}%
  \BibitemOpen
  \bibfield  {author} {\bibinfo {author} {\bibfnamefont {S.}~\bibnamefont
  {Weinberg}},\ }\href {https://doi.org/10.1103/PhysRevLett.40.223} {\bibfield
  {journal} {\bibinfo  {journal} {Phys. Rev. Lett.}\ }\textbf {\bibinfo
  {volume} {40}},\ \bibinfo {pages} {223} (\bibinfo {year} {1978})}\BibitemShut
  {NoStop}%
\bibitem [{\citenamefont {Wilczek}(1978)}]{Wilczek:1977pj}%
  \BibitemOpen
  \bibfield  {author} {\bibinfo {author} {\bibfnamefont {F.}~\bibnamefont
  {Wilczek}},\ }\href {https://doi.org/10.1103/PhysRevLett.40.279} {\bibfield
  {journal} {\bibinfo  {journal} {Phys. Rev. Lett.}\ }\textbf {\bibinfo
  {volume} {40}},\ \bibinfo {pages} {279} (\bibinfo {year} {1978})}\BibitemShut
  {NoStop}%
\bibitem [{\citenamefont {Shifman}\ \emph {et~al.}(1980)\citenamefont
  {Shifman}, \citenamefont {Vainshtein},\ and\ \citenamefont
  {Zakharov}}]{Shifman:1979if}%
  \BibitemOpen
  \bibfield  {author} {\bibinfo {author} {\bibfnamefont {M.~A.}\ \bibnamefont
  {Shifman}}, \bibinfo {author} {\bibfnamefont {A.~I.}\ \bibnamefont
  {Vainshtein}},\ and\ \bibinfo {author} {\bibfnamefont {V.~I.}\ \bibnamefont
  {Zakharov}},\ }\href {https://doi.org/10.1016/0550-3213(80)90209-6}
  {\bibfield  {journal} {\bibinfo  {journal} {Nucl. Phys. B}\ }\textbf
  {\bibinfo {volume} {166}},\ \bibinfo {pages} {493} (\bibinfo {year}
  {1980})}\BibitemShut {NoStop}%
\bibitem [{\citenamefont {Zhitnitsky}(1980)}]{Zhitnitsky:1980tq}%
  \BibitemOpen
  \bibfield  {author} {\bibinfo {author} {\bibfnamefont {A.~R.}\ \bibnamefont
  {Zhitnitsky}},\ }\href@noop {} {\bibfield  {journal} {\bibinfo  {journal}
  {Sov. J. Nucl. Phys.}\ }\textbf {\bibinfo {volume} {31}},\ \bibinfo {pages}
  {260} (\bibinfo {year} {1980})}\BibitemShut {NoStop}%
\bibitem [{\citenamefont {Dine}\ \emph {et~al.}(1981)\citenamefont {Dine},
  \citenamefont {Fischler},\ and\ \citenamefont {Srednicki}}]{Dine:1981rt}%
  \BibitemOpen
  \bibfield  {author} {\bibinfo {author} {\bibfnamefont {M.}~\bibnamefont
  {Dine}}, \bibinfo {author} {\bibfnamefont {W.}~\bibnamefont {Fischler}},\
  and\ \bibinfo {author} {\bibfnamefont {M.}~\bibnamefont {Srednicki}},\ }\href
  {https://doi.org/10.1016/0370-2693(81)90590-6} {\bibfield  {journal}
  {\bibinfo  {journal} {Phys. Lett. B}\ }\textbf {\bibinfo {volume} {104}},\
  \bibinfo {pages} {199} (\bibinfo {year} {1981})}\BibitemShut {NoStop}%
\bibitem [{\citenamefont {Graham}\ \emph {et~al.}(2015)\citenamefont {Graham},
  \citenamefont {Kaplan},\ and\ \citenamefont {Rajendran}}]{Graham:2015cka}%
  \BibitemOpen
  \bibfield  {author} {\bibinfo {author} {\bibfnamefont {P.~W.}\ \bibnamefont
  {Graham}}, \bibinfo {author} {\bibfnamefont {D.~E.}\ \bibnamefont {Kaplan}},\
  and\ \bibinfo {author} {\bibfnamefont {S.}~\bibnamefont {Rajendran}},\ }\href
  {https://doi.org/10.1103/PhysRevLett.115.221801} {\bibfield  {journal}
  {\bibinfo  {journal} {Phys. Rev. Lett.}\ }\textbf {\bibinfo {volume} {115}},\
  \bibinfo {pages} {221801} (\bibinfo {year} {2015})},\ \Eprint
  {https://arxiv.org/abs/1504.07551} {arXiv:1504.07551 [hep-ph]} \BibitemShut
  {NoStop}%
\bibitem [{\citenamefont {Arvanitaki}\ \emph {et~al.}(2017)\citenamefont
  {Arvanitaki}, \citenamefont {Dimopoulos}, \citenamefont {Gorbenko},
  \citenamefont {Huang},\ and\ \citenamefont
  {Van~Tilburg}}]{Arvanitaki:2016xds}%
  \BibitemOpen
  \bibfield  {author} {\bibinfo {author} {\bibfnamefont {A.}~\bibnamefont
  {Arvanitaki}}, \bibinfo {author} {\bibfnamefont {S.}~\bibnamefont
  {Dimopoulos}}, \bibinfo {author} {\bibfnamefont {V.}~\bibnamefont
  {Gorbenko}}, \bibinfo {author} {\bibfnamefont {J.}~\bibnamefont {Huang}},\
  and\ \bibinfo {author} {\bibfnamefont {K.}~\bibnamefont {Van~Tilburg}},\
  }\href {https://doi.org/10.1007/JHEP05(2017)071} {\bibfield  {journal}
  {\bibinfo  {journal} {JHEP}\ }\textbf {\bibinfo {volume} {05}},\ \bibinfo
  {pages} {071}},\ \Eprint {https://arxiv.org/abs/1609.06320} {arXiv:1609.06320
  [hep-ph]} \BibitemShut {NoStop}%
\bibitem [{\citenamefont {Geller}\ \emph {et~al.}(2019)\citenamefont {Geller},
  \citenamefont {Hochberg},\ and\ \citenamefont {Kuflik}}]{Geller:2018xvz}%
  \BibitemOpen
  \bibfield  {author} {\bibinfo {author} {\bibfnamefont {M.}~\bibnamefont
  {Geller}}, \bibinfo {author} {\bibfnamefont {Y.}~\bibnamefont {Hochberg}},\
  and\ \bibinfo {author} {\bibfnamefont {E.}~\bibnamefont {Kuflik}},\ }\href
  {https://doi.org/10.1103/PhysRevLett.122.191802} {\bibfield  {journal}
  {\bibinfo  {journal} {Phys. Rev. Lett.}\ }\textbf {\bibinfo {volume} {122}},\
  \bibinfo {pages} {191802} (\bibinfo {year} {2019})},\ \Eprint
  {https://arxiv.org/abs/1809.07338} {arXiv:1809.07338 [hep-ph]} \BibitemShut
  {NoStop}%
\bibitem [{\citenamefont {Arkani-Hamed}\ \emph {et~al.}(2021)\citenamefont
  {Arkani-Hamed}, \citenamefont {D'Agnolo},\ and\ \citenamefont
  {Kim}}]{Arkani-Hamed:2020yna}%
  \BibitemOpen
  \bibfield  {author} {\bibinfo {author} {\bibfnamefont {N.}~\bibnamefont
  {Arkani-Hamed}}, \bibinfo {author} {\bibfnamefont {R.~T.}\ \bibnamefont
  {D'Agnolo}},\ and\ \bibinfo {author} {\bibfnamefont {H.~D.}\ \bibnamefont
  {Kim}},\ }\href {https://doi.org/10.1103/PhysRevD.104.095014} {\bibfield
  {journal} {\bibinfo  {journal} {Phys. Rev. D}\ }\textbf {\bibinfo {volume}
  {104}},\ \bibinfo {pages} {095014} (\bibinfo {year} {2021})},\ \Eprint
  {https://arxiv.org/abs/2012.04652} {arXiv:2012.04652 [hep-ph]} \BibitemShut
  {NoStop}%
\bibitem [{\citenamefont {Tito~D'Agnolo}\ and\ \citenamefont
  {Teresi}(2022)}]{TitoDAgnolo:2021nhd}%
  \BibitemOpen
  \bibfield  {author} {\bibinfo {author} {\bibfnamefont {R.}~\bibnamefont
  {Tito~D'Agnolo}}\ and\ \bibinfo {author} {\bibfnamefont {D.}~\bibnamefont
  {Teresi}},\ }\href {https://doi.org/10.1103/PhysRevLett.128.021803}
  {\bibfield  {journal} {\bibinfo  {journal} {Phys. Rev. Lett.}\ }\textbf
  {\bibinfo {volume} {128}},\ \bibinfo {pages} {021803} (\bibinfo {year}
  {2022})},\ \Eprint {https://arxiv.org/abs/2106.04591} {arXiv:2106.04591
  [hep-ph]} \BibitemShut {NoStop}%
\bibitem [{\citenamefont {Chatrchyan}\ and\ \citenamefont
  {Servant}(2023)}]{Chatrchyan:2022dpy}%
  \BibitemOpen
  \bibfield  {author} {\bibinfo {author} {\bibfnamefont {A.}~\bibnamefont
  {Chatrchyan}}\ and\ \bibinfo {author} {\bibfnamefont {G.}~\bibnamefont
  {Servant}},\ }\href {https://doi.org/10.1088/1475-7516/2023/06/036}
  {\bibfield  {journal} {\bibinfo  {journal} {JCAP}\ }\textbf {\bibinfo
  {volume} {06}},\ \bibinfo {pages} {036}},\ \Eprint
  {https://arxiv.org/abs/2211.15694} {arXiv:2211.15694 [hep-ph]} \BibitemShut
  {NoStop}%
\bibitem [{\citenamefont {Csaki}\ \emph {et~al.}(2025)\citenamefont {Csaki},
  \citenamefont {Tito~D'Agnolo}, \citenamefont {Kuflik},\ and\ \citenamefont
  {Sesma}}]{Csaki:2024ywk}%
  \BibitemOpen
  \bibfield  {author} {\bibinfo {author} {\bibfnamefont {C.}~\bibnamefont
  {Csaki}}, \bibinfo {author} {\bibfnamefont {R.}~\bibnamefont
  {Tito~D'Agnolo}}, \bibinfo {author} {\bibfnamefont {E.}~\bibnamefont
  {Kuflik}},\ and\ \bibinfo {author} {\bibfnamefont {P.}~\bibnamefont
  {Sesma}},\ }\href {https://doi.org/10.1007/JHEP02(2025)048} {\bibfield
  {journal} {\bibinfo  {journal} {JHEP}\ }\textbf {\bibinfo {volume} {02}},\
  \bibinfo {pages} {048}},\ \Eprint {https://arxiv.org/abs/2411.03438}
  {arXiv:2411.03438 [hep-ph]} \BibitemShut {NoStop}%
\bibitem [{\citenamefont {Preskill}\ \emph {et~al.}(1983)\citenamefont
  {Preskill}, \citenamefont {Wise},\ and\ \citenamefont
  {Wilczek}}]{Preskill:1982cy}%
  \BibitemOpen
  \bibfield  {author} {\bibinfo {author} {\bibfnamefont {J.}~\bibnamefont
  {Preskill}}, \bibinfo {author} {\bibfnamefont {M.~B.}\ \bibnamefont {Wise}},\
  and\ \bibinfo {author} {\bibfnamefont {F.}~\bibnamefont {Wilczek}},\ }\href
  {https://doi.org/10.1016/0370-2693(83)90637-8} {\bibfield  {journal}
  {\bibinfo  {journal} {Phys. Lett. B}\ }\textbf {\bibinfo {volume} {120}},\
  \bibinfo {pages} {127} (\bibinfo {year} {1983})}\BibitemShut {NoStop}%
\bibitem [{\citenamefont {Abbott}\ and\ \citenamefont
  {Sikivie}(1983)}]{Abbott:1982af}%
  \BibitemOpen
  \bibfield  {author} {\bibinfo {author} {\bibfnamefont {L.~F.}\ \bibnamefont
  {Abbott}}\ and\ \bibinfo {author} {\bibfnamefont {P.}~\bibnamefont
  {Sikivie}},\ }\href {https://doi.org/10.1016/0370-2693(83)90638-X} {\bibfield
   {journal} {\bibinfo  {journal} {Phys. Lett. B}\ }\textbf {\bibinfo {volume}
  {120}},\ \bibinfo {pages} {133} (\bibinfo {year} {1983})}\BibitemShut
  {NoStop}%
\bibitem [{\citenamefont {Hu}\ \emph {et~al.}(2000)\citenamefont {Hu},
  \citenamefont {Barkana},\ and\ \citenamefont {Gruzinov}}]{Hu:2000ke}%
  \BibitemOpen
  \bibfield  {author} {\bibinfo {author} {\bibfnamefont {W.}~\bibnamefont
  {Hu}}, \bibinfo {author} {\bibfnamefont {R.}~\bibnamefont {Barkana}},\ and\
  \bibinfo {author} {\bibfnamefont {A.}~\bibnamefont {Gruzinov}},\ }\href
  {https://doi.org/10.1103/PhysRevLett.85.1158} {\bibfield  {journal} {\bibinfo
   {journal} {Phys. Rev. Lett.}\ }\textbf {\bibinfo {volume} {85}},\ \bibinfo
  {pages} {1158} (\bibinfo {year} {2000})},\ \Eprint
  {https://arxiv.org/abs/astro-ph/0003365} {arXiv:astro-ph/0003365}
  \BibitemShut {NoStop}%
\bibitem [{\citenamefont {Goodman}(2000)}]{Goodman:2000tg}%
  \BibitemOpen
  \bibfield  {author} {\bibinfo {author} {\bibfnamefont {J.}~\bibnamefont
  {Goodman}},\ }\href {https://doi.org/10.1016/S1384-1076(00)00015-4}
  {\bibfield  {journal} {\bibinfo  {journal} {New Astron.}\ }\textbf {\bibinfo
  {volume} {5}},\ \bibinfo {pages} {103} (\bibinfo {year} {2000})},\ \Eprint
  {https://arxiv.org/abs/astro-ph/0003018} {arXiv:astro-ph/0003018}
  \BibitemShut {NoStop}%
\bibitem [{\citenamefont {Hui}\ \emph {et~al.}(2017)\citenamefont {Hui},
  \citenamefont {Ostriker}, \citenamefont {Tremaine},\ and\ \citenamefont
  {Witten}}]{Hui:2016ltb}%
  \BibitemOpen
  \bibfield  {author} {\bibinfo {author} {\bibfnamefont {L.}~\bibnamefont
  {Hui}}, \bibinfo {author} {\bibfnamefont {J.~P.}\ \bibnamefont {Ostriker}},
  \bibinfo {author} {\bibfnamefont {S.}~\bibnamefont {Tremaine}},\ and\
  \bibinfo {author} {\bibfnamefont {E.}~\bibnamefont {Witten}},\ }\href
  {https://doi.org/10.1103/PhysRevD.95.043541} {\bibfield  {journal} {\bibinfo
  {journal} {Phys. Rev. D}\ }\textbf {\bibinfo {volume} {95}},\ \bibinfo
  {pages} {043541} (\bibinfo {year} {2017})},\ \Eprint
  {https://arxiv.org/abs/1610.08297} {arXiv:1610.08297 [astro-ph.CO]}
  \BibitemShut {NoStop}%
\bibitem [{\citenamefont {Press}\ and\ \citenamefont
  {Teukolsky}(1972)}]{Press:1972zz}%
  \BibitemOpen
  \bibfield  {author} {\bibinfo {author} {\bibfnamefont {W.~H.}\ \bibnamefont
  {Press}}\ and\ \bibinfo {author} {\bibfnamefont {S.~A.}\ \bibnamefont
  {Teukolsky}},\ }\href {https://doi.org/10.1038/238211a0} {\bibfield
  {journal} {\bibinfo  {journal} {Nature}\ }\textbf {\bibinfo {volume} {238}},\
  \bibinfo {pages} {211} (\bibinfo {year} {1972})}\BibitemShut {NoStop}%
\bibitem [{\citenamefont {Bardeen}\ \emph {et~al.}(1972)\citenamefont
  {Bardeen}, \citenamefont {Press},\ and\ \citenamefont
  {Teukolsky}}]{Bardeen:1972fi}%
  \BibitemOpen
  \bibfield  {author} {\bibinfo {author} {\bibfnamefont {J.~M.}\ \bibnamefont
  {Bardeen}}, \bibinfo {author} {\bibfnamefont {W.~H.}\ \bibnamefont {Press}},\
  and\ \bibinfo {author} {\bibfnamefont {S.~A.}\ \bibnamefont {Teukolsky}},\
  }\href {https://doi.org/10.1086/151796} {\bibfield  {journal} {\bibinfo
  {journal} {Astrophys. J.}\ }\textbf {\bibinfo {volume} {178}},\ \bibinfo
  {pages} {347} (\bibinfo {year} {1972})}\BibitemShut {NoStop}%
\bibitem [{\citenamefont {Yoshino}\ and\ \citenamefont
  {Kodama}(2014)}]{Yoshino:2013ofa}%
  \BibitemOpen
  \bibfield  {author} {\bibinfo {author} {\bibfnamefont {H.}~\bibnamefont
  {Yoshino}}\ and\ \bibinfo {author} {\bibfnamefont {H.}~\bibnamefont
  {Kodama}},\ }\href {https://doi.org/10.1093/ptep/ptu029} {\bibfield
  {journal} {\bibinfo  {journal} {PTEP}\ }\textbf {\bibinfo {volume} {2014}},\
  \bibinfo {pages} {043E02} (\bibinfo {year} {2014})},\ \Eprint
  {https://arxiv.org/abs/1312.2326} {arXiv:1312.2326 [gr-qc]} \BibitemShut
  {NoStop}%
\bibitem [{\citenamefont {Brito}\ \emph {et~al.}(2015)\citenamefont {Brito},
  \citenamefont {Cardoso},\ and\ \citenamefont {Pani}}]{Brito:2014wla}%
  \BibitemOpen
  \bibfield  {author} {\bibinfo {author} {\bibfnamefont {R.}~\bibnamefont
  {Brito}}, \bibinfo {author} {\bibfnamefont {V.}~\bibnamefont {Cardoso}},\
  and\ \bibinfo {author} {\bibfnamefont {P.}~\bibnamefont {Pani}},\ }\href
  {https://doi.org/10.1088/0264-9381/32/13/134001} {\bibfield  {journal}
  {\bibinfo  {journal} {Class. Quant. Grav.}\ }\textbf {\bibinfo {volume}
  {32}},\ \bibinfo {pages} {134001} (\bibinfo {year} {2015})},\ \Eprint
  {https://arxiv.org/abs/1411.0686} {arXiv:1411.0686 [gr-qc]} \BibitemShut
  {NoStop}%
\bibitem [{\citenamefont {Arvanitaki}\ and\ \citenamefont
  {Dubovsky}(2011)}]{Arvanitaki:2010sy}%
  \BibitemOpen
  \bibfield  {author} {\bibinfo {author} {\bibfnamefont {A.}~\bibnamefont
  {Arvanitaki}}\ and\ \bibinfo {author} {\bibfnamefont {S.}~\bibnamefont
  {Dubovsky}},\ }\href {https://doi.org/10.1103/PhysRevD.83.044026} {\bibfield
  {journal} {\bibinfo  {journal} {Phys. Rev. D}\ }\textbf {\bibinfo {volume}
  {83}},\ \bibinfo {pages} {044026} (\bibinfo {year} {2011})},\ \Eprint
  {https://arxiv.org/abs/1004.3558} {arXiv:1004.3558 [hep-th]} \BibitemShut
  {NoStop}%
\bibitem [{\citenamefont {{McClintock}}\ \emph {et~al.}(2014)\citenamefont
  {{McClintock}}, \citenamefont {{Narayan}},\ and\ \citenamefont
  {{Steiner}}}]{McClintock+14}%
  \BibitemOpen
  \bibfield  {author} {\bibinfo {author} {\bibfnamefont {J.~E.}\ \bibnamefont
  {{McClintock}}}, \bibinfo {author} {\bibfnamefont {R.}~\bibnamefont
  {{Narayan}}},\ and\ \bibinfo {author} {\bibfnamefont {J.~F.}\ \bibnamefont
  {{Steiner}}},\ }\href {https://doi.org/10.1007/s11214-013-0003-9} {\bibfield
  {journal} {\bibinfo  {journal} {\ssr}\ }\textbf {\bibinfo {volume} {183}},\
  \bibinfo {pages} {295} (\bibinfo {year} {2014})},\ \Eprint
  {https://arxiv.org/abs/1303.1583} {arXiv:1303.1583 [astro-ph.HE]}
  \BibitemShut {NoStop}%
\bibitem [{\citenamefont {{Reynolds}}(2013)}]{Reynolds13}%
  \BibitemOpen
  \bibfield  {author} {\bibinfo {author} {\bibfnamefont {C.~S.}\ \bibnamefont
  {{Reynolds}}},\ }\href {https://doi.org/10.1088/0264-9381/30/24/244004}
  {\bibfield  {journal} {\bibinfo  {journal} {Classical and Quantum Gravity}\
  }\textbf {\bibinfo {volume} {30}},\ \bibinfo {eid} {244004} (\bibinfo {year}
  {2013})},\ \Eprint {https://arxiv.org/abs/1307.3246} {arXiv:1307.3246
  [astro-ph.HE]} \BibitemShut {NoStop}%
\bibitem [{\citenamefont {Caputo}\ \emph {et~al.}(2026)\citenamefont {Caputo},
  \citenamefont {Franciolini},\ and\ \citenamefont {Witte}}]{Caputo:2025oap}%
  \BibitemOpen
  \bibfield  {author} {\bibinfo {author} {\bibfnamefont {A.}~\bibnamefont
  {Caputo}}, \bibinfo {author} {\bibfnamefont {G.}~\bibnamefont
  {Franciolini}},\ and\ \bibinfo {author} {\bibfnamefont {S.~J.}\ \bibnamefont
  {Witte}},\ }\href {https://doi.org/10.1103/f7w5-36ng} {\bibfield  {journal}
  {\bibinfo  {journal} {Phys. Rev. D}\ }\textbf {\bibinfo {volume} {113}},\
  \bibinfo {pages} {083031} (\bibinfo {year} {2026})},\ \Eprint
  {https://arxiv.org/abs/2507.21788} {arXiv:2507.21788 [hep-ph]} \BibitemShut
  {NoStop}%
\bibitem [{\citenamefont {Aswathi}\ \emph {et~al.}(2025)\citenamefont
  {Aswathi}, \citenamefont {East}, \citenamefont {Siemonsen}, \citenamefont
  {Sun},\ and\ \citenamefont {Jones}}]{Aswathi:2025nxa}%
  \BibitemOpen
  \bibfield  {author} {\bibinfo {author} {\bibfnamefont {P.~S.}\ \bibnamefont
  {Aswathi}}, \bibinfo {author} {\bibfnamefont {W.~E.}\ \bibnamefont {East}},
  \bibinfo {author} {\bibfnamefont {N.}~\bibnamefont {Siemonsen}}, \bibinfo
  {author} {\bibfnamefont {L.}~\bibnamefont {Sun}},\ and\ \bibinfo {author}
  {\bibfnamefont {D.}~\bibnamefont {Jones}},\ }\href
  {https://doi.org/10.1103/n5hr-zljn} {\bibfield  {journal} {\bibinfo
  {journal} {Phys. Rev. D}\ }\textbf {\bibinfo {volume} {112}},\ \bibinfo
  {pages} {123048} (\bibinfo {year} {2025})},\ \Eprint
  {https://arxiv.org/abs/2507.20979} {arXiv:2507.20979 [gr-qc]} \BibitemShut
  {NoStop}%
\bibitem [{\citenamefont {Roy}\ \emph {et~al.}(2026)\citenamefont {Roy},
  \citenamefont {Vicente}, \citenamefont {Aurrekoetxea}, \citenamefont
  {Clough},\ and\ \citenamefont {Ferreira}}]{Roy:2025qaa}%
  \BibitemOpen
  \bibfield  {author} {\bibinfo {author} {\bibfnamefont {S.}~\bibnamefont
  {Roy}}, \bibinfo {author} {\bibfnamefont {R.}~\bibnamefont {Vicente}},
  \bibinfo {author} {\bibfnamefont {J.~C.}\ \bibnamefont {Aurrekoetxea}},
  \bibinfo {author} {\bibfnamefont {K.}~\bibnamefont {Clough}},\ and\ \bibinfo
  {author} {\bibfnamefont {P.~G.}\ \bibnamefont {Ferreira}},\ }\href
  {https://doi.org/10.1103/fv9z-zkxx} {\bibfield  {journal} {\bibinfo
  {journal} {Phys. Rev. Lett.}\ }\textbf {\bibinfo {volume} {136}},\ \bibinfo
  {pages} {191402} (\bibinfo {year} {2026})},\ \Eprint
  {https://arxiv.org/abs/2510.17967} {arXiv:2510.17967 [gr-qc]} \BibitemShut
  {NoStop}%
\bibitem [{\citenamefont {Ning}\ \emph {et~al.}(2026)\citenamefont {Ning},
  \citenamefont {Safdi},\ and\ \citenamefont {Welch}}]{Ning:2026ebu}%
  \BibitemOpen
  \bibfield  {author} {\bibinfo {author} {\bibfnamefont {O.}~\bibnamefont
  {Ning}}, \bibinfo {author} {\bibfnamefont {B.~R.}\ \bibnamefont {Safdi}},\
  and\ \bibinfo {author} {\bibfnamefont {C.}~\bibnamefont {Welch}},\
  }\href@noop {} {\  (\bibinfo {year} {2026})},\ \Eprint
  {https://arxiv.org/abs/2607.01317} {arXiv:2607.01317 [hep-ph]} \BibitemShut
  {NoStop}%
\bibitem [{\citenamefont {Abac}\ \emph
  {et~al.}(2026{\natexlab{a}})\citenamefont {Abac} \emph
  {et~al.}}]{LIGOScientific:2026pwx}%
  \BibitemOpen
  \bibfield  {author} {\bibinfo {author} {\bibfnamefont {A.~G.}\ \bibnamefont
  {Abac}} \emph {et~al.} (\bibinfo {collaboration} {LIGO Scientific, VIRGO,
  KAGRA}),\ }\href@noop {} {\  (\bibinfo {year} {2026}{\natexlab{a}})},\
  \Eprint {https://arxiv.org/abs/2608.11620} {arXiv:2608.11620 [gr-qc]}
  \BibitemShut {NoStop}%
\bibitem [{\citenamefont {Kou}\ \emph {et~al.}(2026)\citenamefont {Kou},
  \citenamefont {Mandic}, \citenamefont {Ding},\ and\ \citenamefont
  {Tian}}]{Kou:2026naz}%
  \BibitemOpen
  \bibfield  {author} {\bibinfo {author} {\bibfnamefont {X.-X.}\ \bibnamefont
  {Kou}}, \bibinfo {author} {\bibfnamefont {V.}~\bibnamefont {Mandic}},
  \bibinfo {author} {\bibfnamefont {R.}~\bibnamefont {Ding}},\ and\ \bibinfo
  {author} {\bibfnamefont {C.}~\bibnamefont {Tian}},\ }\href@noop {} {\
  (\bibinfo {year} {2026})},\ \Eprint {https://arxiv.org/abs/2609.02678}
  {arXiv:2609.02678 [gr-qc]} \BibitemShut {NoStop}%
\bibitem [{\citenamefont {Abac}\ \emph
  {et~al.}(2026{\natexlab{b}})\citenamefont {Abac} \emph
  {et~al.}}]{LIGOScientific:2026wfs}%
  \BibitemOpen
  \bibfield  {author} {\bibinfo {author} {\bibfnamefont {A.~G.}\ \bibnamefont
  {Abac}} \emph {et~al.} (\bibinfo {collaboration} {LIGO Scientific, VIRGO,
  KAGRA}),\ }\href@noop {} {\  (\bibinfo {year} {2026}{\natexlab{b}})},\
  \Eprint {https://arxiv.org/abs/2605.27225} {arXiv:2605.27225 [gr-qc]}
  \BibitemShut {NoStop}%
\bibitem [{\citenamefont {Witte}\ and\ \citenamefont
  {Mummery}(2025)}]{Witte:2024drg}%
  \BibitemOpen
  \bibfield  {author} {\bibinfo {author} {\bibfnamefont {S.~J.}\ \bibnamefont
  {Witte}}\ and\ \bibinfo {author} {\bibfnamefont {A.}~\bibnamefont
  {Mummery}},\ }\href {https://doi.org/10.1103/PhysRevD.111.083044} {\bibfield
  {journal} {\bibinfo  {journal} {Phys. Rev. D}\ }\textbf {\bibinfo {volume}
  {111}},\ \bibinfo {pages} {083044} (\bibinfo {year} {2025})},\ \Eprint
  {https://arxiv.org/abs/2412.03655} {arXiv:2412.03655 [hep-ph]} \BibitemShut
  {NoStop}%
\bibitem [{\citenamefont {Khalaf}\ \emph {et~al.}(2026)\citenamefont {Khalaf},
  \citenamefont {Kuflik}, \citenamefont {Lenoci},\ and\ \citenamefont
  {Stone}}]{Khalaf:2024nwc}%
  \BibitemOpen
  \bibfield  {author} {\bibinfo {author} {\bibfnamefont {M.}~\bibnamefont
  {Khalaf}}, \bibinfo {author} {\bibfnamefont {E.}~\bibnamefont {Kuflik}},
  \bibinfo {author} {\bibfnamefont {A.}~\bibnamefont {Lenoci}},\ and\ \bibinfo
  {author} {\bibfnamefont {N.~C.}\ \bibnamefont {Stone}},\ }\href
  {https://doi.org/10.1103/ww5b-gdkx} {\bibfield  {journal} {\bibinfo
  {journal} {Phys. Rev. D}\ }\textbf {\bibinfo {volume} {113}},\ \bibinfo
  {pages} {043015} (\bibinfo {year} {2026})},\ \Eprint
  {https://arxiv.org/abs/2408.16051} {arXiv:2408.16051 [astro-ph.CO]}
  \BibitemShut {NoStop}%
\bibitem [{\citenamefont {{Alush}}\ and\ \citenamefont
  {{Stone}}(2022)}]{AlushStone22}%
  \BibitemOpen
  \bibfield  {author} {\bibinfo {author} {\bibfnamefont {Y.}~\bibnamefont
  {{Alush}}}\ and\ \bibinfo {author} {\bibfnamefont {N.~C.}\ \bibnamefont
  {{Stone}}},\ }\href {https://doi.org/10.1103/PhysRevD.106.123023} {\bibfield
  {journal} {\bibinfo  {journal} {\prd}\ }\textbf {\bibinfo {volume} {106}},\
  \bibinfo {eid} {123023} (\bibinfo {year} {2022})},\ \Eprint
  {https://arxiv.org/abs/2207.02226} {arXiv:2207.02226 [astro-ph.GA]}
  \BibitemShut {NoStop}%
\bibitem [{\citenamefont {Foschi}\ \emph {et~al.}(2024)\citenamefont {Foschi}
  \emph {et~al.}}]{GRAVITY:2023azi}%
  \BibitemOpen
  \bibfield  {author} {\bibinfo {author} {\bibfnamefont {A.}~\bibnamefont
  {Foschi}} \emph {et~al.} (\bibinfo {collaboration} {GRAVITY}),\ }\href
  {https://doi.org/10.1093/mnras/stae423} {\bibfield  {journal} {\bibinfo
  {journal} {Mon. Not. Roy. Astron. Soc.}\ }\textbf {\bibinfo {volume} {530}},\
  \bibinfo {pages} {3740} (\bibinfo {year} {2024})},\ \Eprint
  {https://arxiv.org/abs/2312.02653} {arXiv:2312.02653 [astro-ph.GA]}
  \BibitemShut {NoStop}%
\bibitem [{\citenamefont {Foschi}\ \emph {et~al.}(2023)\citenamefont {Foschi}
  \emph {et~al.}}]{GRAVITY:2023cjt}%
  \BibitemOpen
  \bibfield  {author} {\bibinfo {author} {\bibfnamefont {A.}~\bibnamefont
  {Foschi}} \emph {et~al.} (\bibinfo {collaboration} {GRAVITY}),\ }\href
  {https://doi.org/10.1093/mnras/stad1939} {\bibfield  {journal} {\bibinfo
  {journal} {Mon. Not. Roy. Astron. Soc.}\ }\textbf {\bibinfo {volume} {524}},\
  \bibinfo {pages} {1075} (\bibinfo {year} {2023})},\ \Eprint
  {https://arxiv.org/abs/2306.17215} {arXiv:2306.17215 [astro-ph.GA]}
  \BibitemShut {NoStop}%
\bibitem [{\citenamefont {{Tomaselli}}\ and\ \citenamefont
  {{Caputo}}(2026)}]{TomaselliCaputo26}%
  \BibitemOpen
  \bibfield  {author} {\bibinfo {author} {\bibfnamefont {G.~M.}\ \bibnamefont
  {{Tomaselli}}}\ and\ \bibinfo {author} {\bibfnamefont {A.}~\bibnamefont
  {{Caputo}}},\ }\href {https://doi.org/10.1103/pm7p-c53w} {\bibfield
  {journal} {\bibinfo  {journal} {\prd}\ }\textbf {\bibinfo {volume} {113}},\
  \bibinfo {eid} {083035} (\bibinfo {year} {2026})},\ \Eprint
  {https://arxiv.org/abs/2509.03568} {arXiv:2509.03568 [astro-ph.GA]}
  \BibitemShut {NoStop}%
\bibitem [{\citenamefont {Abbott}(2022)}]{Abbott22+}%
  \BibitemOpen
  \bibfield  {author} {\bibinfo {author} {\bibfnamefont {R.~e.~a.}\
  \bibnamefont {Abbott}} (\bibinfo {collaboration} {LIGO Scientific
  Collaboration, Virgo Collaboration, and KAGRA Collaboration}),\ }\href
  {https://doi.org/10.1103/PhysRevD.106.042003} {\bibfield  {journal} {\bibinfo
   {journal} {Phys. Rev. D}\ }\textbf {\bibinfo {volume} {106}},\ \bibinfo
  {pages} {042003} (\bibinfo {year} {2022})}\BibitemShut {NoStop}%
\bibitem [{\citenamefont {Ellis}\ \emph {et~al.}(2026)\citenamefont {Ellis},
  \citenamefont {Ning}, \citenamefont {Rodd},\ and\ \citenamefont
  {Sch{\"u}tte-Engel}}]{Ellis:2026gwt}%
  \BibitemOpen
  \bibfield  {author} {\bibinfo {author} {\bibfnamefont {S.~A.~R.}\
  \bibnamefont {Ellis}}, \bibinfo {author} {\bibfnamefont {O.}~\bibnamefont
  {Ning}}, \bibinfo {author} {\bibfnamefont {N.~L.}\ \bibnamefont {Rodd}},\
  and\ \bibinfo {author} {\bibfnamefont {J.}~\bibnamefont
  {Sch{\"u}tte-Engel}},\ }\href@noop {} {\  (\bibinfo {year} {2026})},\ \Eprint
  {https://arxiv.org/abs/2603.15734} {arXiv:2603.15734 [hep-ph]} \BibitemShut
  {NoStop}%
\bibitem [{\citenamefont {Amaro-Seoane}\ \emph {et~al.}(2017)\citenamefont
  {Amaro-Seoane} \emph {et~al.}}]{LISA:2017pwj}%
  \BibitemOpen
  \bibfield  {author} {\bibinfo {author} {\bibfnamefont {P.}~\bibnamefont
  {Amaro-Seoane}} \emph {et~al.} (\bibinfo {collaboration} {LISA}),\
  }\href@noop {} {\bibfield  {journal} {\bibinfo  {journal} {arXiv e-prints}\ }
  (\bibinfo {year} {2017})},\ \Eprint {https://arxiv.org/abs/1702.00786}
  {arXiv:1702.00786 [astro-ph.IM]} \BibitemShut {NoStop}%
\bibitem [{\citenamefont {Seoane}\ \emph {et~al.}(2023)\citenamefont {Seoane}
  \emph {et~al.}}]{LISA:2022yao}%
  \BibitemOpen
  \bibfield  {author} {\bibinfo {author} {\bibfnamefont {P.~A.}\ \bibnamefont
  {Seoane}} \emph {et~al.} (\bibinfo {collaboration} {LISA}),\ }\href
  {https://doi.org/10.1007/s41114-022-00041-y} {\bibfield  {journal} {\bibinfo
  {journal} {Living Rev. Rel.}\ }\textbf {\bibinfo {volume} {26}},\ \bibinfo
  {pages} {2} (\bibinfo {year} {2023})},\ \Eprint
  {https://arxiv.org/abs/2203.06016} {arXiv:2203.06016 [gr-qc]} \BibitemShut
  {NoStop}%
\bibitem [{\citenamefont {Colpi}\ \emph {et~al.}(2024)\citenamefont {Colpi}
  \emph {et~al.}}]{LISA:2024hlh}%
  \BibitemOpen
  \bibfield  {author} {\bibinfo {author} {\bibfnamefont {M.}~\bibnamefont
  {Colpi}} \emph {et~al.} (\bibinfo {collaboration} {LISA}),\ }\href@noop {}
  {\bibfield  {journal} {\bibinfo  {journal} {arXiv e-prints}\ } (\bibinfo
  {year} {2024})},\ \Eprint {https://arxiv.org/abs/2402.07571}
  {arXiv:2402.07571 [astro-ph.CO]} \BibitemShut {NoStop}%
\bibitem [{\citenamefont {Tomaselli}\ \emph
  {et~al.}(2024{\natexlab{a}})\citenamefont {Tomaselli}, \citenamefont
  {Spieksma},\ and\ \citenamefont {Bertone}}]{Tomaselli:2024dbw}%
  \BibitemOpen
  \bibfield  {author} {\bibinfo {author} {\bibfnamefont {G.~M.}\ \bibnamefont
  {Tomaselli}}, \bibinfo {author} {\bibfnamefont {T.~F.~M.}\ \bibnamefont
  {Spieksma}},\ and\ \bibinfo {author} {\bibfnamefont {G.}~\bibnamefont
  {Bertone}},\ }\href {https://doi.org/10.1103/PhysRevLett.133.121402}
  {\bibfield  {journal} {\bibinfo  {journal} {Phys. Rev. Lett.}\ }\textbf
  {\bibinfo {volume} {133}},\ \bibinfo {pages} {121402} (\bibinfo {year}
  {2024}{\natexlab{a}})},\ \Eprint {https://arxiv.org/abs/2407.12908}
  {arXiv:2407.12908 [gr-qc]} \BibitemShut {NoStop}%
\bibitem [{\citenamefont {Tomaselli}\ \emph
  {et~al.}(2024{\natexlab{b}})\citenamefont {Tomaselli}, \citenamefont
  {Spieksma},\ and\ \citenamefont {Bertone}}]{Tomaselli:2024bdd}%
  \BibitemOpen
  \bibfield  {author} {\bibinfo {author} {\bibfnamefont {G.~M.}\ \bibnamefont
  {Tomaselli}}, \bibinfo {author} {\bibfnamefont {T.~F.~M.}\ \bibnamefont
  {Spieksma}},\ and\ \bibinfo {author} {\bibfnamefont {G.}~\bibnamefont
  {Bertone}},\ }\href {https://doi.org/10.1103/PhysRevD.110.064048} {\bibfield
  {journal} {\bibinfo  {journal} {Phys. Rev. D}\ }\textbf {\bibinfo {volume}
  {110}},\ \bibinfo {pages} {064048} (\bibinfo {year} {2024}{\natexlab{b}})},\
  \Eprint {https://arxiv.org/abs/2403.03147} {arXiv:2403.03147 [gr-qc]}
  \BibitemShut {NoStop}%
\bibitem [{\citenamefont {Bo{\v{s}}kovi{\'c}}\ \emph
  {et~al.}(2024)\citenamefont {Bo{\v{s}}kovi{\'c}}, \citenamefont
  {Koschnitzke},\ and\ \citenamefont {Porto}}]{Boskovic:2024fga}%
  \BibitemOpen
  \bibfield  {author} {\bibinfo {author} {\bibfnamefont {M.}~\bibnamefont
  {Bo{\v{s}}kovi{\'c}}}, \bibinfo {author} {\bibfnamefont {M.}~\bibnamefont
  {Koschnitzke}},\ and\ \bibinfo {author} {\bibfnamefont {R.~A.}\ \bibnamefont
  {Porto}},\ }\href {https://doi.org/10.1103/PhysRevLett.133.121401} {\bibfield
   {journal} {\bibinfo  {journal} {Phys. Rev. Lett.}\ }\textbf {\bibinfo
  {volume} {133}},\ \bibinfo {pages} {121401} (\bibinfo {year} {2024})},\
  \Eprint {https://arxiv.org/abs/2403.02415} {arXiv:2403.02415 [gr-qc]}
  \BibitemShut {NoStop}%
\bibitem [{\citenamefont {Bo{\v{s}}kovi{\'c}}\ \emph
  {et~al.}(2026)\citenamefont {Bo{\v{s}}kovi{\'c}}, \citenamefont {Porto},\
  and\ \citenamefont {Koschnitzke}}]{Boskovic:2025ixx}%
  \BibitemOpen
  \bibfield  {author} {\bibinfo {author} {\bibfnamefont {M.}~\bibnamefont
  {Bo{\v{s}}kovi{\'c}}}, \bibinfo {author} {\bibfnamefont {R.~A.}\ \bibnamefont
  {Porto}},\ and\ \bibinfo {author} {\bibfnamefont {M.}~\bibnamefont
  {Koschnitzke}},\ }\href {https://doi.org/10.1103/py8z-2zms} {\bibfield
  {journal} {\bibinfo  {journal} {Phys. Rev. D}\ }\textbf {\bibinfo {volume}
  {113}},\ \bibinfo {pages} {124053} (\bibinfo {year} {2026})},\ \Eprint
  {https://arxiv.org/abs/2512.17887} {arXiv:2512.17887 [gr-qc]} \BibitemShut
  {NoStop}%
\bibitem [{\citenamefont {Kim}\ and\ \citenamefont
  {Lenoci}(2025)}]{Kim:2025wwj}%
  \BibitemOpen
  \bibfield  {author} {\bibinfo {author} {\bibfnamefont {H.}~\bibnamefont
  {Kim}}\ and\ \bibinfo {author} {\bibfnamefont {A.}~\bibnamefont {Lenoci}},\
  }\href {https://doi.org/10.1103/81dj-kxmy} {\bibfield  {journal} {\bibinfo
  {journal} {Phys. Rev. D}\ }\textbf {\bibinfo {volume} {112}},\ \bibinfo
  {pages} {104014} (\bibinfo {year} {2025})},\ \Eprint
  {https://arxiv.org/abs/2508.08367} {arXiv:2508.08367 [gr-qc]} \BibitemShut
  {NoStop}%
\bibitem [{\citenamefont {Bo{\v{s}}kovi{\'c}}\ and\ \citenamefont
  {Savi{\'c}}(2026)}]{Boskovic:2026eth}%
  \BibitemOpen
  \bibfield  {author} {\bibinfo {author} {\bibfnamefont {M.}~\bibnamefont
  {Bo{\v{s}}kovi{\'c}}}\ and\ \bibinfo {author} {\bibfnamefont
  {N.}~\bibnamefont {Savi{\'c}}},\ }\href@noop {} {\  (\bibinfo {year}
  {2026})},\ \Eprint {https://arxiv.org/abs/2607.27174} {arXiv:2607.27174
  [gr-qc]} \BibitemShut {NoStop}%
\bibitem [{\citenamefont {Sesana}(2016)}]{Sesana:2016ljz}%
  \BibitemOpen
  \bibfield  {author} {\bibinfo {author} {\bibfnamefont {A.}~\bibnamefont
  {Sesana}},\ }\href {https://doi.org/10.1103/PhysRevLett.116.231102}
  {\bibfield  {journal} {\bibinfo  {journal} {Phys. Rev. Lett.}\ }\textbf
  {\bibinfo {volume} {116}},\ \bibinfo {pages} {231102} (\bibinfo {year}
  {2016})},\ \Eprint {https://arxiv.org/abs/1602.06951} {arXiv:1602.06951
  [gr-qc]} \BibitemShut {NoStop}%
\bibitem [{\citenamefont {Abac}\ \emph
  {et~al.}(2026{\natexlab{c}})\citenamefont {Abac} \emph
  {et~al.}}]{LIGOScientific:2026ctl}%
  \BibitemOpen
  \bibfield  {author} {\bibinfo {author} {\bibfnamefont {A.~G.}\ \bibnamefont
  {Abac}} \emph {et~al.} (\bibinfo {collaboration} {LIGO Scientific, VIRGO,
  KAGRA}),\ }\href@noop {} {\  (\bibinfo {year} {2026}{\natexlab{c}})},\
  \Eprint {https://arxiv.org/abs/2605.27226} {arXiv:2605.27226 [astro-ph.HE]}
  \BibitemShut {NoStop}%
\bibitem [{\citenamefont {Guttman}\ \emph {et~al.}(2026)\citenamefont
  {Guttman}, \citenamefont {Lasky},\ and\ \citenamefont
  {Thrane}}]{Guttman:2026cnv}%
  \BibitemOpen
  \bibfield  {author} {\bibinfo {author} {\bibfnamefont {N.}~\bibnamefont
  {Guttman}}, \bibinfo {author} {\bibfnamefont {P.~D.}\ \bibnamefont {Lasky}},\
  and\ \bibinfo {author} {\bibfnamefont {E.}~\bibnamefont {Thrane}},\
  }\href@noop {} {\  (\bibinfo {year} {2026})},\ \Eprint
  {https://arxiv.org/abs/2607.22011} {arXiv:2607.22011 [astro-ph.HE]}
  \BibitemShut {NoStop}%
\bibitem [{\citenamefont {Ray}\ \emph {et~al.}(2026)\citenamefont {Ray},
  \citenamefont {Mukherjee}, \citenamefont {Zevin},\ and\ \citenamefont
  {Kalogera}}]{Ray:2026uur}%
  \BibitemOpen
  \bibfield  {author} {\bibinfo {author} {\bibfnamefont {A.}~\bibnamefont
  {Ray}}, \bibinfo {author} {\bibfnamefont {S.}~\bibnamefont {Mukherjee}},
  \bibinfo {author} {\bibfnamefont {M.}~\bibnamefont {Zevin}},\ and\ \bibinfo
  {author} {\bibfnamefont {V.}~\bibnamefont {Kalogera}},\ }\href
  {https://doi.org/10.3847/2041-8213/ae80cb} {\bibfield  {journal} {\bibinfo
  {journal} {Astrophys. J. Lett.}\ }\textbf {\bibinfo {volume} {1005}},\
  \bibinfo {pages} {L55} (\bibinfo {year} {2026})},\ \Eprint
  {https://arxiv.org/abs/2603.17987} {arXiv:2603.17987 [astro-ph.HE]}
  \BibitemShut {NoStop}%
\bibitem [{\citenamefont {Abac}\ \emph {et~al.}(2025)\citenamefont {Abac} \emph
  {et~al.}}]{LIGOScientific:2025rsn}%
  \BibitemOpen
  \bibfield  {author} {\bibinfo {author} {\bibfnamefont {A.~G.}\ \bibnamefont
  {Abac}} \emph {et~al.} (\bibinfo {collaboration} {LIGO Scientific, VIRGO,
  KAGRA}),\ }\href {https://doi.org/10.3847/2041-8213/ae0c9c} {\bibfield
  {journal} {\bibinfo  {journal} {Astrophys. J. Lett.}\ }\textbf {\bibinfo
  {volume} {993}},\ \bibinfo {pages} {L25} (\bibinfo {year} {2025})},\ \Eprint
  {https://arxiv.org/abs/2507.08219} {arXiv:2507.08219 [astro-ph.HE]}
  \BibitemShut {NoStop}%
\bibitem [{\citenamefont {Abbott}\ \emph {et~al.}(2020)\citenamefont {Abbott}
  \emph {et~al.}}]{LIGOScientific:2020zkf}%
  \BibitemOpen
  \bibfield  {author} {\bibinfo {author} {\bibfnamefont {R.}~\bibnamefont
  {Abbott}} \emph {et~al.} (\bibinfo {collaboration} {LIGO Scientific,
  Virgo}),\ }\href {https://doi.org/10.3847/2041-8213/ab960f} {\bibfield
  {journal} {\bibinfo  {journal} {Astrophys. J. Lett.}\ }\textbf {\bibinfo
  {volume} {896}},\ \bibinfo {pages} {L44} (\bibinfo {year} {2020})},\ \Eprint
  {https://arxiv.org/abs/2006.12611} {arXiv:2006.12611 [astro-ph.HE]}
  \BibitemShut {NoStop}%
\bibitem [{\citenamefont {{Yang}}\ \emph {et~al.}(2019)\citenamefont {{Yang}},
  \citenamefont {{Bartos}}, \citenamefont {{Haiman}}, \citenamefont {{Kocsis}},
  \citenamefont {{M{\'a}rka}}, \citenamefont {{Stone}},\ and\ \citenamefont
  {{M{\'a}rka}}}]{Yang+19}%
  \BibitemOpen
  \bibfield  {author} {\bibinfo {author} {\bibfnamefont {Y.}~\bibnamefont
  {{Yang}}}, \bibinfo {author} {\bibfnamefont {I.}~\bibnamefont {{Bartos}}},
  \bibinfo {author} {\bibfnamefont {Z.}~\bibnamefont {{Haiman}}}, \bibinfo
  {author} {\bibfnamefont {B.}~\bibnamefont {{Kocsis}}}, \bibinfo {author}
  {\bibfnamefont {Z.}~\bibnamefont {{M{\'a}rka}}}, \bibinfo {author}
  {\bibfnamefont {N.~C.}\ \bibnamefont {{Stone}}},\ and\ \bibinfo {author}
  {\bibfnamefont {S.}~\bibnamefont {{M{\'a}rka}}},\ }\href
  {https://doi.org/10.3847/1538-4357/ab16e3} {\bibfield  {journal} {\bibinfo
  {journal} {\apj}\ }\textbf {\bibinfo {volume} {876}},\ \bibinfo {eid} {122}
  (\bibinfo {year} {2019})},\ \Eprint {https://arxiv.org/abs/1903.01405}
  {arXiv:1903.01405 [astro-ph.HE]} \BibitemShut {NoStop}%
\bibitem [{\citenamefont {{Tagawa}}\ \emph {et~al.}(2021)\citenamefont
  {{Tagawa}}, \citenamefont {{Haiman}}, \citenamefont {{Bartos}}, \citenamefont
  {{Kocsis}},\ and\ \citenamefont {{Omukai}}}]{Tagawa+21}%
  \BibitemOpen
  \bibfield  {author} {\bibinfo {author} {\bibfnamefont {H.}~\bibnamefont
  {{Tagawa}}}, \bibinfo {author} {\bibfnamefont {Z.}~\bibnamefont {{Haiman}}},
  \bibinfo {author} {\bibfnamefont {I.}~\bibnamefont {{Bartos}}}, \bibinfo
  {author} {\bibfnamefont {B.}~\bibnamefont {{Kocsis}}},\ and\ \bibinfo
  {author} {\bibfnamefont {K.}~\bibnamefont {{Omukai}}},\ }\href
  {https://doi.org/10.1093/mnras/stab2315} {\bibfield  {journal} {\bibinfo
  {journal} {\mnras}\ }\textbf {\bibinfo {volume} {507}},\ \bibinfo {pages}
  {3362} (\bibinfo {year} {2021})},\ \Eprint {https://arxiv.org/abs/2104.09510}
  {arXiv:2104.09510 [astro-ph.HE]} \BibitemShut {NoStop}%
\bibitem [{\citenamefont {Vaccaro}\ \emph {et~al.}(2026)\citenamefont
  {Vaccaro}, \citenamefont {Mapelli}, \citenamefont {Trani},\ and\
  \citenamefont {Liu}}]{Vaccaro:2026nkp}%
  \BibitemOpen
  \bibfield  {author} {\bibinfo {author} {\bibfnamefont {M.~P.}\ \bibnamefont
  {Vaccaro}}, \bibinfo {author} {\bibfnamefont {M.}~\bibnamefont {Mapelli}},
  \bibinfo {author} {\bibfnamefont {A.~A.}\ \bibnamefont {Trani}},\ and\
  \bibinfo {author} {\bibfnamefont {B.}~\bibnamefont {Liu}},\ }\href@noop {} {\
   (\bibinfo {year} {2026})},\ \Eprint {https://arxiv.org/abs/2606.10823}
  {arXiv:2606.10823 [astro-ph.GA]} \BibitemShut {NoStop}%
\bibitem [{\citenamefont {O'Leary}\ \emph {et~al.}(2009)\citenamefont
  {O'Leary}, \citenamefont {Kocsis},\ and\ \citenamefont
  {Loeb}}]{OLeary:2008myb}%
  \BibitemOpen
  \bibfield  {author} {\bibinfo {author} {\bibfnamefont {R.~M.}\ \bibnamefont
  {O'Leary}}, \bibinfo {author} {\bibfnamefont {B.}~\bibnamefont {Kocsis}},\
  and\ \bibinfo {author} {\bibfnamefont {A.}~\bibnamefont {Loeb}},\ }\href
  {https://doi.org/10.1111/j.1365-2966.2009.14653.x} {\bibfield  {journal}
  {\bibinfo  {journal} {Mon. Not. Roy. Astron. Soc.}\ }\textbf {\bibinfo
  {volume} {395}},\ \bibinfo {pages} {2127} (\bibinfo {year} {2009})},\ \Eprint
  {https://arxiv.org/abs/0807.2638} {arXiv:0807.2638 [astro-ph]} \BibitemShut
  {NoStop}%
\bibitem [{\citenamefont {Gond{\'a}n}\ \emph {et~al.}(2018)\citenamefont
  {Gond{\'a}n}, \citenamefont {Kocsis}, \citenamefont {Raffai},\ and\
  \citenamefont {Frei}}]{Gondan:2017wzd}%
  \BibitemOpen
  \bibfield  {author} {\bibinfo {author} {\bibfnamefont {L.}~\bibnamefont
  {Gond{\'a}n}}, \bibinfo {author} {\bibfnamefont {B.}~\bibnamefont {Kocsis}},
  \bibinfo {author} {\bibfnamefont {P.}~\bibnamefont {Raffai}},\ and\ \bibinfo
  {author} {\bibfnamefont {Z.}~\bibnamefont {Frei}},\ }\href
  {https://doi.org/10.3847/1538-4357/aabfee} {\bibfield  {journal} {\bibinfo
  {journal} {Astrophys. J.}\ }\textbf {\bibinfo {volume} {860}},\ \bibinfo
  {pages} {5} (\bibinfo {year} {2018})},\ \Eprint
  {https://arxiv.org/abs/1711.09989} {arXiv:1711.09989 [astro-ph.HE]}
  \BibitemShut {NoStop}%
\bibitem [{\citenamefont {Rasskazov}\ and\ \citenamefont
  {Kocsis}(2019)}]{Rasskazov:2019gjw}%
  \BibitemOpen
  \bibfield  {author} {\bibinfo {author} {\bibfnamefont {A.}~\bibnamefont
  {Rasskazov}}\ and\ \bibinfo {author} {\bibfnamefont {B.}~\bibnamefont
  {Kocsis}},\ }\href {https://doi.org/10.3847/1538-4357/ab2c74} {\bibfield
  {journal} {\bibinfo  {journal} {Astrophys. J.}\ }\textbf {\bibinfo {volume}
  {881}},\ \bibinfo {pages} {20} (\bibinfo {year} {2019})},\ \Eprint
  {https://arxiv.org/abs/1902.03242} {arXiv:1902.03242 [astro-ph.HE]}
  \BibitemShut {NoStop}%
\bibitem [{\citenamefont {{Samsing}}\ \emph {et~al.}(2022)\citenamefont
  {{Samsing}}, \citenamefont {{Bartos}}, \citenamefont {{D'Orazio}},
  \citenamefont {{Haiman}}, \citenamefont {{Kocsis}}, \citenamefont {{Leigh}},
  \citenamefont {{Liu}}, \citenamefont {{Pessah}},\ and\ \citenamefont
  {{Tagawa}}}]{Samsing+22}%
  \BibitemOpen
  \bibfield  {author} {\bibinfo {author} {\bibfnamefont {J.}~\bibnamefont
  {{Samsing}}}, \bibinfo {author} {\bibfnamefont {I.}~\bibnamefont {{Bartos}}},
  \bibinfo {author} {\bibfnamefont {D.~J.}\ \bibnamefont {{D'Orazio}}},
  \bibinfo {author} {\bibfnamefont {Z.}~\bibnamefont {{Haiman}}}, \bibinfo
  {author} {\bibfnamefont {B.}~\bibnamefont {{Kocsis}}}, \bibinfo {author}
  {\bibfnamefont {N.~W.~C.}\ \bibnamefont {{Leigh}}}, \bibinfo {author}
  {\bibfnamefont {B.}~\bibnamefont {{Liu}}}, \bibinfo {author} {\bibfnamefont
  {M.~E.}\ \bibnamefont {{Pessah}}},\ and\ \bibinfo {author} {\bibfnamefont
  {H.}~\bibnamefont {{Tagawa}}},\ }\href
  {https://doi.org/10.1038/s41586-021-04333-1} {\bibfield  {journal} {\bibinfo
  {journal} {\nat}\ }\textbf {\bibinfo {volume} {603}},\ \bibinfo {pages} {237}
  (\bibinfo {year} {2022})},\ \Eprint {https://arxiv.org/abs/2010.09765}
  {arXiv:2010.09765 [astro-ph.HE]} \BibitemShut {NoStop}%
\bibitem [{\citenamefont {{Romero-Shaw}}\ \emph {et~al.}(2020)\citenamefont
  {{Romero-Shaw}}, \citenamefont {{Lasky}}, \citenamefont {{Thrane}},\ and\
  \citenamefont {{Calder{\'o}n Bustillo}}}]{RomeroShaw+20}%
  \BibitemOpen
  \bibfield  {author} {\bibinfo {author} {\bibfnamefont {I.}~\bibnamefont
  {{Romero-Shaw}}}, \bibinfo {author} {\bibfnamefont {P.~D.}\ \bibnamefont
  {{Lasky}}}, \bibinfo {author} {\bibfnamefont {E.}~\bibnamefont {{Thrane}}},\
  and\ \bibinfo {author} {\bibfnamefont {J.}~\bibnamefont {{Calder{\'o}n
  Bustillo}}},\ }\href {https://doi.org/10.3847/2041-8213/abbe26} {\bibfield
  {journal} {\bibinfo  {journal} {\apjl}\ }\textbf {\bibinfo {volume} {903}},\
  \bibinfo {eid} {L5} (\bibinfo {year} {2020})},\ \Eprint
  {https://arxiv.org/abs/2009.04771} {arXiv:2009.04771 [astro-ph.HE]}
  \BibitemShut {NoStop}%
\bibitem [{\citenamefont {Romero-Shaw}\ \emph {et~al.}(2022)\citenamefont
  {Romero-Shaw}, \citenamefont {Lasky},\ and\ \citenamefont
  {Thrane}}]{Romero-Shaw:2022xko}%
  \BibitemOpen
  \bibfield  {author} {\bibinfo {author} {\bibfnamefont {I.~M.}\ \bibnamefont
  {Romero-Shaw}}, \bibinfo {author} {\bibfnamefont {P.~D.}\ \bibnamefont
  {Lasky}},\ and\ \bibinfo {author} {\bibfnamefont {E.}~\bibnamefont
  {Thrane}},\ }\href {https://doi.org/10.3847/1538-4357/ac9798} {\bibfield
  {journal} {\bibinfo  {journal} {Astrophys. J.}\ }\textbf {\bibinfo {volume}
  {940}},\ \bibinfo {pages} {171} (\bibinfo {year} {2022})},\ \Eprint
  {https://arxiv.org/abs/2206.14695} {arXiv:2206.14695 [astro-ph.HE]}
  \BibitemShut {NoStop}%
\bibitem [{\citenamefont {{Gupte}}\ \emph {et~al.}(2025)\citenamefont
  {{Gupte}}, \citenamefont {{Ramos-Buades}}, \citenamefont {{Buonanno}},
  \citenamefont {{Gair}}, \citenamefont {{Coleman Miller}}, \citenamefont
  {{Dax}}, \citenamefont {{Green}}, \citenamefont {{P{\"u}rrer}}, \citenamefont
  {{Wildberger}}, \citenamefont {{Macke}}, \citenamefont {{Romero-Shaw}},\ and\
  \citenamefont {{Sch{\"o}lkopf}}}]{Gupte+25}%
  \BibitemOpen
  \bibfield  {author} {\bibinfo {author} {\bibfnamefont {N.}~\bibnamefont
  {{Gupte}}}, \bibinfo {author} {\bibfnamefont {A.}~\bibnamefont
  {{Ramos-Buades}}}, \bibinfo {author} {\bibfnamefont {A.}~\bibnamefont
  {{Buonanno}}}, \bibinfo {author} {\bibfnamefont {J.}~\bibnamefont {{Gair}}},
  \bibinfo {author} {\bibfnamefont {M.}~\bibnamefont {{Coleman Miller}}},
  \bibinfo {author} {\bibfnamefont {M.}~\bibnamefont {{Dax}}}, \bibinfo
  {author} {\bibfnamefont {S.~R.}\ \bibnamefont {{Green}}}, \bibinfo {author}
  {\bibfnamefont {M.}~\bibnamefont {{P{\"u}rrer}}}, \bibinfo {author}
  {\bibfnamefont {J.}~\bibnamefont {{Wildberger}}}, \bibinfo {author}
  {\bibfnamefont {J.}~\bibnamefont {{Macke}}}, \bibinfo {author} {\bibfnamefont
  {I.~M.}\ \bibnamefont {{Romero-Shaw}}},\ and\ \bibinfo {author}
  {\bibfnamefont {B.}~\bibnamefont {{Sch{\"o}lkopf}}},\ }\href
  {https://doi.org/10.1103/vpyp-nvfp} {\bibfield  {journal} {\bibinfo
  {journal} {\prd}\ }\textbf {\bibinfo {volume} {112}},\ \bibinfo {eid}
  {104045} (\bibinfo {year} {2025})},\ \Eprint
  {https://arxiv.org/abs/2404.14286} {arXiv:2404.14286 [gr-qc]} \BibitemShut
  {NoStop}%
\bibitem [{\citenamefont {Zhang}\ and\ \citenamefont
  {Yang}(2020)}]{Zhang:2019eid}%
  \BibitemOpen
  \bibfield  {author} {\bibinfo {author} {\bibfnamefont {J.}~\bibnamefont
  {Zhang}}\ and\ \bibinfo {author} {\bibfnamefont {H.}~\bibnamefont {Yang}},\
  }\href {https://doi.org/10.1103/PhysRevD.101.043020} {\bibfield  {journal}
  {\bibinfo  {journal} {Phys. Rev. D}\ }\textbf {\bibinfo {volume} {101}},\
  \bibinfo {pages} {043020} (\bibinfo {year} {2020})},\ \Eprint
  {https://arxiv.org/abs/1907.13582} {arXiv:1907.13582 [gr-qc]} \BibitemShut
  {NoStop}%
\bibitem [{\citenamefont {Tomaselli}\ \emph {et~al.}(2023)\citenamefont
  {Tomaselli}, \citenamefont {Spieksma},\ and\ \citenamefont
  {Bertone}}]{Tomaselli:2023ysb}%
  \BibitemOpen
  \bibfield  {author} {\bibinfo {author} {\bibfnamefont {G.~M.}\ \bibnamefont
  {Tomaselli}}, \bibinfo {author} {\bibfnamefont {T.~F.~M.}\ \bibnamefont
  {Spieksma}},\ and\ \bibinfo {author} {\bibfnamefont {G.}~\bibnamefont
  {Bertone}},\ }\href {https://doi.org/10.1088/1475-7516/2023/07/070}
  {\bibfield  {journal} {\bibinfo  {journal} {JCAP}\ }\textbf {\bibinfo
  {volume} {07}},\ \bibinfo {pages} {070}},\ \Eprint
  {https://arxiv.org/abs/2305.15460} {arXiv:2305.15460 [gr-qc]} \BibitemShut
  {NoStop}%
\bibitem [{\citenamefont {Cardoso}\ \emph {et~al.}(2018)\citenamefont
  {Cardoso}, \citenamefont {Dias}, \citenamefont {Hartnett}, \citenamefont
  {Middleton}, \citenamefont {Pani},\ and\ \citenamefont
  {Santos}}]{Cardoso:2018tly}%
  \BibitemOpen
  \bibfield  {author} {\bibinfo {author} {\bibfnamefont {V.}~\bibnamefont
  {Cardoso}}, \bibinfo {author} {\bibfnamefont {{\'O}.~J.~C.}\ \bibnamefont
  {Dias}}, \bibinfo {author} {\bibfnamefont {G.~S.}\ \bibnamefont {Hartnett}},
  \bibinfo {author} {\bibfnamefont {M.}~\bibnamefont {Middleton}}, \bibinfo
  {author} {\bibfnamefont {P.}~\bibnamefont {Pani}},\ and\ \bibinfo {author}
  {\bibfnamefont {J.~E.}\ \bibnamefont {Santos}},\ }\href
  {https://doi.org/10.1088/1475-7516/2018/03/043} {\bibfield  {journal}
  {\bibinfo  {journal} {JCAP}\ }\textbf {\bibinfo {volume} {03}},\ \bibinfo
  {pages} {043}},\ \Eprint {https://arxiv.org/abs/1801.01420} {arXiv:1801.01420
  [gr-qc]} \BibitemShut {NoStop}%
\bibitem [{\citenamefont {Dolan}(2007)}]{Dolan:2007mj}%
  \BibitemOpen
  \bibfield  {author} {\bibinfo {author} {\bibfnamefont {S.~R.}\ \bibnamefont
  {Dolan}},\ }\href {https://doi.org/10.1103/PhysRevD.76.084001} {\bibfield
  {journal} {\bibinfo  {journal} {Phys. Rev. D}\ }\textbf {\bibinfo {volume}
  {76}},\ \bibinfo {pages} {084001} (\bibinfo {year} {2007})},\ \Eprint
  {https://arxiv.org/abs/0705.2880} {arXiv:0705.2880 [gr-qc]} \BibitemShut
  {NoStop}%
\bibitem [{\citenamefont {Baumann}\ \emph
  {et~al.}(2019{\natexlab{a}})\citenamefont {Baumann}, \citenamefont {Chia},
  \citenamefont {Stout},\ and\ \citenamefont {ter Haar}}]{Baumann:2019eav}%
  \BibitemOpen
  \bibfield  {author} {\bibinfo {author} {\bibfnamefont {D.}~\bibnamefont
  {Baumann}}, \bibinfo {author} {\bibfnamefont {H.~S.}\ \bibnamefont {Chia}},
  \bibinfo {author} {\bibfnamefont {J.}~\bibnamefont {Stout}},\ and\ \bibinfo
  {author} {\bibfnamefont {L.}~\bibnamefont {ter Haar}},\ }\href
  {https://doi.org/10.1088/1475-7516/2019/12/006} {\bibfield  {journal}
  {\bibinfo  {journal} {JCAP}\ }\textbf {\bibinfo {volume} {12}},\ \bibinfo
  {pages} {006}},\ \Eprint {https://arxiv.org/abs/1908.10370} {arXiv:1908.10370
  [gr-qc]} \BibitemShut {NoStop}%
\bibitem [{\citenamefont {Baryakhtar}\ \emph {et~al.}(2021)\citenamefont
  {Baryakhtar}, \citenamefont {Galanis}, \citenamefont {Lasenby},\ and\
  \citenamefont {Simon}}]{Baryakhtar:2020gao}%
  \BibitemOpen
  \bibfield  {author} {\bibinfo {author} {\bibfnamefont {M.}~\bibnamefont
  {Baryakhtar}}, \bibinfo {author} {\bibfnamefont {M.}~\bibnamefont {Galanis}},
  \bibinfo {author} {\bibfnamefont {R.}~\bibnamefont {Lasenby}},\ and\ \bibinfo
  {author} {\bibfnamefont {O.}~\bibnamefont {Simon}},\ }\href
  {https://doi.org/10.1103/PhysRevD.103.095019} {\bibfield  {journal} {\bibinfo
   {journal} {Phys. Rev. D}\ }\textbf {\bibinfo {volume} {103}},\ \bibinfo
  {pages} {095019} (\bibinfo {year} {2021})},\ \Eprint
  {https://arxiv.org/abs/2011.11646} {arXiv:2011.11646 [hep-ph]} \BibitemShut
  {NoStop}%
\bibitem [{\citenamefont {Peters}\ and\ \citenamefont
  {Mathews}(1963)}]{Peters:1963ux}%
  \BibitemOpen
  \bibfield  {author} {\bibinfo {author} {\bibfnamefont {P.~C.}\ \bibnamefont
  {Peters}}\ and\ \bibinfo {author} {\bibfnamefont {J.}~\bibnamefont
  {Mathews}},\ }\href {https://doi.org/10.1103/PhysRev.131.435} {\bibfield
  {journal} {\bibinfo  {journal} {Phys. Rev.}\ }\textbf {\bibinfo {volume}
  {131}},\ \bibinfo {pages} {435} (\bibinfo {year} {1963})}\BibitemShut
  {NoStop}%
\bibitem [{\citenamefont {{Turner}}(1977)}]{Turner:1977ab}%
  \BibitemOpen
  \bibfield  {author} {\bibinfo {author} {\bibfnamefont {M.}~\bibnamefont
  {{Turner}}},\ }\href {https://doi.org/10.1086/155501} {\bibfield  {journal}
  {\bibinfo  {journal} {\apj}\ }\textbf {\bibinfo {volume} {216}},\ \bibinfo
  {pages} {610} (\bibinfo {year} {1977})}\BibitemShut {NoStop}%
\bibitem [{\citenamefont {Guo}\ \emph {et~al.}(2026)\citenamefont {Guo},
  \citenamefont {Zhong}, \citenamefont {Chen}, \citenamefont {Cardoso},
  \citenamefont {Ikeda},\ and\ \citenamefont {Zhou}}]{Guo:2025pea}%
  \BibitemOpen
  \bibfield  {author} {\bibinfo {author} {\bibfnamefont {Y.}~\bibnamefont
  {Guo}}, \bibinfo {author} {\bibfnamefont {Z.}~\bibnamefont {Zhong}}, \bibinfo
  {author} {\bibfnamefont {Y.}~\bibnamefont {Chen}}, \bibinfo {author}
  {\bibfnamefont {V.}~\bibnamefont {Cardoso}}, \bibinfo {author} {\bibfnamefont
  {T.}~\bibnamefont {Ikeda}},\ and\ \bibinfo {author} {\bibfnamefont
  {L.}~\bibnamefont {Zhou}},\ }\href
  {https://doi.org/10.1088/1475-7516/2026/08/065} {\bibfield  {journal}
  {\bibinfo  {journal} {JCAP}\ }\textbf {\bibinfo {volume} {08}},\ \bibinfo
  {pages} {065}},\ \Eprint {https://arxiv.org/abs/2509.09643} {arXiv:2509.09643
  [gr-qc]} \BibitemShut {NoStop}%
\bibitem [{\citenamefont {{Bahcall}}\ and\ \citenamefont
  {{Wolf}}(1977)}]{Bahcall:1977ab}%
  \BibitemOpen
  \bibfield  {author} {\bibinfo {author} {\bibfnamefont {J.~N.}\ \bibnamefont
  {{Bahcall}}}\ and\ \bibinfo {author} {\bibfnamefont {R.~A.}\ \bibnamefont
  {{Wolf}}},\ }\href {https://doi.org/10.1086/155534} {\bibfield  {journal}
  {\bibinfo  {journal} {\apj}\ }\textbf {\bibinfo {volume} {216}},\ \bibinfo
  {pages} {883} (\bibinfo {year} {1977})}\BibitemShut {NoStop}%
\bibitem [{\citenamefont {Freitag}\ \emph {et~al.}(2006)\citenamefont
  {Freitag}, \citenamefont {Amaro-Seoane},\ and\ \citenamefont
  {Kalogera}}]{Freitag:2006qf}%
  \BibitemOpen
  \bibfield  {author} {\bibinfo {author} {\bibfnamefont {M.}~\bibnamefont
  {Freitag}}, \bibinfo {author} {\bibfnamefont {P.}~\bibnamefont
  {Amaro-Seoane}},\ and\ \bibinfo {author} {\bibfnamefont {V.}~\bibnamefont
  {Kalogera}},\ }\href {https://doi.org/10.1086/506193} {\bibfield  {journal}
  {\bibinfo  {journal} {Astrophys. J.}\ }\textbf {\bibinfo {volume} {649}},\
  \bibinfo {pages} {91} (\bibinfo {year} {2006})},\ \Eprint
  {https://arxiv.org/abs/astro-ph/0603280} {arXiv:astro-ph/0603280}
  \BibitemShut {NoStop}%
\bibitem [{\citenamefont {{Hannah}}\ \emph {et~al.}(2024)\citenamefont
  {{Hannah}}, \citenamefont {{Seth}}, \citenamefont {{Stone}},\ and\
  \citenamefont {{van Velzen}}}]{Hannah+24}%
  \BibitemOpen
  \bibfield  {author} {\bibinfo {author} {\bibfnamefont {C.~H.}\ \bibnamefont
  {{Hannah}}}, \bibinfo {author} {\bibfnamefont {A.~C.}\ \bibnamefont
  {{Seth}}}, \bibinfo {author} {\bibfnamefont {N.~C.}\ \bibnamefont
  {{Stone}}},\ and\ \bibinfo {author} {\bibfnamefont {S.}~\bibnamefont {{van
  Velzen}}},\ }\href {https://doi.org/10.3847/1538-3881/ad630a} {\bibfield
  {journal} {\bibinfo  {journal} {\aj}\ }\textbf {\bibinfo {volume} {168}},\
  \bibinfo {eid} {137} (\bibinfo {year} {2024})},\ \Eprint
  {https://arxiv.org/abs/2407.10911} {arXiv:2407.10911 [astro-ph.GA]}
  \BibitemShut {NoStop}%
\bibitem [{\citenamefont {{Thomas}}\ \emph {et~al.}(2016)\citenamefont
  {{Thomas}}, \citenamefont {{Ma}}, \citenamefont {{McConnell}}, \citenamefont
  {{Greene}}, \citenamefont {{Blakeslee}},\ and\ \citenamefont
  {{Janish}}}]{Thomas+16}%
  \BibitemOpen
  \bibfield  {author} {\bibinfo {author} {\bibfnamefont {J.}~\bibnamefont
  {{Thomas}}}, \bibinfo {author} {\bibfnamefont {C.-P.}\ \bibnamefont {{Ma}}},
  \bibinfo {author} {\bibfnamefont {N.~J.}\ \bibnamefont {{McConnell}}},
  \bibinfo {author} {\bibfnamefont {J.~E.}\ \bibnamefont {{Greene}}}, \bibinfo
  {author} {\bibfnamefont {J.~P.}\ \bibnamefont {{Blakeslee}}},\ and\ \bibinfo
  {author} {\bibfnamefont {R.}~\bibnamefont {{Janish}}},\ }\href
  {https://doi.org/10.1038/nature17197} {\bibfield  {journal} {\bibinfo
  {journal} {\nat}\ }\textbf {\bibinfo {volume} {532}},\ \bibinfo {pages} {340}
  (\bibinfo {year} {2016})},\ \Eprint {https://arxiv.org/abs/1604.01400}
  {arXiv:1604.01400 [astro-ph.GA]} \BibitemShut {NoStop}%
\bibitem [{\citenamefont {Miralda-Escude}\ and\ \citenamefont
  {Gould}(2000)}]{Miralda-Escude:2000kqv}%
  \BibitemOpen
  \bibfield  {author} {\bibinfo {author} {\bibfnamefont {J.}~\bibnamefont
  {Miralda-Escude}}\ and\ \bibinfo {author} {\bibfnamefont {A.}~\bibnamefont
  {Gould}},\ }\href {https://doi.org/10.1086/317837} {\bibfield  {journal}
  {\bibinfo  {journal} {Astrophys. J.}\ }\textbf {\bibinfo {volume} {545}},\
  \bibinfo {pages} {847} (\bibinfo {year} {2000})},\ \Eprint
  {https://arxiv.org/abs/astro-ph/0003269} {arXiv:astro-ph/0003269}
  \BibitemShut {NoStop}%
\bibitem [{\citenamefont {Alexander}\ and\ \citenamefont
  {Hopman}(2009)}]{Alexander:2008tq}%
  \BibitemOpen
  \bibfield  {author} {\bibinfo {author} {\bibfnamefont {T.}~\bibnamefont
  {Alexander}}\ and\ \bibinfo {author} {\bibfnamefont {C.}~\bibnamefont
  {Hopman}},\ }\href {https://doi.org/10.1088/0004-637X/697/2/1861} {\bibfield
  {journal} {\bibinfo  {journal} {Astrophys. J.}\ }\textbf {\bibinfo {volume}
  {697}},\ \bibinfo {pages} {1861} (\bibinfo {year} {2009})},\ \Eprint
  {https://arxiv.org/abs/0808.3150} {arXiv:0808.3150 [astro-ph]} \BibitemShut
  {NoStop}%
\bibitem [{\citenamefont {{{\"O}zel}}\ \emph {et~al.}(2010)\citenamefont
  {{{\"O}zel}}, \citenamefont {{Psaltis}}, \citenamefont {{Narayan}},\ and\
  \citenamefont {{McClintock}}}]{Ozel+10}%
  \BibitemOpen
  \bibfield  {author} {\bibinfo {author} {\bibfnamefont {F.}~\bibnamefont
  {{{\"O}zel}}}, \bibinfo {author} {\bibfnamefont {D.}~\bibnamefont
  {{Psaltis}}}, \bibinfo {author} {\bibfnamefont {R.}~\bibnamefont
  {{Narayan}}},\ and\ \bibinfo {author} {\bibfnamefont {J.~E.}\ \bibnamefont
  {{McClintock}}},\ }\href {https://doi.org/10.1088/0004-637X/725/2/1918}
  {\bibfield  {journal} {\bibinfo  {journal} {\apj}\ }\textbf {\bibinfo
  {volume} {725}},\ \bibinfo {pages} {1918} (\bibinfo {year} {2010})},\ \Eprint
  {https://arxiv.org/abs/1006.2834} {arXiv:1006.2834 [astro-ph.GA]}
  \BibitemShut {NoStop}%
\bibitem [{Note1()}]{Note1}%
  \BibitemOpen
  \bibinfo {note} {The exponent may also be lower due to the top-heavy nature
  of \protect \textit {in situ} star formation in the GN~\cite
  {Lu+13}.}\BibitemShut {Stop}%
\bibitem [{\citenamefont {{Gonglewski}}\ \emph {et~al.}(2026)\citenamefont
  {{Gonglewski}}, \citenamefont {{Secunda}}, \citenamefont {{Mac Low}},
  \citenamefont {{Ford}}, \citenamefont {{McKernan}},\ and\ \citenamefont
  {{Schneider}}}]{Gonglewski+26}%
  \BibitemOpen
  \bibfield  {author} {\bibinfo {author} {\bibfnamefont {K.~L.}\ \bibnamefont
  {{Gonglewski}}}, \bibinfo {author} {\bibfnamefont {A.}~\bibnamefont
  {{Secunda}}}, \bibinfo {author} {\bibfnamefont {M.-M.}\ \bibnamefont {{Mac
  Low}}}, \bibinfo {author} {\bibfnamefont {K.~E.~S.}\ \bibnamefont {{Ford}}},
  \bibinfo {author} {\bibfnamefont {B.}~\bibnamefont {{McKernan}}},\ and\
  \bibinfo {author} {\bibfnamefont {F.~R.~N.}\ \bibnamefont {{Schneider}}},\
  }\href {https://doi.org/10.48550/arXiv.2608.13641} {\bibfield  {journal}
  {\bibinfo  {journal} {arXiv e-prints}\ ,\ \bibinfo {eid} {arXiv:2608.13641}}
  (\bibinfo {year} {2026})},\ \Eprint {https://arxiv.org/abs/2608.13641}
  {arXiv:2608.13641 [astro-ph.GA]} \BibitemShut {NoStop}%
\bibitem [{\citenamefont {Tomaselli}(2025)}]{Tomaselli:2025jfo}%
  \BibitemOpen
  \bibfield  {author} {\bibinfo {author} {\bibfnamefont {G.~M.}\ \bibnamefont
  {Tomaselli}},\ }\href {https://doi.org/10.1103/h3fy-fyrx} {\bibfield
  {journal} {\bibinfo  {journal} {Phys. Rev. D}\ }\textbf {\bibinfo {volume}
  {112}},\ \bibinfo {pages} {063033} (\bibinfo {year} {2025})},\ \Eprint
  {https://arxiv.org/abs/2507.15110} {arXiv:2507.15110 [gr-qc]} \BibitemShut
  {NoStop}%
\bibitem [{\citenamefont {{Bar-Or}}\ \emph {et~al.}(2013)\citenamefont
  {{Bar-Or}}, \citenamefont {{Kupi}},\ and\ \citenamefont
  {{Alexander}}}]{BarOr:2013ab}%
  \BibitemOpen
  \bibfield  {author} {\bibinfo {author} {\bibfnamefont {B.}~\bibnamefont
  {{Bar-Or}}}, \bibinfo {author} {\bibfnamefont {G.}~\bibnamefont {{Kupi}}},\
  and\ \bibinfo {author} {\bibfnamefont {T.}~\bibnamefont {{Alexander}}},\
  }\href {https://doi.org/10.1088/0004-637X/764/1/52} {\bibfield  {journal}
  {\bibinfo  {journal} {\apj}\ }\textbf {\bibinfo {volume} {764}},\ \bibinfo
  {eid} {52} (\bibinfo {year} {2013})},\ \Eprint
  {https://arxiv.org/abs/1209.4594} {arXiv:1209.4594 [astro-ph.GA]}
  \BibitemShut {NoStop}%
\bibitem [{\citenamefont {Aller}\ and\ \citenamefont
  {Richstone}(2002)}]{Aller:2002rp}%
  \BibitemOpen
  \bibfield  {author} {\bibinfo {author} {\bibfnamefont {M.~C.}\ \bibnamefont
  {Aller}}\ and\ \bibinfo {author} {\bibfnamefont {D.}~\bibnamefont
  {Richstone}},\ }\href {https://doi.org/10.1086/344484} {\bibfield  {journal}
  {\bibinfo  {journal} {Astron. J.}\ }\textbf {\bibinfo {volume} {124}},\
  \bibinfo {pages} {3035} (\bibinfo {year} {2002})},\ \Eprint
  {https://arxiv.org/abs/astro-ph/0210573} {arXiv:astro-ph/0210573}
  \BibitemShut {NoStop}%
\bibitem [{\citenamefont {{Stone}}\ and\ \citenamefont {{van
  Velzen}}(2016)}]{vanVelzen16}%
  \BibitemOpen
  \bibfield  {author} {\bibinfo {author} {\bibfnamefont {N.~C.}\ \bibnamefont
  {{Stone}}}\ and\ \bibinfo {author} {\bibfnamefont {S.}~\bibnamefont {{van
  Velzen}}},\ }\href {https://doi.org/10.3847/2041-8205/825/1/L14} {\bibfield
  {journal} {\bibinfo  {journal} {\apjl}\ }\textbf {\bibinfo {volume} {825}},\
  \bibinfo {eid} {L14} (\bibinfo {year} {2016})},\ \Eprint
  {https://arxiv.org/abs/1604.02056} {arXiv:1604.02056 [astro-ph.GA]}
  \BibitemShut {NoStop}%
\bibitem [{\citenamefont {Aghanim}\ \emph {et~al.}(2020)\citenamefont {Aghanim}
  \emph {et~al.}}]{Planck:2018vyg}%
  \BibitemOpen
  \bibfield  {author} {\bibinfo {author} {\bibfnamefont {N.}~\bibnamefont
  {Aghanim}} \emph {et~al.} (\bibinfo {collaboration} {Planck}),\ }\href
  {https://doi.org/10.1051/0004-6361/201833910} {\bibfield  {journal} {\bibinfo
   {journal} {Astron. Astrophys.}\ }\textbf {\bibinfo {volume} {641}},\
  \bibinfo {pages} {A6} (\bibinfo {year} {2020})},\ \bibinfo {note} {[Erratum:
  Astron.Astrophys. 652, C4 (2021)]},\ \Eprint
  {https://arxiv.org/abs/1807.06209} {arXiv:1807.06209 [astro-ph.CO]}
  \BibitemShut {NoStop}%
\bibitem [{\citenamefont {{Lu}}\ \emph {et~al.}(2013)\citenamefont {{Lu}},
  \citenamefont {{Do}}, \citenamefont {{Ghez}}, \citenamefont {{Morris}},
  \citenamefont {{Yelda}},\ and\ \citenamefont {{Matthews}}}]{Lu+13}%
  \BibitemOpen
  \bibfield  {author} {\bibinfo {author} {\bibfnamefont {J.~R.}\ \bibnamefont
  {{Lu}}}, \bibinfo {author} {\bibfnamefont {T.}~\bibnamefont {{Do}}}, \bibinfo
  {author} {\bibfnamefont {A.~M.}\ \bibnamefont {{Ghez}}}, \bibinfo {author}
  {\bibfnamefont {M.~R.}\ \bibnamefont {{Morris}}}, \bibinfo {author}
  {\bibfnamefont {S.}~\bibnamefont {{Yelda}}},\ and\ \bibinfo {author}
  {\bibfnamefont {K.}~\bibnamefont {{Matthews}}},\ }\href
  {https://doi.org/10.1088/0004-637X/764/2/155} {\bibfield  {journal} {\bibinfo
   {journal} {\apj}\ }\textbf {\bibinfo {volume} {764}},\ \bibinfo {eid} {155}
  (\bibinfo {year} {2013})},\ \Eprint {https://arxiv.org/abs/1301.0540}
  {arXiv:1301.0540 [astro-ph.SR]} \BibitemShut {NoStop}%
\bibitem [{\citenamefont {{Zel'Dovich}}(1971)}]{Zeld1971}%
  \BibitemOpen
  \bibfield  {author} {\bibinfo {author} {\bibfnamefont {Y.~B.}\ \bibnamefont
  {{Zel'Dovich}}},\ }\href@noop {} {\bibfield  {journal} {\bibinfo  {journal}
  {JETP Lett.}\ }\textbf {\bibinfo {volume} {14}},\ \bibinfo {pages} {180}
  (\bibinfo {year} {1971})}\BibitemShut {NoStop}%
\bibitem [{\citenamefont {Unruh}(1974)}]{Unruh:1974bw}%
  \BibitemOpen
  \bibfield  {author} {\bibinfo {author} {\bibfnamefont {W.~G.}\ \bibnamefont
  {Unruh}},\ }\href {https://doi.org/10.1103/PhysRevD.10.3194} {\bibfield
  {journal} {\bibinfo  {journal} {Phys. Rev. D}\ }\textbf {\bibinfo {volume}
  {10}},\ \bibinfo {pages} {3194} (\bibinfo {year} {1974})}\BibitemShut
  {NoStop}%
\bibitem [{\citenamefont {Yoshino}\ and\ \citenamefont
  {Kodama}(2012)}]{Yoshino:2012kn}%
  \BibitemOpen
  \bibfield  {author} {\bibinfo {author} {\bibfnamefont {H.}~\bibnamefont
  {Yoshino}}\ and\ \bibinfo {author} {\bibfnamefont {H.}~\bibnamefont
  {Kodama}},\ }\href {https://doi.org/10.1143/PTP.128.153} {\bibfield
  {journal} {\bibinfo  {journal} {Prog. Theor. Phys.}\ }\textbf {\bibinfo
  {volume} {128}},\ \bibinfo {pages} {153} (\bibinfo {year} {2012})},\ \Eprint
  {https://arxiv.org/abs/1203.5070} {arXiv:1203.5070 [gr-qc]} \BibitemShut
  {NoStop}%
\bibitem [{\citenamefont {Baumann}\ \emph
  {et~al.}(2019{\natexlab{b}})\citenamefont {Baumann}, \citenamefont {Chia},\
  and\ \citenamefont {Porto}}]{Baumann:2018vus}%
  \BibitemOpen
  \bibfield  {author} {\bibinfo {author} {\bibfnamefont {D.}~\bibnamefont
  {Baumann}}, \bibinfo {author} {\bibfnamefont {H.~S.}\ \bibnamefont {Chia}},\
  and\ \bibinfo {author} {\bibfnamefont {R.~A.}\ \bibnamefont {Porto}},\ }\href
  {https://doi.org/10.1103/PhysRevD.99.044001} {\bibfield  {journal} {\bibinfo
  {journal} {Phys. Rev. D}\ }\textbf {\bibinfo {volume} {99}},\ \bibinfo
  {pages} {044001} (\bibinfo {year} {2019}{\natexlab{b}})},\ \Eprint
  {https://arxiv.org/abs/1804.03208} {arXiv:1804.03208 [gr-qc]} \BibitemShut
  {NoStop}%
\bibitem [{\citenamefont {{Neumayer}}\ \emph {et~al.}(2020)\citenamefont
  {{Neumayer}}, \citenamefont {{Seth}},\ and\ \citenamefont
  {{B{\"o}ker}}}]{Neumayer+20}%
  \BibitemOpen
  \bibfield  {author} {\bibinfo {author} {\bibfnamefont {N.}~\bibnamefont
  {{Neumayer}}}, \bibinfo {author} {\bibfnamefont {A.}~\bibnamefont {{Seth}}},\
  and\ \bibinfo {author} {\bibfnamefont {T.}~\bibnamefont {{B{\"o}ker}}},\
  }\href {https://doi.org/10.1007/s00159-020-00125-0} {\bibfield  {journal}
  {\bibinfo  {journal} {\aapr}\ }\textbf {\bibinfo {volume} {28}},\ \bibinfo
  {eid} {4} (\bibinfo {year} {2020})},\ \Eprint
  {https://arxiv.org/abs/2001.03626} {arXiv:2001.03626 [astro-ph.GA]}
  \BibitemShut {NoStop}%
\bibitem [{\citenamefont {Bahcall}\ and\ \citenamefont
  {Wolf}(1976)}]{Bahcall:1976aa}%
  \BibitemOpen
  \bibfield  {author} {\bibinfo {author} {\bibfnamefont {J.~N.}\ \bibnamefont
  {Bahcall}}\ and\ \bibinfo {author} {\bibfnamefont {R.~A.}\ \bibnamefont
  {Wolf}},\ }\href {https://doi.org/10.1086/154711} {\bibfield  {journal}
  {\bibinfo  {journal} {Astrophys. J.}\ }\textbf {\bibinfo {volume} {209}},\
  \bibinfo {pages} {214} (\bibinfo {year} {1976})}\BibitemShut {NoStop}%
\bibitem [{\citenamefont {{Cohn}}\ and\ \citenamefont
  {{Kulsrud}}(1978)}]{CohnKulsrud78}%
  \BibitemOpen
  \bibfield  {author} {\bibinfo {author} {\bibfnamefont {H.}~\bibnamefont
  {{Cohn}}}\ and\ \bibinfo {author} {\bibfnamefont {R.~M.}\ \bibnamefont
  {{Kulsrud}}},\ }\href {https://doi.org/10.1086/156685} {\bibfield  {journal}
  {\bibinfo  {journal} {\apj}\ }\textbf {\bibinfo {volume} {226}},\ \bibinfo
  {pages} {1087} (\bibinfo {year} {1978})}\BibitemShut {NoStop}%
\bibitem [{\citenamefont {{Lezhnin}}\ and\ \citenamefont
  {{Vasiliev}}(2015)}]{LezhninVasiliev15}%
  \BibitemOpen
  \bibfield  {author} {\bibinfo {author} {\bibfnamefont {K.}~\bibnamefont
  {{Lezhnin}}}\ and\ \bibinfo {author} {\bibfnamefont {E.}~\bibnamefont
  {{Vasiliev}}},\ }\href {https://doi.org/10.1088/2041-8205/808/1/L5}
  {\bibfield  {journal} {\bibinfo  {journal} {\apjl}\ }\textbf {\bibinfo
  {volume} {808}},\ \bibinfo {eid} {L5} (\bibinfo {year} {2015})},\ \Eprint
  {https://arxiv.org/abs/1506.01717} {arXiv:1506.01717 [astro-ph.GA]}
  \BibitemShut {NoStop}%
\bibitem [{\citenamefont {Keshet}\ \emph {et~al.}(2009)\citenamefont {Keshet},
  \citenamefont {Hopman},\ and\ \citenamefont {Alexander}}]{Keshet:2009vu}%
  \BibitemOpen
  \bibfield  {author} {\bibinfo {author} {\bibfnamefont {U.}~\bibnamefont
  {Keshet}}, \bibinfo {author} {\bibfnamefont {C.}~\bibnamefont {Hopman}},\
  and\ \bibinfo {author} {\bibfnamefont {T.}~\bibnamefont {Alexander}},\ }\href
  {https://doi.org/10.1088/0004-637X/698/1/L64} {\bibfield  {journal} {\bibinfo
   {journal} {Astrophys. J. Lett.}\ }\textbf {\bibinfo {volume} {698}},\
  \bibinfo {pages} {L64} (\bibinfo {year} {2009})},\ \Eprint
  {https://arxiv.org/abs/0901.4343} {arXiv:0901.4343 [astro-ph.GA]}
  \BibitemShut {NoStop}%
\bibitem [{\citenamefont {{French}}\ \emph {et~al.}(2016)\citenamefont
  {{French}}, \citenamefont {{Arcavi}},\ and\ \citenamefont
  {{Zabludoff}}}]{French16}%
  \BibitemOpen
  \bibfield  {author} {\bibinfo {author} {\bibfnamefont {K.~D.}\ \bibnamefont
  {{French}}}, \bibinfo {author} {\bibfnamefont {I.}~\bibnamefont {{Arcavi}}},\
  and\ \bibinfo {author} {\bibfnamefont {A.}~\bibnamefont {{Zabludoff}}},\
  }\href {https://doi.org/10.3847/2041-8205/818/1/L21} {\bibfield  {journal}
  {\bibinfo  {journal} {\apjl}\ }\textbf {\bibinfo {volume} {818}},\ \bibinfo
  {eid} {L21} (\bibinfo {year} {2016})},\ \Eprint
  {https://arxiv.org/abs/1601.04705} {arXiv:1601.04705 [astro-ph.GA]}
  \BibitemShut {NoStop}%
\bibitem [{\citenamefont {{Wevers}}\ \emph {et~al.}(2024)\citenamefont
  {{Wevers}}, \citenamefont {{French}}, \citenamefont {{Zabludoff}},
  \citenamefont {{Fischer}}, \citenamefont {{Rowlands}}, \citenamefont
  {{Guolo}}, \citenamefont {{Dalla Barba}}, \citenamefont {{Arcodia}},
  \citenamefont {{Berton}}, \citenamefont {{Bian}}, \citenamefont {{Linial}},
  \citenamefont {{Miniutti}},\ and\ \citenamefont {{Pasham}}}]{Wevers24}%
  \BibitemOpen
  \bibfield  {author} {\bibinfo {author} {\bibfnamefont {T.}~\bibnamefont
  {{Wevers}}}, \bibinfo {author} {\bibfnamefont {K.~D.}\ \bibnamefont
  {{French}}}, \bibinfo {author} {\bibfnamefont {A.~I.}\ \bibnamefont
  {{Zabludoff}}}, \bibinfo {author} {\bibfnamefont {T.~C.}\ \bibnamefont
  {{Fischer}}}, \bibinfo {author} {\bibfnamefont {K.}~\bibnamefont
  {{Rowlands}}}, \bibinfo {author} {\bibfnamefont {M.}~\bibnamefont {{Guolo}}},
  \bibinfo {author} {\bibfnamefont {B.}~\bibnamefont {{Dalla Barba}}}, \bibinfo
  {author} {\bibfnamefont {R.}~\bibnamefont {{Arcodia}}}, \bibinfo {author}
  {\bibfnamefont {M.}~\bibnamefont {{Berton}}}, \bibinfo {author}
  {\bibfnamefont {F.}~\bibnamefont {{Bian}}}, \bibinfo {author} {\bibfnamefont
  {I.}~\bibnamefont {{Linial}}}, \bibinfo {author} {\bibfnamefont
  {G.}~\bibnamefont {{Miniutti}}},\ and\ \bibinfo {author} {\bibfnamefont
  {D.~R.}\ \bibnamefont {{Pasham}}},\ }\href
  {https://doi.org/10.3847/2041-8213/ad5f1b} {\bibfield  {journal} {\bibinfo
  {journal} {\apjl}\ }\textbf {\bibinfo {volume} {970}},\ \bibinfo {eid} {L23}
  (\bibinfo {year} {2024})},\ \Eprint {https://arxiv.org/abs/2406.02678}
  {arXiv:2406.02678 [astro-ph.HE]} \BibitemShut {NoStop}%
\bibitem [{\citenamefont {Abac}\ \emph
  {et~al.}(2026{\natexlab{d}})\citenamefont {Abac} \emph
  {et~al.}}]{LIGOScientific:2026sit}%
  \BibitemOpen
  \bibfield  {author} {\bibinfo {author} {\bibfnamefont {N.}~\bibnamefont
  {Abac}} \emph {et~al.} (\bibinfo {collaboration} {LIGO Scientific, VIRGO,
  KAGRA}),\ }\href@noop {} {\  (\bibinfo {year} {2026}{\natexlab{d}})},\
  \Eprint {https://arxiv.org/abs/2605.27223} {arXiv:2605.27223 [gr-qc]}
  \BibitemShut {NoStop}%
\bibitem [{\citenamefont {{Preto}}\ and\ \citenamefont
  {{Amaro-Seoane}}(2010)}]{Preto+10}%
  \BibitemOpen
  \bibfield  {author} {\bibinfo {author} {\bibfnamefont {M.}~\bibnamefont
  {{Preto}}}\ and\ \bibinfo {author} {\bibfnamefont {P.}~\bibnamefont
  {{Amaro-Seoane}}},\ }\href {https://doi.org/10.1088/2041-8205/708/1/L42}
  {\bibfield  {journal} {\bibinfo  {journal} {\apjl}\ }\textbf {\bibinfo
  {volume} {708}},\ \bibinfo {pages} {L42} (\bibinfo {year} {2010})},\ \Eprint
  {https://arxiv.org/abs/0910.3206} {arXiv:0910.3206 [astro-ph.GA]}
  \BibitemShut {NoStop}%
\bibitem [{\citenamefont {{Linial}}\ and\ \citenamefont
  {{Sari}}(2022)}]{LinialSari22}%
  \BibitemOpen
  \bibfield  {author} {\bibinfo {author} {\bibfnamefont {I.}~\bibnamefont
  {{Linial}}}\ and\ \bibinfo {author} {\bibfnamefont {R.}~\bibnamefont
  {{Sari}}},\ }\href {https://doi.org/10.3847/1538-4357/ac9bfd} {\bibfield
  {journal} {\bibinfo  {journal} {\apj}\ }\textbf {\bibinfo {volume} {940}},\
  \bibinfo {eid} {101} (\bibinfo {year} {2022})},\ \Eprint
  {https://arxiv.org/abs/2206.14817} {arXiv:2206.14817 [astro-ph.GA]}
  \BibitemShut {NoStop}%
\bibitem [{\citenamefont {{Spera}}\ and\ \citenamefont
  {{Mapelli}}(2017)}]{SperaMapelli17}%
  \BibitemOpen
  \bibfield  {author} {\bibinfo {author} {\bibfnamefont {M.}~\bibnamefont
  {{Spera}}}\ and\ \bibinfo {author} {\bibfnamefont {M.}~\bibnamefont
  {{Mapelli}}},\ }\href {https://doi.org/10.1093/mnras/stx1576} {\bibfield
  {journal} {\bibinfo  {journal} {\mnras}\ }\textbf {\bibinfo {volume} {470}},\
  \bibinfo {pages} {4739} (\bibinfo {year} {2017})},\ \Eprint
  {https://arxiv.org/abs/1706.06109} {arXiv:1706.06109 [astro-ph.SR]}
  \BibitemShut {NoStop}%
\bibitem [{\citenamefont {{Ugolini}}\ \emph {et~al.}(2025)\citenamefont
  {{Ugolini}}, \citenamefont {{Limongi}}, \citenamefont {{Schneider}},
  \citenamefont {{Chieffi}}, \citenamefont {{Di Carlo}},\ and\ \citenamefont
  {{Spera}}}]{Ugolini+25}%
  \BibitemOpen
  \bibfield  {author} {\bibinfo {author} {\bibfnamefont {C.}~\bibnamefont
  {{Ugolini}}}, \bibinfo {author} {\bibfnamefont {M.}~\bibnamefont
  {{Limongi}}}, \bibinfo {author} {\bibfnamefont {R.}~\bibnamefont
  {{Schneider}}}, \bibinfo {author} {\bibfnamefont {A.}~\bibnamefont
  {{Chieffi}}}, \bibinfo {author} {\bibfnamefont {U.~N.}\ \bibnamefont {{Di
  Carlo}}},\ and\ \bibinfo {author} {\bibfnamefont {M.}~\bibnamefont
  {{Spera}}},\ }\href {https://doi.org/10.1051/0004-6361/202451483} {\bibfield
  {journal} {\bibinfo  {journal} {\aap}\ }\textbf {\bibinfo {volume} {695}},\
  \bibinfo {eid} {A122} (\bibinfo {year} {2025})},\ \Eprint
  {https://arxiv.org/abs/2501.18689} {arXiv:2501.18689 [astro-ph.HE]}
  \BibitemShut {NoStop}%
\bibitem [{\citenamefont {{Rakavy}}\ and\ \citenamefont
  {{Shaviv}}(1967)}]{RakavyShaviv67}%
  \BibitemOpen
  \bibfield  {author} {\bibinfo {author} {\bibfnamefont {G.}~\bibnamefont
  {{Rakavy}}}\ and\ \bibinfo {author} {\bibfnamefont {G.}~\bibnamefont
  {{Shaviv}}},\ }\href {https://doi.org/10.1086/149204} {\bibfield  {journal}
  {\bibinfo  {journal} {\apj}\ }\textbf {\bibinfo {volume} {148}},\ \bibinfo
  {pages} {803} (\bibinfo {year} {1967})}\BibitemShut {NoStop}%
\bibitem [{\citenamefont {{Barkat}}\ \emph {et~al.}(1967)\citenamefont
  {{Barkat}}, \citenamefont {{Rakavy}},\ and\ \citenamefont
  {{Sack}}}]{Barkat+67}%
  \BibitemOpen
  \bibfield  {author} {\bibinfo {author} {\bibfnamefont {Z.}~\bibnamefont
  {{Barkat}}}, \bibinfo {author} {\bibfnamefont {G.}~\bibnamefont {{Rakavy}}},\
  and\ \bibinfo {author} {\bibfnamefont {N.}~\bibnamefont {{Sack}}},\ }\href
  {https://doi.org/10.1103/PhysRevLett.18.379} {\bibfield  {journal} {\bibinfo
  {journal} {\prl}\ }\textbf {\bibinfo {volume} {18}},\ \bibinfo {pages} {379}
  (\bibinfo {year} {1967})}\BibitemShut {NoStop}%
\bibitem [{\citenamefont {{Farmer}}\ \emph {et~al.}(2019)\citenamefont
  {{Farmer}}, \citenamefont {{Renzo}}, \citenamefont {{de Mink}}, \citenamefont
  {{Marchant}},\ and\ \citenamefont {{Justham}}}]{Farmer+19}%
  \BibitemOpen
  \bibfield  {author} {\bibinfo {author} {\bibfnamefont {R.}~\bibnamefont
  {{Farmer}}}, \bibinfo {author} {\bibfnamefont {M.}~\bibnamefont {{Renzo}}},
  \bibinfo {author} {\bibfnamefont {S.~E.}\ \bibnamefont {{de Mink}}}, \bibinfo
  {author} {\bibfnamefont {P.}~\bibnamefont {{Marchant}}},\ and\ \bibinfo
  {author} {\bibfnamefont {S.}~\bibnamefont {{Justham}}},\ }\href
  {https://doi.org/10.3847/1538-4357/ab518b} {\bibfield  {journal} {\bibinfo
  {journal} {\apj}\ }\textbf {\bibinfo {volume} {887}},\ \bibinfo {eid} {53}
  (\bibinfo {year} {2019})},\ \Eprint {https://arxiv.org/abs/1910.12874}
  {arXiv:1910.12874 [astro-ph.SR]} \BibitemShut {NoStop}%
\bibitem [{\citenamefont {{Gerosa}}\ and\ \citenamefont
  {{Fishbach}}(2021)}]{Gerosa21}%
  \BibitemOpen
  \bibfield  {author} {\bibinfo {author} {\bibfnamefont {D.}~\bibnamefont
  {{Gerosa}}}\ and\ \bibinfo {author} {\bibfnamefont {M.}~\bibnamefont
  {{Fishbach}}},\ }\href {https://doi.org/10.1038/s41550-021-01398-w}
  {\bibfield  {journal} {\bibinfo  {journal} {Nature Astronomy}\ }\textbf
  {\bibinfo {volume} {5}},\ \bibinfo {pages} {749} (\bibinfo {year} {2021})},\
  \Eprint {https://arxiv.org/abs/2105.03439} {arXiv:2105.03439 [astro-ph.HE]}
  \BibitemShut {NoStop}%
\bibitem [{\citenamefont {{Levin}}\ and\ \citenamefont
  {{Beloborodov}}(2003)}]{Levin03}%
  \BibitemOpen
  \bibfield  {author} {\bibinfo {author} {\bibfnamefont {Y.}~\bibnamefont
  {{Levin}}}\ and\ \bibinfo {author} {\bibfnamefont {A.~M.}\ \bibnamefont
  {{Beloborodov}}},\ }\href {https://doi.org/10.1086/376675} {\bibfield
  {journal} {\bibinfo  {journal} {\apjl}\ }\textbf {\bibinfo {volume} {590}},\
  \bibinfo {pages} {L33} (\bibinfo {year} {2003})},\ \Eprint
  {https://arxiv.org/abs/astro-ph/0303436} {arXiv:astro-ph/0303436 [astro-ph]}
  \BibitemShut {NoStop}%
\end{thebibliography}%
